\documentclass[preprint,aps,prc,superscriptaddress,showpacs,nofootinbib,secnumarabic,floatfix]{revtex4-1}

\usepackage{graphicx}
\usepackage{bm}
\usepackage{color}
\usepackage{gensymb}
\usepackage{amsmath}
\usepackage{slashed}
\usepackage{appendix}
\usepackage{hyperref}

\newcommand{\ltapprox}{\protect\raisebox{-0.5ex}{$\:\stackrel{\textstyle <}
	{\approx}\:$}}
\newcommand{\gtapprox}{\protect\raisebox{-0.5ex}{$\:\stackrel{\textstyle >}
	{\approx}\:$}}

\newcommand{\bvec}[1]{\ensuremath{\boldsymbol{#1}}}

\begin{document}

\title{Dilepton production in microscopic transport theory with in-medium $\rho$-meson
       spectral function}  
       
\newcommand{\JLU}{Institut f\"ur Theoretische Physik, Justus-Liebig-Universit\"at, 35392 Giessen, Germany}
\newcommand{\HFHF}{Helmholtz Research Academy Hesse for FAIR (HFHF), Campus Giessen, 35392 Giessen, Germany}

\author{A.B. Larionov}
\email{Corresponding author:\\[-6pt] Alexei.Larionov@theo.physik.uni-giessen.de}
\affiliation{\JLU}

\author{U. Mosel}
\affiliation{\JLU}

\author{L. von Smekal}
\affiliation{\JLU}
\affiliation{\HFHF}

\begin{abstract}
  We use the microscopic GiBUU transport model to calculate dilepton ($e^+e^-$) production in heavy-ion collisions at SIS18 energies focusing on the effect of collisional broadening of the $\rho$-meson.
  The collisional width of the $\rho$-meson at finite temperature and baryon density in nuclear matter is calculated on the basis of the collision integral of the GiBUU model.
  A systematic comparison with HADES data on dilepton production in heavy-ion collisions is performed. 
  The collisional broadening of the $\rho$ improves the agreement between theory and experiment for the dilepton invariant-mass distributions near the $\rho$ pole mass and for the excess radiation in Au+Au at $1.23 A$~GeV.
  We furthermore show that some remaining underprediction of the experimental dilepton spectra in C+C at $1 A$~GeV and Au+Au at $1.23 A$~GeV
  at intermediate invariant masses $0.2-0.4$~GeV can be accounted for by adjusting the $pn$ bremsstrahlung cross section in a way to agree with the inclusive dilepton spectrum from $dp$ collisions at $1.25 A$~GeV.  
\end{abstract}

\maketitle

\tableofcontents

\section{Introduction}
\label{intro}
The study of in-medium properties of hadrons has been an active field of research over the last 30 years. The EMC experiment had already shown that the electromagnetic structure functions of nucleons change when these are bound in a nucleus. Also the spectral functions of bound nucleons differ from those of free ones thus reflecting the complicated in-medium interactions with other nucleons. On the basis of QCD sum rules Hatsuda and Lee \cite{Hatsuda:1991ez} predicted that the masses of vector mesons should drop significantly inside the nuclear medium, a prediction that agreed with a similar one by Brown and Rho \cite{Brown:1991kk}. In these predictions the mass drop was due to the disappearance of the $q\bar q$ scalar condensate with increasing nucleon density.
On the other hand, it has been shown that the QCD sum rules could be fulfilled as well by a significant broadening of the vector-meson spectral function in the medium \cite{Leupold:1997dg,Leupold:1998bt} without any significant mass shift. 
This agrees with calculations based on the generalized Nambu-Jona-Lasinio model which predicted that the masses and coupling constants of the vector $\rho$ and $\omega$ mesons stay constant up to the critical density \cite{Bernard:1988db}.

The measurements of the low-mass $e^+e^-$ spectra in Pb+Au collisions at $40 A$~GeV
\cite{Adamova:2002kf} and $158 A$~GeV \cite{Agakichiev:2005ai} at CERN/SPS did not allow to make a clear distinction between the in-medium mass shift and
a possible broadening of the $\rho$ meson. The later precision measurements of the low-mass $\mu^+ \mu^-$ spectra in In+In collisions at $158 A$~GeV by NA60 \cite{Arnaldi:2006jq,Arnaldi:2008fw} yielded definite evidence for a $\rho$ broadening without any noticeable mass shift which has been successfully described
within the thermal fireball model of dimuon radiation \cite{vanHees:2006ng}. 

The in-medium spectral function of the $\rho$-meson has been calculated in quite different approaches. In Refs.~\cite{Rapp:1997fs,Peters:1997va,Post:2000qi,Post:2003hu,Muehlich:2006nn} purely hadronic resonance models were developed that do not include
chiral symmetry restoration in the nuclear medium but could describe the observed broadening reasonably well. On the other hand, in Refs.~\cite{Jung:2016yxl,Jung:2019nnr} a chirally gauged linear sigma model with quarks was used to calculate chirally consistent $\rho$ and $a_1$ spectral functions from analytically continued Functional Renormalization Group (aFRG) flow equations at finite temperature and density. These studies demonstrated the degeneracy of the spectral functions of the chiral partners in a way that further supports the $\rho$ broadening essentially without mass shift, as chiral symmetry gets gradually restored at finite temperature and, in particular, in the vicinity of a chiral critical endpoint at finite chemical potential. The outstanding theoretical question then was if experiments could be used to distinguish between explanations of the observed broadening in terms of collisional broadening on one hand, and chiral symmetry restoration, on the other.

Early on dileptons have been used as probes for these in-medium changes of vector mesons because dilepton ($e^+ e^-$ or $\mu^+\mu^-$) decays of hadrons are not distorted by hadronic final-state interactions and thus open a window to study the decays of short-lived hadronic resonances inside the nuclear medium. These experiments are summarized and reviewed in \cite{Rapp:1999ej,Hayano:2008vn,Leupold:2009kz} and most recently in \cite{Salabura:2020tou}.

Measuring $e^+ e^-$ pairs from heavy-ion collisions is in the focus of the experimental program of the HADES collaboration
\cite{Agakichiev:2006tg,Agakishiev:2007ts,Agakishiev:2011vf,Adamczewski-Musch:2019byl}.
To disentangle the various contributions to the measured dilepton spectra remains a major challenge in the extraction and interpretation of dilepton signals from nuclear matter, however. 
The decays of long-lived particles, most importantly the $\pi^0$ and $\eta$ Dalitz decays ($\pi^0, \eta \to \gamma e^+ e^-$), which occur long after the breakup of the compressed nuclear configuration, are relatively well known experimentally. More difficult is the evaluation of the bremsstrahlung, $pn \to pn e^+ e^-$ and $\pi^\pm N \to \pi^\pm e^+ e^-$, where one has to rely mostly on theory. The two most prominent dilepton signals from nuclear matter are the  $P_{33}$ $\Delta(1232)$ Dalitz decay, $\Delta \to N e^+ e^-$, and the $\rho$-meson direct decay, $\rho \to e^+ e^-$. The latter is of special interest due to the possibility to probe through this decay the $\rho$-meson spectral function in the nuclear medium.

The HADES results seem to indicate a significant broadening of the $\rho$ meson in Ar+KCl at $1.756 A$~GeV and Au+Au at $1.23 A$~GeV. On the basis of a so-called
coarse-grained transport model \cite{Huovinen:2002im,Endres:2015fna,Galatyuk:2015pkq,Staudenmaier:2017vtq}, in which local thermal equilibrium is assumed,
it was speculated that these new experimental results could signal the onset of chiral symmetry restoration \cite{Seck:2017zjr}.

In the present paper we instead  perform microscopic transport simulations of the dilepton production in heavy-ion collisions at SIS18 energies, without invoking thermal equilibrium and thermal radiation which can be questionable at these rather low energies \cite{Lang:1991qa}. Also, no quark-gluon degrees of freedom or explicit chiral symmetry effects are contained in the present calculations, so that a comparison with data can give some insight into the question if dilepton production data can indeed be interpreted in terms of a chiral symmetry restoration, as suggested, e.g., in \cite{Adamczewski-Musch:2019byl}, or 'the unleashing of quark-gluon degrees of freedom' \cite{Rapp:2019rr}.
The calculations are based on the Giessen Boltzmann-Uehling-Uhlenbeck (GiBUU) microscopic transport model \cite{Buss:2011mx}. The focus of our present study is on the effect of collisional broadening of the $\rho$-meson. We calculate the collisional width of the $\rho$-meson in excited nuclear matter by using the collision term of the transport equation. This width is then added to the free $\rho$ decay width and used to evaluate the $\rho$ spectral function in nuclear matter that is included in the transport simulations. 

We analyse the detailed composition of the dilepton spectra, and present the time evolution of the different components. We discuss the interplay between the collisional broadening of the $\rho$-meson and its off-shell transport. We provide  a systematic comparison of the GiBUU calculations with available HADES data for the invariant mass, rapidity, and transverse momentum distributions of the dileptons, and we also present an analysis of the dilepton excess radiation. 

The structure of our paper is as follows: In Sec.~\ref{model} we briefly describe the GiBUU transport model with particular emphasis on the off-shell propagation of the $\rho$ meson and the dilepton production channels. Sec.~\ref{rhomed} contains the formalism used to describe the $\rho$ spectral function. We first demonstrate how the spectral function emerges in the splitting of production and decay processes of the $\rho$ meson. We then discuss the collisional broadening of the $\rho$ meson caused by the resonance production on the nucleons of the Fermi sea, $\rho N \to R$. 
We present our results from the GiBUU transport simulations in Sec.~\ref{results}, starting with the time evolution of the density, the temperature, the invariant mass distribution of the $\rho$-meson, and the different components of the dilepton invariant mass spectrum. We then compare our calculations with HADES data
for the dilepton observables in $p+p$ collisions at beam energies of 1.25~GeV, 2.2~GeV and 3.5~GeV, $d+p$ at $1.25 A$~GeV, C+C at $1 A$~GeV and $2 A$~GeV, Ar+KCl at $1.76 A$~GeV,
and Au+Au at $1.23 A$~GeV. For the Au+Au system, we also compare the calculated particle multiplicities with experimental data.
The predictions for the dilepton invariant mass spectrum from Ag+Ag at $1.58 A$~GeV are given.
Finally, our summary and conclusions together with a brief outlook are provided in Sec.~\ref{summary}.
App.~\ref{hadronnumbers} addresses the calculated hadron multiplicities.  
App.~\ref{uncert} contains a discussion of the various uncertainties which may influence our results.  

\section{The model}
\label{model}

The GiBUU transport model \cite{Buss:2011mx} is built on the solution of the coupled set of semi-classical quantum-kinetic equations for the baryons ($N,~\Delta,~N^*,~Y,\ldots$), respective antibaryons
($\bar N,~\bar\Delta,~\bar N^*,~\bar Y,\ldots$) and mesons ($\pi,~\eta,~\rho,~\omega,~K,~K^*,\ldots$). In 
relativistic kinematics \cite{Blaettel:1993uz} the kinetic equation for the nucleons with fixed isospin projection reads,
\begin{eqnarray}
\lefteqn{(p^{* 0})^{-1}
  \left[ p^{* \mu} \partial_\mu + ( p_{\mu}^* {\cal F}^{\alpha\mu}
    + m^* \partial^\alpha  m^* )
    \frac{\partial}{\partial p^{* \alpha}} \right]
 f^*(x,\bvec{p}^*) =} &  &                                                              \nonumber \\
& = & \int\, \frac{g_s d^3 p_2^* }{(2\pi)^3}\, v_{12} \int d\Omega\, \frac{d\sigma_{12\to34}}{d\Omega}\,
      ( f_3^* f_4^* \bar f_1^* \bar f_2^* - f_1^* f_2^* \bar f_3^* \bar f_4^*)~,      \label{BUUstar}
\end{eqnarray}
where $\alpha=1,2,3$ and $\mu=0,1,2,3$.
The left side of the kinetic equations describes particle propagation in a self-consistent relativistic mean field (RMF) potential that includes scalar ($S$) and vector ($V$) nuclear potentials as well as the Coulomb potential. The RMF is included only for the baryons while the mesons feel only the Coulomb potential. This difference is justified because the potential for the baryons is necessary for the nuclear binding. Also, in earlier work it has been shown that in-medium properties of the $\Delta$ and the pion affect the dilepton yield only moderately \cite{Ehehalt:1993px}. The collision integrals on the right side of the kinetic equations describe two and three-body collisions as well as resonance decays. Here, we have explicitly included only the expression for the elastic two-body collision integral, for simplicity, where $f^*(x,\bvec{p}^*)$ with $x \equiv (t,\bvec{r})$ is the distribution function of the nucleons in the kinetic phase space $(\bvec{r},\bvec{p}^*)$.
It is  defined such that $f^*(x,\bvec{p}^*) \, g_s d^3 r d^3 p^*/(2\pi)^3$ equals the number of particles in the phase space element $d^3 r d^3 p^*$;
$g_s=2$ is the nucleon spin degeneracy, $p^{*\mu} = p^\mu - V^\mu$ its kinetic four-momentum satisfying the mass-shell condition $p^{*\mu}p^*_\mu = m^{*2}$, where $m^*=m_N+S$ is the Dirac mass of the nucleon with mass $m_N=0.938$~GeV in the vacuum. 
${\cal F}^{\mu\nu} \equiv \partial^\mu V^\nu - \partial^\nu V^\mu$ is the field-strength tensor of the vector potential $V$.

In the collision integral, we have introduced short-hand notations
$f_n^* \equiv f^*(x,\bvec{p}_n^*),~\bar f_n^* \equiv 1 - f_n^*$ ($n=1,2,3,4$), $\bvec{p}_1^* \equiv \bvec{p}^*$.
The relative velocity of the colliding particles is defined as $v_{12}=I_{12}/p_1^{*0}p_2^{*0}$
where $I_{12}=\sqrt{(p_1^* p_2^*)^2 - (m_1^* m_2^*)^2}$ is the M\"oller flux factor.
The angular differential cross section is defined by ${d\sigma_{12\to34}}$ for scattering in the solid angle element ${d\Omega}=\sin\Theta\,d\Theta\,d\phi$ 
with polar, $\Theta$, and azimuthal, $\phi$, scattering  angles in the center-of-mass (c.m.) frame.
The extension of the collision integral to include inelastic channels and broad particles is rather straightforward
(see Sec.~3.3 of Ref.~\cite{Buss:2011mx} for details), and all these features are also included in our present GiBUU simulations.

The scalar, $S$, and vector, $V^\mu$, mean fields are obtained from the Dirac equation for the nucleon,
\begin{equation}
    [\gamma^\mu(i\partial_\mu-V_\mu) - (m_N+S)] \psi(x) = 0~,     \label{DiracEq}
\end{equation}
coupled to the scalar-isoscalar $\sigma$-meson, the vector-isoscalar $\omega$-meson, and the electromagnetic field $A^\mu$, via
\begin{eqnarray}
   & & S=g_{\sigma N} \sigma,       \label{S} \\
   & & V^\mu = g_{\omega N} \omega^\mu + \frac{e}{2} (1+\tau^3) A^\mu~,~~~\mu=0,\dots 3,                   \label{V^mu}
\end{eqnarray}
where $\tau^3=+(-)1$ for the proton (neutron); $e=1/\sqrt{137}$ (in natural units with $\hbar=c=1$);
and the coupling constants to the meson fields are those of the non-linear Walecka model in the version NL2
of Ref.~\cite{Lang:1992jz}, i.e.~$g_{\sigma N}=8.5$, $g_{\omega N}=7.54$.  The mesonic mean fields are calculated by solving
the Lagrange equations of motion with source terms provided by the baryon densities and currents (see Sec.~3.1.3
of Ref.~\cite{Buss:2011mx} for details).

The numerical solution of Eq.~(\ref{BUUstar}) is based on the test-particle representation of the distribution function
\begin{equation}
  f^*(x,\bvec{p}^*) = \frac{ (2\pi)^3 }{ g_s N }
  \sum_{n=1}^{N_{\rm phys} N} \delta( \bvec{r} - \bvec{r}_n(t) )
                            \delta( \bvec{p}^* - \bvec{p}_n^*(t) )~,      \label{testPart^star}
\end{equation}
where $N_{\rm phys}$ is the number of physical particles, while $N$ is the number of test particles per physical one.  
Turning off the interaction terms, by setting the right side in Eq.~(\ref{BUUstar}) to zero (the Vlasov limit), one obtains the equations of motion for the centroids of the $\delta$-functions $\bvec{r}_n(t),\bvec{p}_n^*(t)$,
\begin{eqnarray}
& & \dot{\bvec{r}}_n =  \frac{\bvec{p}_n^*}{p_n^{* 0}}~,                     \label{rDotStar} \\
& & \dot{p}^{*\alpha}_n = \frac{p_{n\mu}^*}{p_n^{* 0}} {\cal F}_n^{\alpha\mu}
                              + \frac{m_n^*}{p_n^{* 0}} 
                                \frac{\partial m_n^*}{\partial r_\alpha}~,     \label{pDotStar}
\end{eqnarray}
where $\alpha=1,2,3$ and $\mu=0,\dots 3$. It can be shown that Eqs.~(\ref{rDotStar}) and (\ref{pDotStar}) are equivalent
to the Hamiltonian equations
\begin{eqnarray}
& & \dot{\bvec{r}}_n =  \frac{\partial \varepsilon(\bvec{r}_n,\bvec{p}_n,t)}{\partial \bvec{p}_n}~,  \label{rDot} \\
& & \dot{\bvec{p}}_n = -\frac{\partial \varepsilon(\bvec{r}_n,\bvec{p}_n,t)}{\partial \bvec{r}_n}~,  \label{pDot}
\end{eqnarray}
with the single-particle energy defined as
\begin{equation}
   \varepsilon = V^0 + \sqrt{(\bvec{p}^*)^2+(m^*)^2}~.  \label{disp_rel_RMF}
\end{equation}

Particles in the medium can be collision-broadened and/or have already a decay width in vacuum. For the propagation of such broad particles, one has to use the off-shell transport implemented in GiBUU. This is based on using a generalized distribution function that also includes
the particle energy as an independent variable \cite{Effenberger:1999uv,Cassing:1999wx,Cassing:1999mh,Leupold:1999ga},
\begin{equation}
  F(x,p) = \frac{(2\pi)^4}{N} \sum_{n=1}^{N_{\rm phys} N} \delta( \bvec{r} - \bvec{r}_n(t) )
                              \delta( \bvec{p} - \bvec{p}_n(t) ) \delta( p^0 - \varepsilon_n(t) )~.   \label{testPart^osp}
\end{equation}
The time evolution of the centroids  $\bvec{r}_n(t),\bvec{p}_n(t),\varepsilon_n(t)$ is given by the so-called off-shell potential (OSP) ansatz
\cite{Buss:2011mx} which is based on the following equations:
\begin{eqnarray}
  & & \dot{\bvec{r}}_n = \left(1-\frac{\partial H_n}{\partial \varepsilon_n}\right)^{-1} \frac{\partial H_n}{\partial \bvec{p}_n}~,  \label{rDotOSP}\\
  & & \dot{\bvec{p}}_n = -\left(1-\frac{\partial H_n}{\partial \varepsilon_n}\right)^{-1} \frac{\partial H_n}{\partial \bvec{r}_n}~,  \label{pDotOSP}\\
  & & \dot{\varepsilon}_n = \left(1-\frac{\partial H_n}{\partial \varepsilon_n}\right)^{-1} \frac{\partial H_n}{\partial t}~.         \label{eDotOSP}
\end{eqnarray}
Here $H_n(\varepsilon_n,\bvec{p}_n,t,\bvec{r}_n)$ is a (generalized) single-particle Hamilton function as defined below.
Eqs.~(\ref{rDotOSP}) and (\ref{pDotOSP}) are obtained by expressing the partial derivatives of the single-particle energy 
$\varepsilon_n \equiv \varepsilon(\bvec{r}_n,\bvec{p}_n,t)$ in Eqs.~(\ref{rDot}) and (\ref{pDot}) in terms of those of $H_n$ using the self-consistency condition, 
\begin{equation}
    \varepsilon_n = H_n(\varepsilon_n,\bvec{p}_n,t,\bvec{r}_n)~.             \label{selfCons}
\end{equation}
Eq.~(\ref{eDotOSP}) then follows directly from this self-consistency condition.

The single-particle Hamilton function $H_n$ is defined such that particle $n$ can be arbitrarily far off shell, when its in-medium width $\Gamma_n > 0$, but becomes an on-shell particle for $\Gamma_n = 0$. The simplest form of $H_n$ that satisfies these requirements is:
\begin{equation}
  H_n = \sqrt{m_\mathrm{phys}^2+\mbox{Re}\,\Pi+\Delta m^2_n+\bvec{p}_n^2}~,~~~\Delta m^2_n= -\chi_n \mbox{Im}\,\Pi~,        \label{H_n}
\end{equation}
where $m_\mathrm{phys}$ is the vacuum mass of the physical on-shell particle, $\Pi(\varepsilon_n,\bvec{p}_n,t,\bvec{r}_n)$  is a retarded self-energy, and $\chi_n$ is a constant fixed from the initial conditions at the production time
of the test particle $n$. 
In the most general case the long-range potential and short-range collisional interactions of the particle in the nuclear medium modify,
respectively, the real and imaginary parts of the particle self-energy. In particular, the imaginary part of self-energy is related
to the width by a usual bosonic formula:
\begin{equation}
  \mbox{Im}\Pi = - \sqrt{p_n^2} \,\Gamma_n~,    \label{ImPi}
\end{equation}
where $p_n^2=\varepsilon_n^2-\bvec{p}_n^2$.
Since the collisional width of the particle is roughly proportional to the nucleon density, Eqs.~(\ref{H_n}) and (\ref{ImPi}) imply that for the particle with small 
natural decay width (e.g.~a pion) the deviation of the particle-mass squared from its on-shell value, i.e.\ $\Delta m^2_n$ scales with the nucleon density.
On the other hand, for particles with a large natural decay width (e.g.~the  $\rho$),
the quantity $\Delta m^2_n$ becomes constant when the particle is emitted to the vacuum,
since the decay width depends only on the particle invariant mass 
(cf.~Eq.~(\ref{Gamma_rho2pipi}) below). In-particular, this means that the OSP ansatz without collisional widths is equivalent to the treatment of broad particles with off-shell masses chosen according to their vacuum spectral functions.

With the Hamilton function of Eq.~(\ref{H_n}), the OSP ansatz is equivalent
to solving the test-particle equations of motion for relativistic off-shell bosons as derived from the retarded Green function formalism
\cite{Cassing:1999mh}.\footnote{The off-shell dynamics of vector mesons has first been discussed in Ref.~\cite{Effenberger:1999ay} where an ad hoc form
of the OSP ansatz with a scalar off-shell potential was introduced to bring off-shell particles back on-shell when they leave the nucleus.} 

In the present work, the OSP ansatz is applied to describe the dynamics of the $\rho$ meson in the nuclear medium; we set $\mbox{Re}\,\Pi=0$ for simplicity. After time stepping according to Eqs.~(\ref{rDotOSP}) and (\ref{pDotOSP}), the single-particle energy $\varepsilon_n$ is obtained at the new time step by solving Eq.~(\ref{selfCons}) for fixed three-momentum $\bvec{p}_n$. 

The collision term on the right of the transport equation (\ref{BUUstar}) is modeled geometrically: 
when the two test particles 1 and 2 are approaching their minimum distance $b_{12}$ their collision is simulated by
Monte-Carlo provided $b_{12} < \sqrt{\sigma_{12}/\pi}$ where $\sigma_{12}$ is the total interaction
cross section. To approximate Lorentz covariance, the minimum distance $b_{12}$ is calculated in the c.m. frame of the colliding particles assuming
straight-line trajectories:
\begin{equation}
    b_{12}^2 = (\bvec{x}_{12}^{cm})^2 - \frac{(\bvec{x}_{12}^{cm} \cdot \bvec{\beta}_{12}^{cm})^2}{(\bvec{\beta}_{12}^{cm})^2}~,   \label{kodama_position} 
\end{equation}
where 
\begin{equation}
   \bvec{x}_{12}^{cm} = \bvec{x}_{1}^{cm} - \bvec{x}_{2}^{cm} - \bvec{\beta}_{1}^{cm} t_1^{cm} + \bvec{\beta}_{2}^{cm} t_2^{cm}   \label{x_12^cm}
\end{equation}
is the relative position vector at zero time; $\bvec{\beta}_{1}^{cm}$ and $\bvec{\beta}_{2}^{cm}$ are the particle velocities. 
All quantities in Eqs.~(\ref{kodama_position}), (\ref{x_12^cm}) refer to the c.m.~frame of particles 1 and 2. The four vectors $(t_i^{cm},\bvec{x}_i^{cm})$, $i=1,2$
are obtained by Lorentz transformation from their four-vectors $(t,\bvec{x}_i)$ in the computational frame to the c.m.~frame of the pair.\footnote{For heavy-ion collisions, the c.m.~frame of colliding nuclei is normally chosen as the computational frame.  
The choice of the time $t$ (it was set to zero in actual calculations) does not influence the minimum distance, Eq.~(\ref{kodama_position}),
since it leads only to the change of the component of $\bvec{x}_{12}^{cm}$ along the relative velocity $\bvec{\beta}_{2}^{cm}-\bvec{\beta}_{1}^{cm}$.}
This exactly corresponds to the covariant prescription given by Eq.~(2b) of 
Ref.~\cite{Kodama:1983yk}. The determination of the collision time instant 
is a more delicate problem. One can not naively use the collision instant (i.e.~the time of closest approach) in the c.m.~frame since this would result
in different collision instants for particles 1 and 2 in the computational frame. It has been shown in Ref.~\cite{Kodama:1983yk} that
the causality in a collision sequence can be preserved by using collision proper times. The proper time difference between the collision instant
and the current instant for particle 1 is 
\begin{equation}
    \Delta\tau_1 = \frac{\tilde{\bvec{x}}_{12} \cdot \tilde{\bvec{\beta}}_2}{(\tilde{\bvec{\beta}}_2)^2} - \tilde{t}_1~,      \label{Delta_tau_1}
\end{equation}
where the relative position vector at the zero time is now
\begin{equation}
    \tilde{\bvec{x}}_{12} = \tilde{\bvec{x}}_{1} - \tilde{\bvec{x}}_{2} + \tilde{\bvec{\beta}}_{2} \tilde{t}_2~.   \label{tilde_x_12}
\end{equation}
All quantities with tilde refer to the rest frame of particle 1. The condition that particles 1 and 2 pass their closest-approach distance
during the time interval $[t-\Delta t/2;t+\Delta t/2]$,  where $\Delta t$ is the time step in the computational frame, then can be written as
\begin{equation}
    |\Delta\tau_1\gamma_1 + \Delta\tau_2\gamma_2| < \Delta t~,       \label{kodama_time}
\end{equation}
where $\gamma_{1,2}$ are the Lorentz factors of particles 1, 2 in the computational frame; and $\Delta\tau_2$ is the proper time difference for particle 2
defined in its rest frame by Eqs.~(\ref{Delta_tau_1}), (\ref{tilde_x_12}) with replacement $1 \leftrightarrow 2$ (for more details see Ref.~\cite{Wolf:1993cj} and Appendix B of Ref.~\cite{EffenbergerPhD}).

Baryon-baryon (baryon-meson) collisions at $\sqrt{s} < 4\, (2.2)$ GeV are simulated within the resonance model which treats the interactions of nucleons, their resonances and mesons explicitly. At higher invariant energies the PYTHIA model \cite{Sjostrand:2006za} is applied. 
Pauli blocking is applied for collisions with nucleons in the final state.

The model also accounts for meson-meson collisions, in particular, $\pi \pi \to \rho$ which are
of importance for $\rho$ production at large invariant masses (see Fig.\ \ref{fig:dNdM_res_woFilt_rmf_rho} below). 
Other meson-meson, e.g. $\pi \rho$, collisions are infrequent in the baryon-dominated matter created at SIS18 energies.

\newpage
The channels of dilepton production are:
\begin{itemize}
\item  Direct decays of vector mesons, $V \to e^+ e^-$, with $V=\rho, \omega, \phi$: 
For vector mesons this partial width is calculated based on the vector dominance model (VDM) \cite{Nambu:1962zz,Dumbrajs:1983jd}, 
\begin{equation}   
  \Gamma_{V \to e^+ e^-}(m) = C_V \frac{m_V^4}{m^3} (1+2m_e^2/m^2) \sqrt{1-4m_e^2/m^2}~,    \label{Gamma_V2ee}
\end{equation}
where $m_V$ and $m$ are the on-shell and off-shell mass of the decaying meson, respectively, and  $C_V = 4\pi\alpha^2/3f_V^2$.
The numerical values used in GiBUU are $C_\rho = 9.078 \cdot 10^{-6}$, $C_\omega = 7.666 \cdot 10^{-7}$,
and $C_\phi = 1.234 \cdot 10^{-6}$  \cite{Weil:2012ji}.
These values have been fit to reproduce the empirical partial widths $V \to e^+ e^-$ for the on-shell meson masses. For a collision-broadened meson with a very small off-shell mass $m$, the applicability of this expression with constant $C_V$ becomes questionable, however. In particular, far below mass shell, $\Gamma_V$ from Eq.~(\ref{Gamma_V2ee}) grows artificially strongly with decreasing $m$ until it reaches 
an unphysically sharp peak very close to the $e^+e^-$ threshold where it vanishes.

\item Direct $\eta \to e^+ e^-$ decay: For the partial
decay width $\Gamma_{\eta \to e^+ e^-} = \Gamma_{\eta}^{\rm tot} \, \mbox{BR}_{\eta \to e^+ e^-}$, where $\Gamma_{\eta}^{\rm tot}=1.3$ keV is the total width of the $\eta$,  
the phenomenological upper limit of the branching ratio $\mbox{BR}_{\eta \to e^+ e^-} = 7 \cdot 10^{-7}$ is adopted \cite{Tanabashi:2018oca}.

\item Dalitz decays $A \to B e^+ e^-$: Using a factorization prescription (c.f.~\cite{Krivoruchenko:2001hs}) the following expression
for the partial width can be obtained,
\begin{equation}
  \frac{d\Gamma_{A \to B e^+ e^-}}{dm^2} = \Gamma_{A \to B \gamma^*} \frac{\alpha}{3 \pi m^2} (1+2m_e^2/m^2) \sqrt{1-4m_e^2/m^2}~,   \label{dGammaDaldm2}
\end{equation}
where $m$ is the invariant mass of the dilepton pair, and $\Gamma_{A \to B \gamma^*}$ is the decay width to the virtual photon in the final state.

\underline{Meson Dalitz decays}: In the case of the pseudoscalar meson $P=\pi^0, \eta, \eta^\prime$ decays $P \to \gamma e^+ e^-$ and the
$\omega$-meson decay $\omega \to \pi^0 e^+ e^-$ the decay width to virtual photon is proportional to the decay width to the real photon \cite{Landsberg:1986fd}:
\begin{equation}
  \Gamma_{A \to B \gamma^*}(m^2) = \eta_{\rm sym} \Gamma_{A \to B \gamma} \left(\frac{q_{B \gamma^*}(m^2)}{q_{B \gamma^*}(0)}\right)^3 
   |F_{AB}(m^2)|^2~,
                                                                                   \label{Gamma_A2BgammaSt}
\end{equation}
where $\eta_{\rm sym}=2$ for pseudoscalar meson decays and $\eta_{\rm sym}=1$ for $\omega$-meson decay is the symmetry factor,
\begin{equation}
  q_{B \gamma^*}(m^2) = \frac{m_A^2-m_B^2}{2m_A}
                     \left[ \left(1+\frac{m^2}{m_A^2-m_B^2}\right)^2 - \frac{4m_A^2 m^2}{(m_A^2-m_B^2)^2} \right]^{1/2}
\end{equation}
is the c.m.~momentum of $B$ and $\gamma^*$,
and $F_{AB}(m^2)$ is the $A \to B$ transition form factor.
The term $(q_{B \gamma^*}(m^2)/q_{B \gamma^*(0)})^3$ reflects the $p$-wave coupling
arising in the effective Lagrangian description of meson decays
\cite{Dalitz:1951aj,Faessler:1999de}.\footnote{An $s$-wave coupling is forbidden due to the presence of the pseudoscalar meson either in the initial or in the final state of the
  $A \to B \gamma^*$ decay.} 
  For the partial decay widths to the real photon the following values are used:
$\Gamma_{\pi^0  \to \gamma \gamma} = 0.98823 \Gamma_{\pi^0}^{\rm tot}$ with  $\Gamma_{\pi^0}^{\rm tot}=7.836$ eV,
$\Gamma_{\eta \to \gamma \gamma} = 511$ eV, $\Gamma_{\eta^\prime \to \gamma \gamma} = 4.3$ keV, $\Gamma_{\omega \to \pi^0 \gamma} =703$ keV.

For the form factor of the $\pi^0 \to \gamma \gamma^*$ vertex it is enough to use the linear approximation in $m^2$:
\begin{equation}
   F_{\pi^0 \gamma}(m^2) = 1 + b_\pi m^2
\end{equation}
with $b_\pi = 5.5$ GeV$^{-2}$ \cite{Landsberg:1986fd}.
In the $\eta \to \gamma \gamma^*$ vertex, the pole approximation is used:  
\begin{equation}
   F_{\eta \gamma}(m^2) = (1-m^2/\Lambda_{\eta}^2)^{-1}
\end{equation}
with $\Lambda_{\eta}^{-2}=1.95$ GeV$^{-2}$ as determined in the NA60 measurements of the low-mass $\mu^+ \mu^-$ pairs in 158~A~GeV In+In collisions \cite{Arnaldi:2009aa}.
The $\eta^\prime \to \gamma \gamma^*$ vertex form factor is neglected.
The form factor of the $\omega \to \pi^0 \gamma^*$ vertex is adopted from \cite{Bratkovskaya:1996qe}:
\begin{equation}
   F_{\omega\pi^0}(m^2) = \frac{\Lambda_{\omega}^2}{[(\Lambda_{\omega}^2-m^2)^2+\Lambda_{\omega}^2 \Gamma_{\omega}^2]^{1/2}}~, \label{F_omega_pi0}
\end{equation}
where $\Lambda_{\omega}=0.65$ GeV and $\Gamma_{\omega}=75$ MeV. Eq.~(\ref{F_omega_pi0}) is also in good agreement with the NA60 data
\cite{Arnaldi:2009aa}. 

\underline{$\Delta$ Dalitz decay}: In the case of the Delta resonance Dalitz decay, $\Delta(1232) \to N e^+ e^-$, we apply Eq. (\ref{dGammaDaldm2}) with
the partial decay width of Ref.~\cite{Krivoruchenko:2001hs},
\begin{equation}
  \Gamma_{\Delta \to N \gamma^*}(m^2) = \frac{\alpha}{16} \frac{(m_\Delta+m_N)^2}{m_\Delta^3 m_N^2} [(m_\Delta+m_N)^2-m^2]^{1/2}
        [(m_\Delta-m_N)^2-m^2]^{3/2} |F_{\Delta N}(m^2)|^2~,   \label{Gamma_D2NgammaSt}
\end{equation}
where the form factor is set to be constant, $F_{\Delta N}(m^2)\equiv F_{\Delta N}(0) =3.029$, obtained from the real photon decay width $\Gamma_{\Delta \to N \gamma^*}(0) =  0.66$ MeV.
Note that the $\Delta \to N \gamma^*$ magnetic dipole transition form factor is still under discussion \cite{Ramalho:2015qna}.

\item Bremsstrahlung: The $pn \to pn  e^+ e^-$ and  $pp \to pp  e^+ e^-$ bremsstrahlung is included in the form of the boson exchange model of Ref.~\cite{Shyam:2010vr}. This model takes into account the $e^+ e^-$ emission from the internal charged pion exchange line in the $pn$ scattering.\footnote{We apply the version with the pion electromagnetic form factor taking into account the direct coupling
  of the photon to the quark content of the pion, apart from the photon coupling to the $\rho^0$ meson
  (the FF2 parameterization of \cite{Shyam:2010vr}).} However, in contrast to Ref.~\cite{Shyam:2010vr} we add the contribution of the
$NN \to N\Delta$ reaction followed by $\Delta \to N e^+ e^-$ incoherently. The $N^*(1520)$ Dalitz decays are
effectively included in our calculations via the two-step decay $N^* \to \rho N$, $\rho \to e^+ e^-$.

For the charged pion bremsstrahlung, $\pi^\pm N \to \pi^\pm N e^+ e^-$, the soft-photon approximation (SPA) \cite{Gale:1987ki,Wolf:1990ur}
is applied,
\begin{equation}
  E \, \frac{d\sigma_{e^+e^-}}{d^3pdm} = \frac{\alpha^2}{6\pi^3} \frac{\overline{\sigma}_{\rm el}(s)}{mE^2}
  \frac{R_2(s_2)}{R_2(s)}~,     \label{SPA}
\end{equation}
with
\begin{equation}
  \overline{\sigma}_{\rm el}(s) = \int\limits_{-|t|_{\rm max}}^0 dt \frac{-t}{m_\pi^2} \frac{d\sigma_{\rm el}(s,t)}{dt}
  \approx \frac{2q_{\rm cm}^2(s)}{m_\pi^2} \sigma_{\rm el}(s)~,     \label{overline_sigma}
\end{equation}
where $d\sigma_{\rm el}(s,t)/dt$ is the differential elastic scattering cross section, and
$q_{\rm cm}(s)=[(s+m_\pi^2-m_N^2)^2/4s - m_\pi^2]^{1/2}$ is the c.m.~momentum of pion and nucleon.
$(E,\bvec{p})$ is the four-momentum of the $e^+ e^-$ pair in the c.m.~system of the colliding pion and nucleon.
The reduction of the two-body phase space available for the outgoing pion and nucleon is included in Eq.~(\ref{SPA}) via the ratio
of phase space volumes at the invariant energy squared with ($s_2$) and without ($s$) emission of the $e^+ e^-$ pair, where
\begin{equation}
  s_2 = s + m^2 - 2\sqrt{s}E~,     \label{s_2}
\end{equation}
and 
\begin{equation}
    R_2(s) = 2 q_{\rm cm}(s)/\sqrt{s}~.    \label{R_2}
\end{equation}
In the last step of Eq.~(\ref{overline_sigma}) an isotropic cross section is assumed for simplicity.
Pion-nucleon scattering in heavy-ion collision processes at $1-2 A$~GeV is mostly mediated by the $\Delta(1232)$ resonance,
i.e.~$\pi N \to \Delta \to \pi N$. In this case the angular distribution is forward-backward peaked due to the dominant
$p$-wave. Thus, the approximation of Eq.~(\ref{overline_sigma}) is quite rough. However, given a large overall uncertainty
of the SPA at $e^+ e^-$ invariant masses above $100-200$~MeV \cite{Gale:1987ki} the approximation of Eq.~(\ref{overline_sigma}) seems still reasonable.

Since the cross section for $e^+ e^-$ production in pp collisions at the beam energy of 1-3~GeV is quite small, on the $\mu b$ level,
the direct calculation of dilepton production in heavy-ion collisions would be extremely time consuming. Therefore, a so-called shining method
\cite{Weil:2013mya} is applied. In a given time step $dt$, the probability of the dilepton decay of a resonance is ${\cal P}=\Gamma_{e^+ e^-} dt/\gamma$
where $\Gamma_{e^+ e^-}$ is the partial decay width $R \to X e^+ e^-$ in the resonance rest frame and $\gamma$ is the Lorentz factor of the resonance.
In the shining method, the dilepton decay of every relevant resonance at every time step is simulated and the produced $e^+ e^-$ pair carries
the weight ${\cal P}$ which is then used to fill various statistical distributions.
Note that the actual state of the resonance is not changed after the dilepton emission, i.e. the resonance is further propagated according
to the test particle equations of motion, Eqs.~(\ref{rDotOSP}), (\ref{pDotOSP}), and participates in collision and decay processes.
If the resonance survives until the end of time evolution, the 
decay dilepton will have the weight ${\cal P}=\Gamma_{e^+ e^-}/\Gamma$ where $\Gamma$ is the total in-medium width of the resonance. 

The $e^+ e^-$ bremsstrahlung in $pn$, $pp$, and $\pi^\pm N$ collisions is simulated in the following way: If the collision takes place, the
probability of the $e^+ e^-$ emission ${\cal P}=\sigma_{e^+e^-}/\sigma$ is calculated where $\sigma$ is the total interaction cross section of colliding
particles and $\sigma_{e^+e^-}$ is the partial cross section of dilepton emission. The produced $e^+ e^-$ pair has the weight ${\cal P}$ while the actual
two-body collision is simulated neglecting the dilepton emission. 
\end{itemize}

In the case of the pseudoscalar meson Dalitz decays, the polar angle distribution of the outgoing $e^-$ in the rest frame
of $\gamma^*$ is sampled according to the distribution $d{\cal P}/d\cos\Theta \propto 1 + \cos^2\Theta$
where the $z$-axis is chosen along the three-momentum of $\gamma^*$ \cite{Bratkovskaya:1995kh}.
For all other dilepton decays an isotropic angular distribution is used, just as for bremsstrahlung in the c.m.~system of the colliding pair.

\section{Spectral function of the $\rho$-meson in the nuclear medium}
\label{rhomed}

The retarded $\rho$-meson propagator has the following spectral representation \cite{Jung:2019nnr},
\begin{equation}
        G_{\mu\nu}^R(q) = \int_0^\infty ds\,  \frac{\mathcal A(s)}{s} \, \frac{q^2 g_{\mu \nu}  - q_\mu q_\nu}{ (q_0 + i 0)^2 -\vec q^2 - s}~,   \label{specG}
\end{equation}
where $q$ is the four-momentum of the $\rho$-meson,
so that the spin-averaged spectral function of the $\rho$-meson is given by
\begin{equation}
  \mathrm{sgn} (q_0) \,  {\cal A}(q^2) = - \frac{1}{\pi}\, \mbox{Im}\, G^R(q)~,     \label{calAdef}
\end{equation}
where
\begin{equation}
  G^R(q) = \frac{1}{3}\Big(g^{\mu \nu} - \frac{q^\mu q^\nu}{q^2}\Big)\, G_{\mu\nu}^R(q)    \label{G}
\end{equation}
is the corresponding spin-averaged propagator.\footnote{This definition is equivalent to setting ${\cal A}(q) = \frac{1}{3}(2A^{T}(q)+A^{L}(q))$, where
  $A^{T}(q)$ and $A^{L}(q)$ are, respectively, the transverse and longitudinal spectral functions defined
  in Ref.~\cite{Peters:1997va}.}
  The tensor structure in Eq.~(\ref{specG}) represents an off-shell extension of the sum over the polarization states of an on-shell $\rho$-meson with pole mass $m_\rho$,
\begin{equation}
  \sum_{\lambda=0,\pm1} \varepsilon^{(\lambda)}_\mu \varepsilon^{*(\lambda)}_\nu
  = -g_{\mu \nu} +\frac{q_\mu q_\nu}{m_\rho^2}~.    \label{polar_sum}
\end{equation}
Neglecting a possible $q^2$-dependence in the real part of the $\rho$-meson self-energy $\Pi(q^2)$ (the more general case, including polarization dependence of the self-energy is discussed, e.g.~in Refs.~\cite{Rapp:1997fs,Peters:1997va,Post:2000qi}), and writing $\mbox{Im}\, \, \Pi(q^2) = - \sqrt{q^2} \, \Gamma(\sqrt{q^2}) $, on the other hand, we can make the following Ansatz for the spin-averaged Feynman propagator of the $\rho$-meson in the nuclear medium,
\begin{equation}
    G^F(q) = \frac{1}{q^2-m_\rho^2+i\sqrt{q^2}\,\Gamma(\sqrt{q^2})}~.   \label{GF}
\end{equation}
Here, $\Gamma(\sqrt{q^2}) $ defines the off-shell decay width of the $\rho$-meson (see below) which vanishes below the $e^+e^-$ threshold, for $q^2 < 4m_e^2$. Including the sign function in the corresponding  retarded self-energy, $\mbox{Im}\, \, \Pi^R(q^2) = - \mathrm{sgn} (q_0)\sqrt{q^2} \,\Gamma(\sqrt{q^2}) $ in (\ref{calAdef}), this amounts to using the following form for the spectral function,
\begin{equation}
    {\cal A}(q^2) =     \frac{ \sqrt{q^2}\, \Gamma(\sqrt{q^2}) /\pi}{(q^2-m_\rho^2)^2 + q^2\Gamma^2(\sqrt{q^2})}~,    \label{calA}
\end{equation}
with the normalization condition,
\begin{equation}
    \int\limits_{4m_e^2}^{\infty} dq^2 \, {\mathcal A}(q^2) = 1~.                 \label{norm_cond}
\end{equation}
For later convenience, we also note that for $q^2 > 4 m_e^2$ we may write,
\begin{equation}
    |G^F(q)|^2 = \frac{\pi\,  \mathcal A(q^2) }{\sqrt{q^2}\, \Gamma(\sqrt{q^2})} \, . \label{42}
\end{equation}

In order to illustrate the use of the spectral function for our purposes consider, for example, the process $a + b \to \rho \to X$.
The invariant matrix element of this process is
\begin{equation}
   M_{X;ab} = - \sum_{\lambda=0,\pm1} \frac{M_{X;\rho^{(\lambda)}} M_{\rho^{(\lambda)};ab}}{q^2-m_\rho^2+i\sqrt{q^2}\Gamma}~,  \label{M_Xab}
\end{equation}
where we used the decomposition over polarizations, Eq.(\ref{polar_sum}). Neglecting the interference terms with different $\rho$
polarizations in the direct and conjugated amplitudes and performing the independent summations over $\lambda$ in the $\rho$-production
and decay amplitudes squared, we obtain
\begin{equation}
  \overline{|M_{X;ab}|^2} = \frac{\overline{|M_{X;\rho^{(\lambda)}}|^2}~\overline{|M_{\rho^{(\lambda)};ab}|^2}}{|q^2-m_\rho^2+i\sqrt{q^2}\Gamma|^2}~,
  \label{M_Xab^2}
\end{equation}
where an overline means the sum over polarizations of final states and the averaging over polarizations of initial states.   
The differential cross section of the process $a + b \to \rho \to X$ is given by the following expression
\begin{equation}
  d\sigma_{a + b \to X} = \frac{(2\pi)^4 \overline{|M_{X;ab}|^2} d\Phi_X}{4I_{ab}}~,     \label{dsigma_abX}
\end{equation}
where
\begin{equation}
  I_{ab}=\sqrt{(p_ap_b)^2-p_a^2p_b^2}                   \label{I_ab}
\end{equation}
is the M\"oller flux factor and
\begin{equation}
   d\Phi_X = \delta^{(4)}(p_a+p_b-\sum_{i=1}^{n_X} p_i) \prod_{i=1}^{n_X} \frac{d^3p_i}{(2\pi)^3 2E_i}     \label{dPhi_X}
\end{equation}
is the invariant phase space volume element of the final state. Using the formula for the partial decay width
$\rho \to X$ in the rest frame of the $\rho$-meson,
\begin{equation}
   d\Gamma_{\rho \to X} = \frac{(2\pi)^4 \overline{|M_{X;\rho^{(\lambda)}}|^2} d\Phi_X}{2 \sqrt{q^2}}~,
                                   \label{dGamma_rho2X}
\end{equation}
we can now rewrite Eq.(\ref{dsigma_abX}) in a factorized form:
\begin{equation}
    d\sigma_{a + b \to X} = \frac{d\Gamma_{\rho \to X}}{\Gamma}\, \sigma_{ab \to \rho}~,     \label{dsigma_abX_fact} 
\end{equation}
so that with Eq.~(\ref{42}) we obtain from Eq.~(\ref{M_Xab}),
\begin{equation}
   \sigma_{ab \to \rho} = 2\pi\, {\cal A}\big((p_a+p_b)^2\big)\,  \frac{\overline{|M_{\rho^{(\lambda)};ab}|^2}}{4I_{ab}}~.  \label{sigma_abrho} 
\end{equation}
Here we have thus defined a quantity which has the meaning of a cross section for the production of an off-shell $\rho$-meson in the collision of particles $a$ and $b$. Indeed, Eq. (\ref{sigma_abrho}) can be obtained from the standard formula for
an $a b \to \rho$ process where $\rho$ is treated as a fictitious ``on-shell'' particle with a mass of $\sqrt{s}$,
\begin{equation}
  \sigma_{ab \to \rho}^{on-shell} = 2\pi \delta(s-(p_a+p_b)^2) \frac{\overline{|M_{\rho^{(\lambda)};ab}|^2}}{4I_{ab}}~,
        \label{sigma_abrho_ons}                          
\end{equation}
which is then multiplied with a weight given by ${\cal A}(s)$, and integrated over $s$. Thus, the square of the invariant mass of the intermediate $\rho$-meson is distributed according to the spectral function ${\cal A}(s)$. Similar relations can be readily derived for any other process mediated by an off-shell $\rho$-meson.
Moreover, using detailed balance,
\begin{equation}
   \overline{|M_{ab;\rho^{(\lambda)}}|^2} = \frac{(2S_a+1)(2S_b+1)}{2S_\rho+1}\, \overline{|M_{\rho^{(\lambda)};ab}|^2}~,   \label{detbal}
\end{equation}
Eq.~(\ref{sigma_abrho}) can be transformed into a relativistic Breit-Wigner formula,
\begin{equation}
  \sigma_{ab \to \rho} = \frac{2S_\rho+1}{(2S_a+1)(2S_b+1)} \, \frac{4\pi}{q_{ab}^2}\, 
  \frac{q^2 \Gamma \Gamma_{\rho \to ab}}{(q^2-m_\rho^2)^2 + q^2\Gamma^2}~,                 \label{relBW}
\end{equation}
where $S_{a,b,\rho}$ are the spins of the involved particles, $q_{ab}=I_{ab}/\sqrt{q^2}$ is the center-of-mass (c.m.) momentum of the colliding particles $a$ and $b$, and
$\Gamma_{\rho \to ab} =  \overline{|M_{ab;\rho^{(\lambda)}}|^2}\, q_{ab} / 8\pi q^2$ denotes the $\rho \to ab$ decay width.

The total width of $\rho$-meson in its rest frame is the sum of the vacuum decay width and the collisional width in the nuclear medium:
\begin{equation}
  \Gamma = \Gamma_{\rm dec} + \Gamma_{\rm coll}~.                  \label{Gamma}   
\end{equation}
The off-shell decay width $\Gamma_{\rm dec}(m)= \Gamma_{\rho \to \pi \pi}(m)+\Gamma_{\rho \to e^+ e^-}(m)$,
where $m \equiv \sqrt{q^2}$ is the off-shell $\rho$-meson mass parameter, is dominated by the $\rho \to \pi \pi$ channel, for which
\begin{equation}
  \Gamma_{\rho \to \pi \pi}(m) = \Gamma_{\rho \to \pi \pi}^0 \left(\frac{q_{\rm cm}(m)}{q_{\rm cm}(m_\rho)}\right)^3 \frac{m_\rho}{m}
                             \frac{1+(q_{\rm cm}(m_\rho)R)^2}{1+(q_{\rm cm}(m)R)^2}~,                     \label{Gamma_rho2pipi}
\end{equation}
where $q_{\rm cm}(m)=\sqrt{m^2/4-m_\pi^2}$ is the c.m.~momentum of the decay pions.
In the present calculations the following values of the constants are used: $m_\rho=775.5$~MeV, $\Gamma_{\rho \to \pi \pi}^0=149.1$~MeV.
The value of the parameter $R=1$~fm (see \cite{Manley:1992yb}) is universally set for all resonance decay widths (see Eq.(\ref{Gamma_R2rhoN}) below).

The collisional width of the $\rho$-meson in the nuclear medium is calculated in the semi-classical approximation, i.e.\ by using the loss 
term of the collision integral. For simplicity, isospin-symmetric nuclear matter is assumed. This leads to the following expression for the width in the $\rho$ meson's rest frame:
\begin{equation}
  \Gamma_{\rm coll} = \gamma_{\rm Lor} \langle v_{\rho N} \sigma_{\rho N} \rangle \rho_N~,              \label{Gamma_coll}
\end{equation}
where $\rho_N=\rho_n+\rho_p$ is the total nucleon density, $\sigma_{\rho N}=(\sigma_{\rho n}+\sigma_{\rho p})/2$ is the isospin-averaged
total $\rho$-meson nucleon cross section, $v_{\rho N}=I_{\rho N} / q^0 p^0_N$ is the relative velocity of the $\rho$-meson and the nucleon,
and $\gamma_{\rm Lor}=q^0/m$ is the Lorentz factor of the $\rho$-meson. $\langle \ldots \rangle$ denotes the averaging over nucleon Fermi
motion. The collisional width just defined is given in the so-called low-density approximation which is reflected in the linear dependence on density. 

At $\sqrt{s} \ltapprox 2$ GeV, the total $\rho N$ cross section is saturated by 
the resonance production channels.\footnote{See Table A.3 in Ref.~\cite{Buss:2011mx} for the full list of non strange resonances included in GiBUU.
The $N^*$ ($I=1/2$) resonances having rating of only one '*' were excluded from the calculations. 
There are also the non-resonant contributions of strangeness production $\rho N \to Y K$ included according to Ref.~\cite{Larionov:2011fs}.
They are, however, of minor importance for the present study.}  
Similar to Eq.(\ref{relBW}), the corresponding partial resonance cross sections  $\rho N \to {\cal R}$ are also given by the relativistic Breit-Wigner formula:
\begin{equation}
  \sigma_{\rho N \to {\cal R}} = \frac{2S_{\cal R}+1}{6} \, \frac{4\pi}{q_{\rho N}^2}\, 
  \frac{s \Gamma_{\cal R}\Gamma_{{\cal R} \to \rho N}}{(s-m_{\cal R}^2)^2 + s\Gamma_{\cal R}^2}~,                 \label{relBW_rhoN2R}
\end{equation}
where $S_{\cal R}$, $m_{\cal R}$ and $\Gamma_{\cal R}$ are, respectively, the spin, pole mass and the total width of the resonance ${\cal R}$; 
$s=(q+p_N)^2$ is the squared c.m. energy of $\rho$ and $N$;
$q_{\rho N} = [(s+q^2-m_N^2)^2/4s - q^2]^{1/2}$ is the c.m. momentum of $\rho$ and $N$; $\Gamma_{{\cal R}\to \rho N}$ is the partial ${\cal R} \to \rho N$ decay width
for the off-shell $\rho$ mass $\sqrt{q^2}$ (so-called 'in-width') for which the parameterization of Ref.~\cite{Manley:1992yb} is used:
\begin{equation}
    \Gamma_{{\cal R} \to \rho N}(\sqrt{s}) = \sum_l \Gamma_l^0 \frac{q_{\rho N} B_l^2(q_{\rho N}R)}{\sqrt{s} \rho_l(m_{\cal R})}~,   \label{Gamma_R2rhoN}
\end{equation}
where $\Gamma_l^0$ is the partial decay width evaluated at $\sqrt{s}=m_{\cal R}$ for the relative orbital angular momentum $l$ of $\rho$ and $N$;
$B_l(x)$ is a Blatt-Weisskopf barrier-penetration factor \cite{Manley:1984jz}; and
\begin{equation}
    \rho_l(\sqrt{s}) = \int dq^2 {\mathcal A}(q^2) \frac{q_{\rho N}}{\sqrt{s}} B_l^2(q_{\rho N}R)~.    \label{rho_l}
\end{equation}
Here ${\mathcal A}(q^2)$ is the $\rho$ spectral function, Eq.(\ref{calA}), calculated without collisional broadening. Note that for simplicity we also neglect the
in-medium effects on the baryon resonance ${\cal R}$ in Eq.(\ref{relBW_rhoN2R}) by using the total width $\Gamma_{\cal R}$ in vacuum.

At $\sqrt{s} \approx 2$ GeV, the sum of resonance cross sections 
starts to underestimate the total phenomenological high-energy cross section (see, e.g.~\cite{Donnachie:1992ny,Sjostrand:2001yu,Falter:2004uc}). A smooth transition to the high-energy regime described by PYTHIA is reached by including the $\rho N \to \pi N$ background cross section which absorbs the missing 
part of the total phenomenological $\rho N$ cross section.

In order to explore the dependence of the collisional width on the density and the excitation energy 
of the average $\langle \ldots \rangle$ in Eq.~(\ref{Gamma_coll}), the nucleon momentum $\bvec{p}$ has been sampled by Monte Carlo
according to a probability distribution $dP \propto  n_{\bvec{p}} d^3p$ where $n_{\bvec{p}}$ is the Fermi distribution at some finite temperature:
\begin{equation}
    n_{\bvec{p}} = \frac{1}{\exp((E_{\bvec{p}}-\mu)/T)+1}~,                      \label{Fermi}
\end{equation}
with $E_{\bvec{p}}=\sqrt{\bvec{p}^2+m_N^2}$. The chemical potential $\mu$ for the given values of the nucleon density $\rho_N$ and temperature $T$ has been
determined from
\begin{equation}
   \rho_N = \int \frac{4d^3p}{(2\pi)^3} \, n_{\bvec{p}}~.                \label{rho_N}
\end{equation}
We note that the `equivalent temperature' here is introduced only as a parameter to characterize the excitation energy. 

\begin{figure}
  \includegraphics[scale = 0.40]{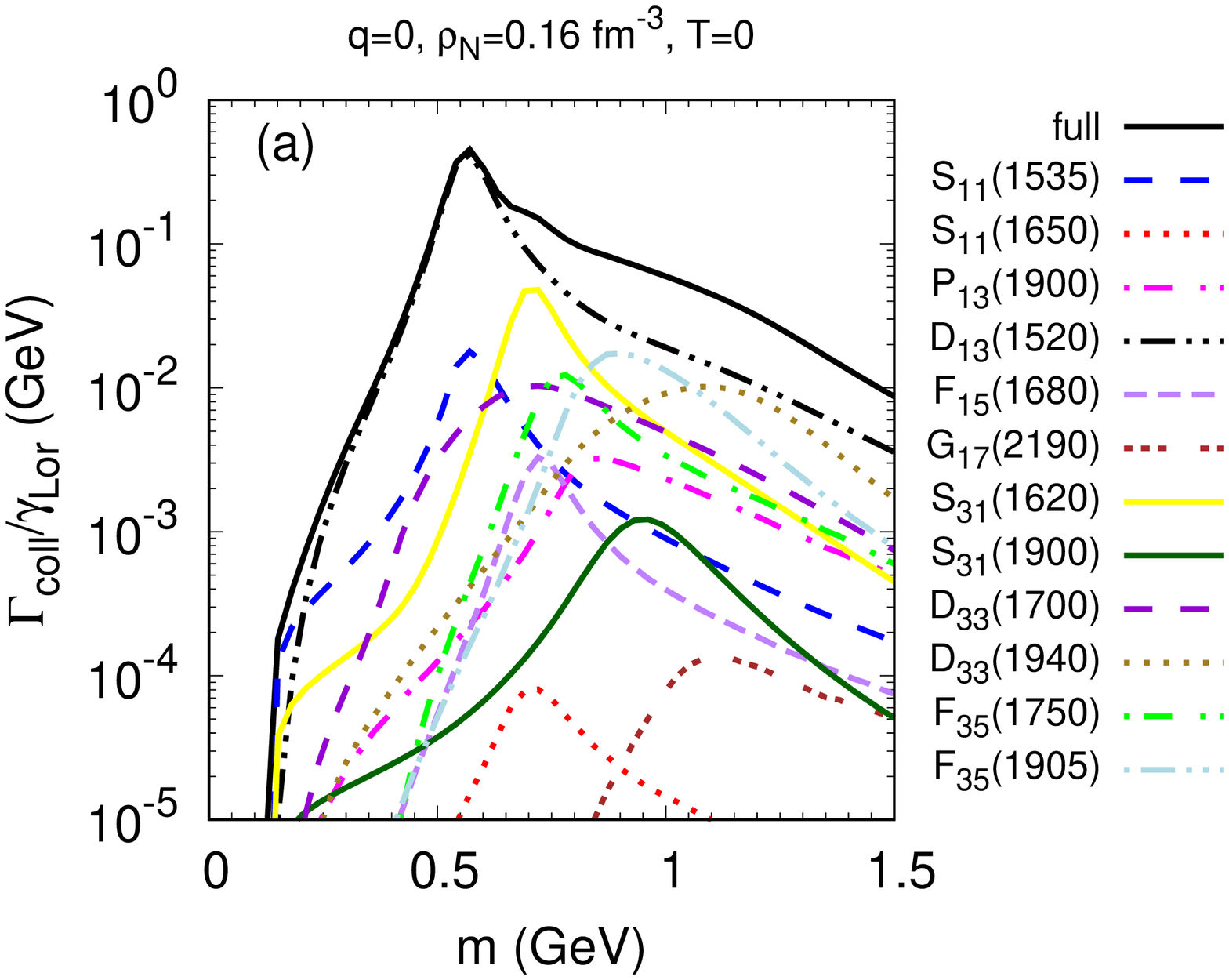}
  \includegraphics[scale = 0.40]{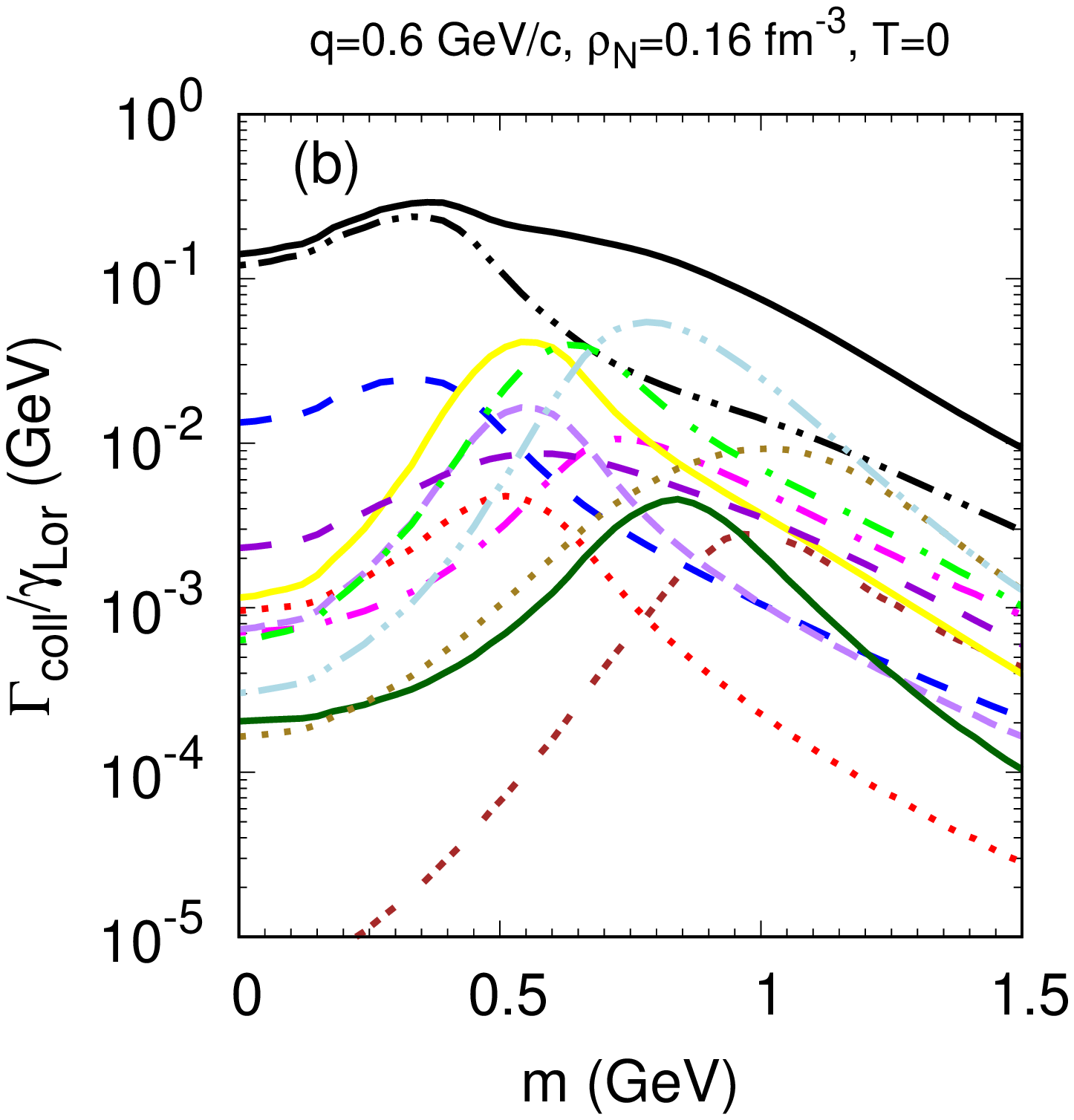}
  \includegraphics[scale = 0.40]{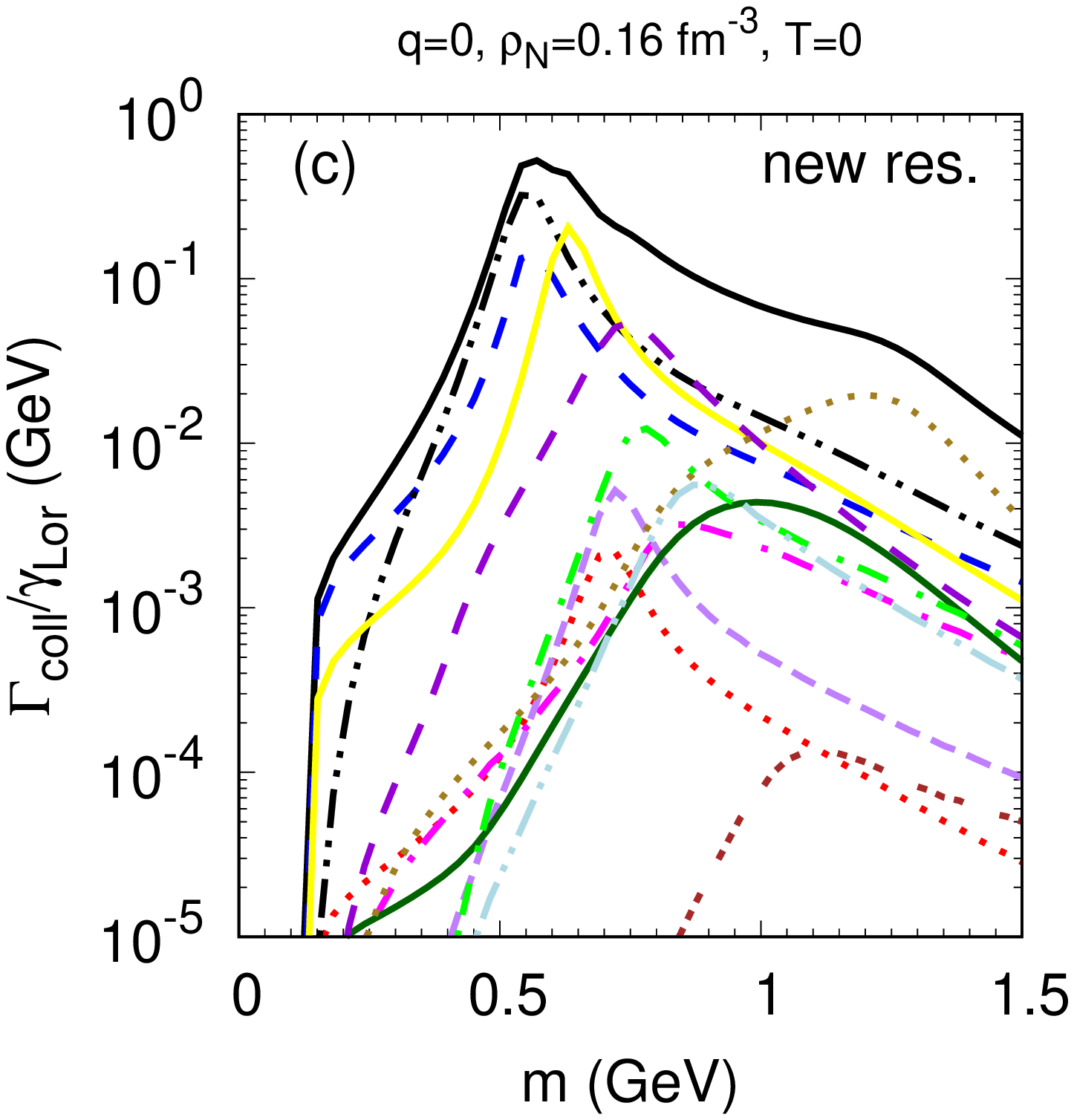}
  \includegraphics[scale = 0.40]{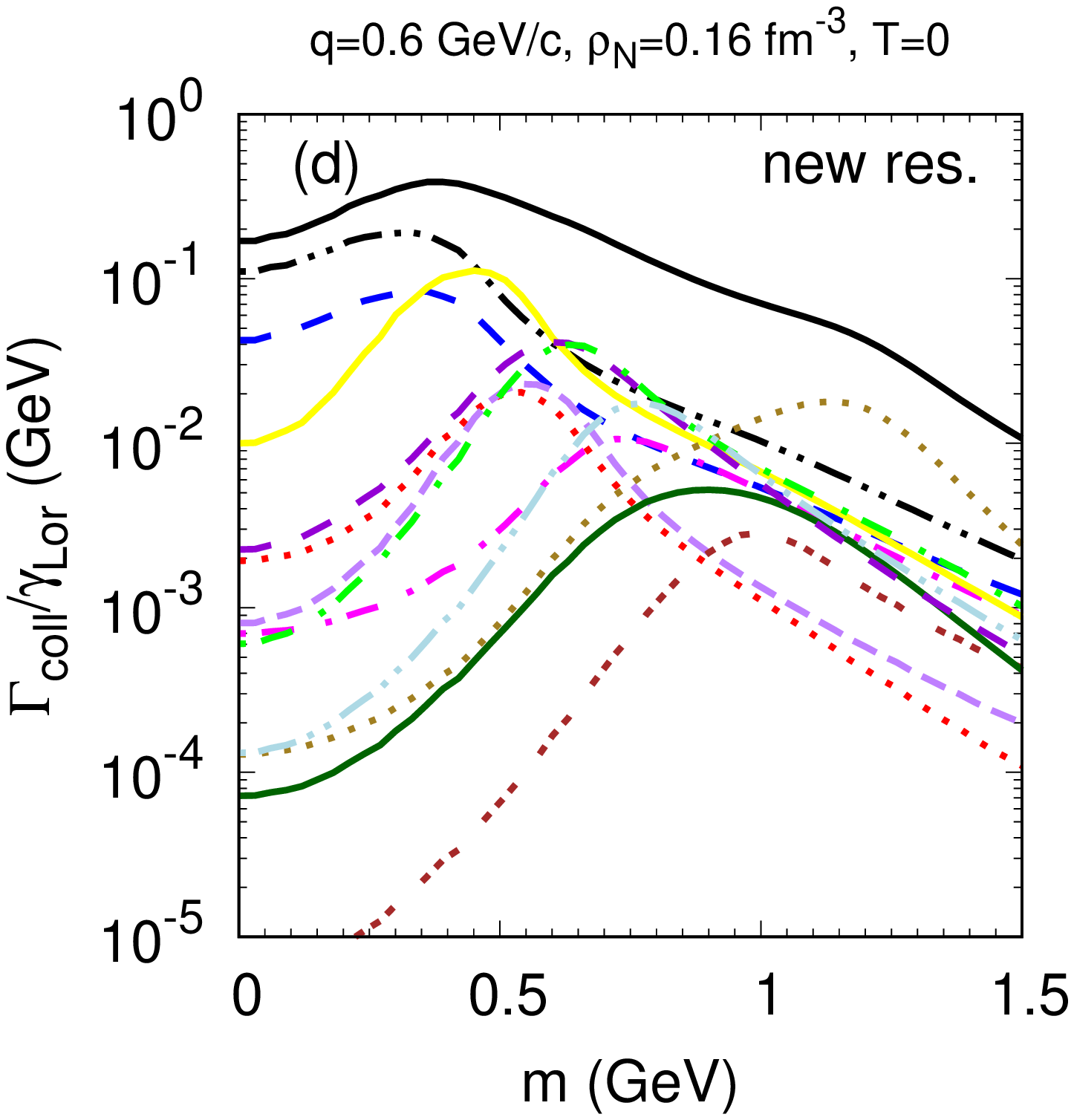}
  \caption{\label{fig:GamCollres} Collisional width of the $\rho$-meson in nuclear matter at saturation density 
    and zero temperature as a function of the meson mass at momentum $q=0$ (panels (a), (c)) and $q=0.6$ GeV/c (panels (b), (d)).
    In panels (a) and (b) the default resonance parameters of GiBUU (cf.~Ref.~\cite{Manley:1992yb}) are used.
    Panels (c) and (d) show the results with the updated resonance parameters of Ref.~\cite{Hunt:2018wqz}.  
    The width is calculated by summing partial resonance contributions. The full result is shown as the thick (black) solid line.
    Other lines show the partial contributions of the specific baryonic resonances as indicated. The Lorentz factor of the $\rho$-meson is divided out, i.e.~the plotted width is given in the rest frame of nuclear matter.}
\end{figure}

Figs.~\ref{fig:GamCollres} (a) and (b) show the collisional width of the $\rho$-meson calculated in ground-state nuclear matter within the baryon resonance model using the 
resonance parameters of Ref.~\cite{Manley:1992yb}, i.e.~those used in default GiBUU calculations.
The set of resonance parameters \cite{Manley:1992yb} was obtained within the multichannel unitarity analysis of $\pi N$ scattering data.
At $q=0$, the dominant contributions are given by the resonances with large $s$-wave couplings to the $\rho N$ state:
$D_{13}$ $N^*(1520)$ ($\Gamma_{\rho N}^{L=0}/\Gamma_{\rm tot}=21\%$) and $S_{31}$ $\Delta^*(1620)$ ($\Gamma_{\rho N}^{L=0}/\Gamma_{\rm tot}=25\%$).
As expected, at finite momentum of the $\rho$-meson, the resonances with non-zero angular momentum coupling to the $\rho N$ state
grow in importance. At $q=0.6$ GeV/c, the contribution of $N^*(1520)$ remains dominant, however, also the resonances
with $p$-wave coupling to the $\rho N$ state, i.e.\ $F_{35}$ $\Delta^*(1750)$ ($\Gamma_{\rho N}^{L=1}/\Gamma_{\rm tot}=22\%$)
and $F_{35}$ $\Delta^*(1905)$ ($\Gamma_{\rho N}^{L=1}/\Gamma_{\rm tot}=87\%$) contribute significantly. We also observe strongly increased
contributions of $S_{11}$ $N^*(1650)$ ($\Gamma_{\rho N}^{L=2}/\Gamma_{\rm tot}=3\%$) and $G_{17}$ $N^*(2190)$ ($\Gamma_{\rho N}^{L=2}/\Gamma_{\rm tot}=29\%$)
at finite momentum of $\rho$. 

In the recent Ref.~\cite{Hunt:2018wqz}, the nucleon resonance parameters have been updated including both $\pi N$ and $\gamma N$ scattering data.
In order to assess the influence of these updates, 
we have also used the new parameters of the resonances coupled to the $\rho N $ channel according to Ref.~\cite{Hunt:2018wqz}. The collisional width of the $\rho$-meson with the updated resonance parameters 
is shown in Figs.~\ref{fig:GamCollres} (c) and (d). As compared to the default parameters, the most pronounced changes occur with the new ones for $D_{13}(1520)$, $S_{11}(1535)$, and $S_{31}(1620)$. The branching ratio of the $(\rho N)_S$ channel decreases from $21\%$ to $14\%$ for $D_{13}(1520)$,
while it increases from $2\%$ to $14\%$ for $S_{11}(1535)$. The mass of $S_{31}(1620)$ decreases from 1672 MeV to 1589 MeV, while the total width
decreases from 154 MeV to 107 MeV. As a result, the collisional width of $\rho$ meson is slightly larger at small masses with the new resonance parameters.

\begin{figure}
  \includegraphics[scale = 0.50]{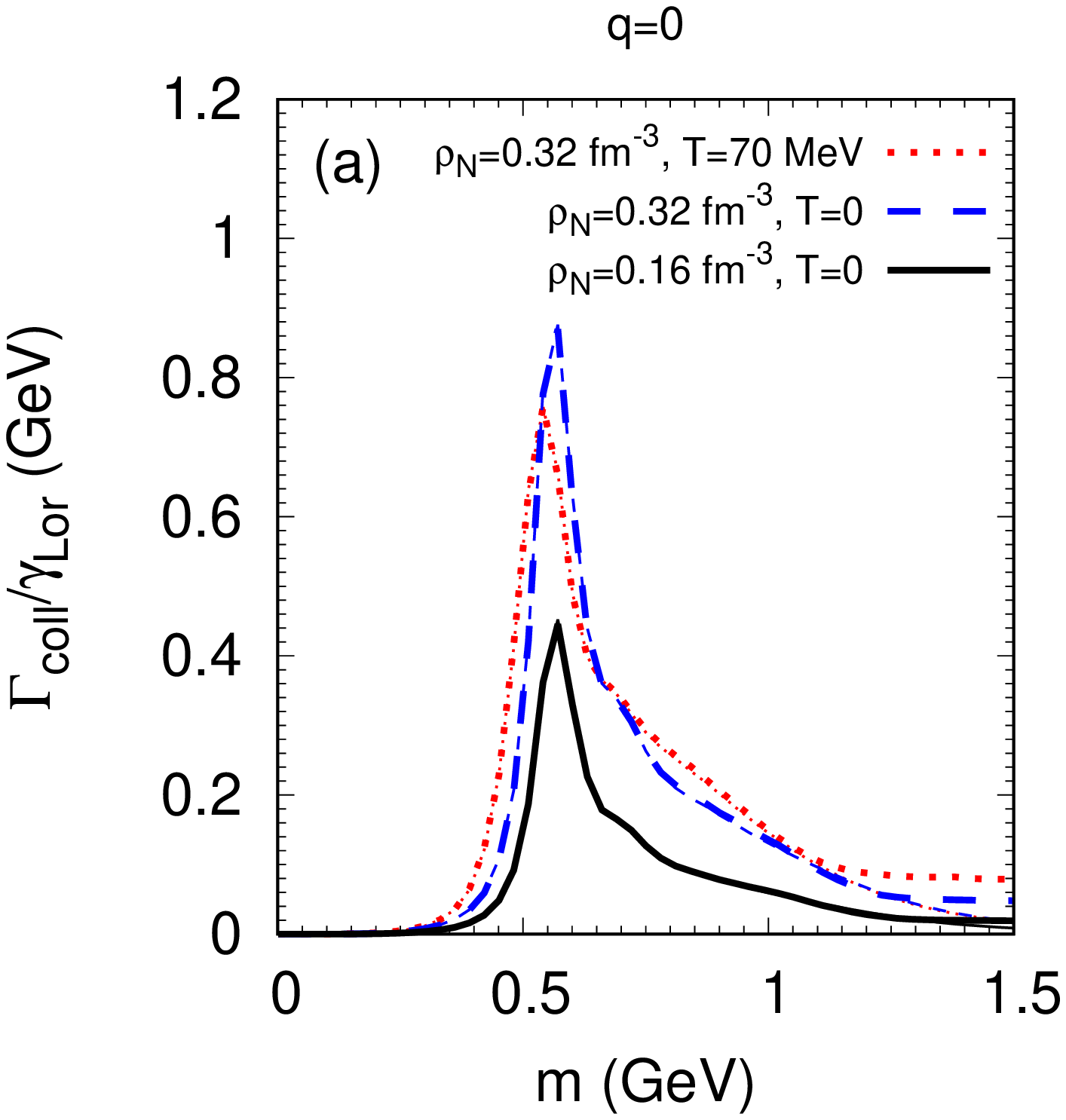}
  \includegraphics[scale = 0.50]{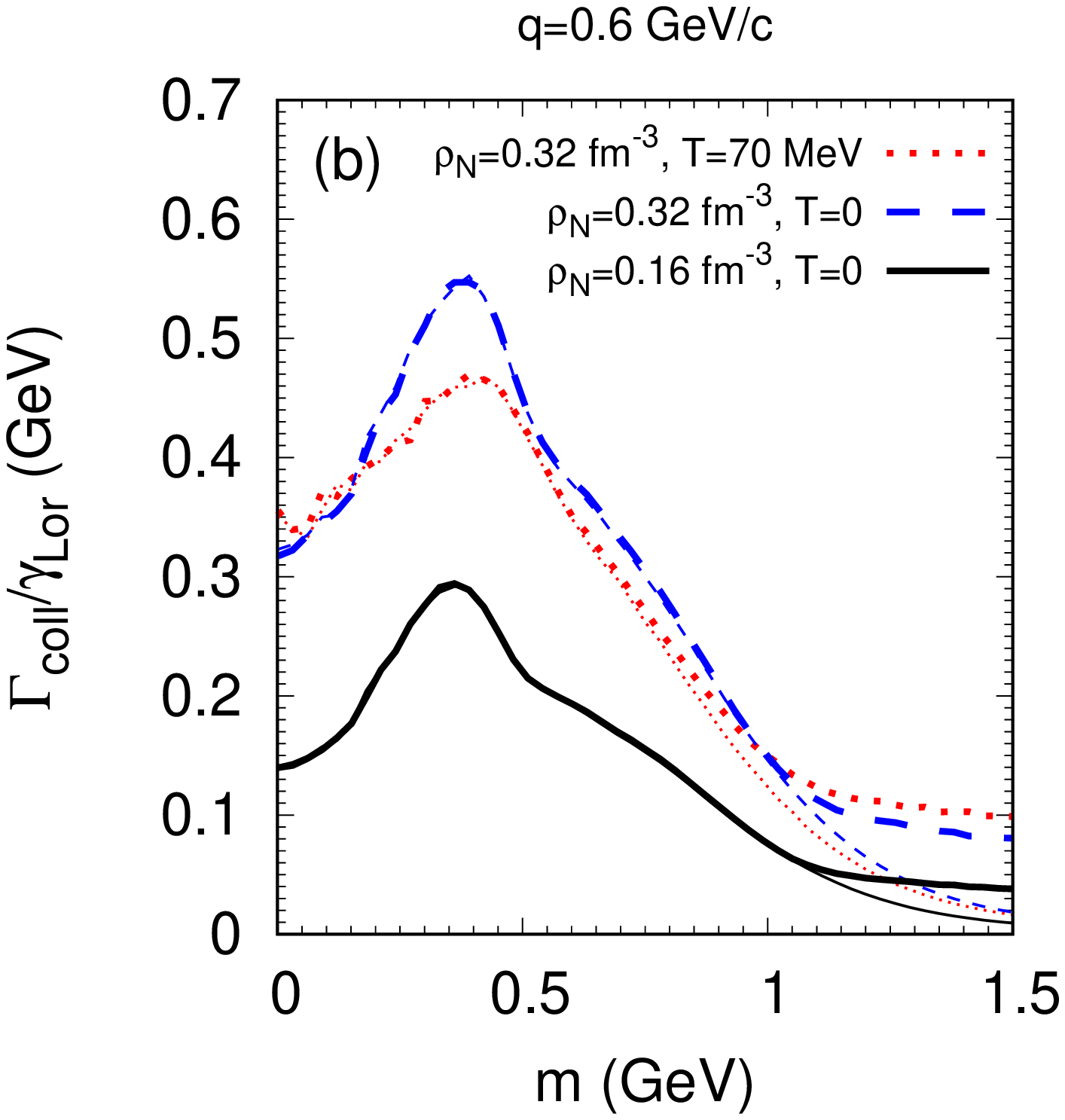}
  \caption{\label{fig:GamColl} Collisional width of the $\rho$-meson as a function of invariant mass in nuclear matter at different densities and temperatures, solid (black): $\rho_N=0.16$ fm$^{-3}$, $T=0$; long-dashed (blue): $\rho_N=0.32$ fm$^{-3}$, $T=0$;
    and short-dashed (red): $\rho_N=0.32$ fm$^{-3}$, $T=70$~MeV,
      as a function of the meson mass at momentum $q=0$ (a) and $q=0.6$ GeV/c (b).
    Thick and thin lines correspond to resonance + high energy
    and pure resonance model calculations, respectively.
    The Lorentz factor of the $\rho$-meson is divided out.}
\end{figure}
Fig.~\ref{fig:GamColl} shows the invariant-mass dependence of the collisional width of the  $\rho$-meson. 
At $T=0$ we observe an approximate scaling $\Gamma_{\rm coll} \propto \rho_N$. Increasing the temperature leads to a smearing of the mass dependence. This is because at finite $T$ the range of $\sqrt{s}$ of the colliding $\rho N$ pair
becomes broader for a fixed four momentum of the $\rho$. This leads to less pronounced baryon resonance structures.
Overall, the collisional width is comparable to or even larger then the vacuum $\rho$ width ($\approx 149$ MeV). Thus, significant modifications of the $\rho$ spectral function due to collisional broadening can be expected. 

\begin{figure}
  \includegraphics[scale = 0.50]{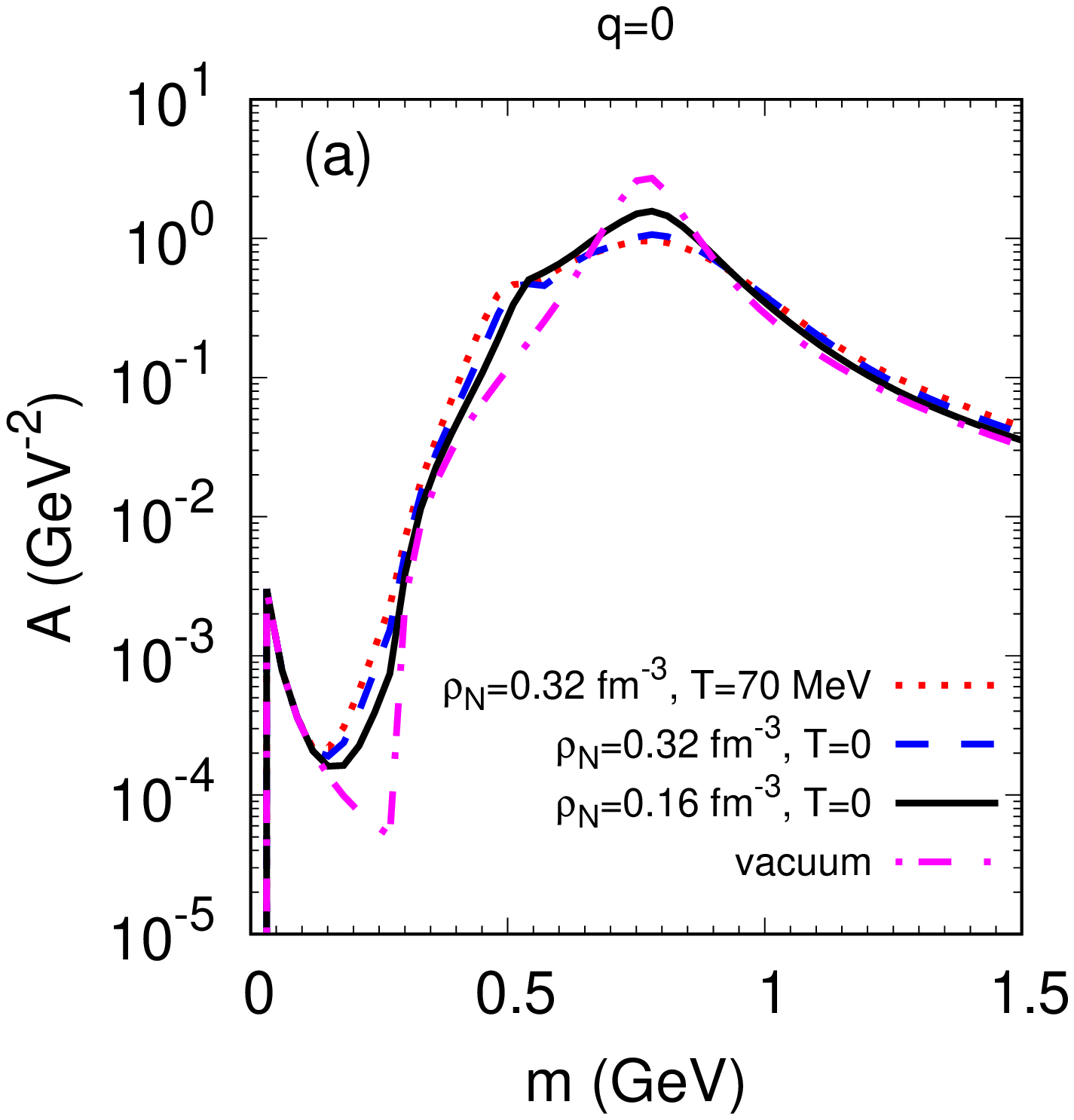}
  \includegraphics[scale = 0.50]{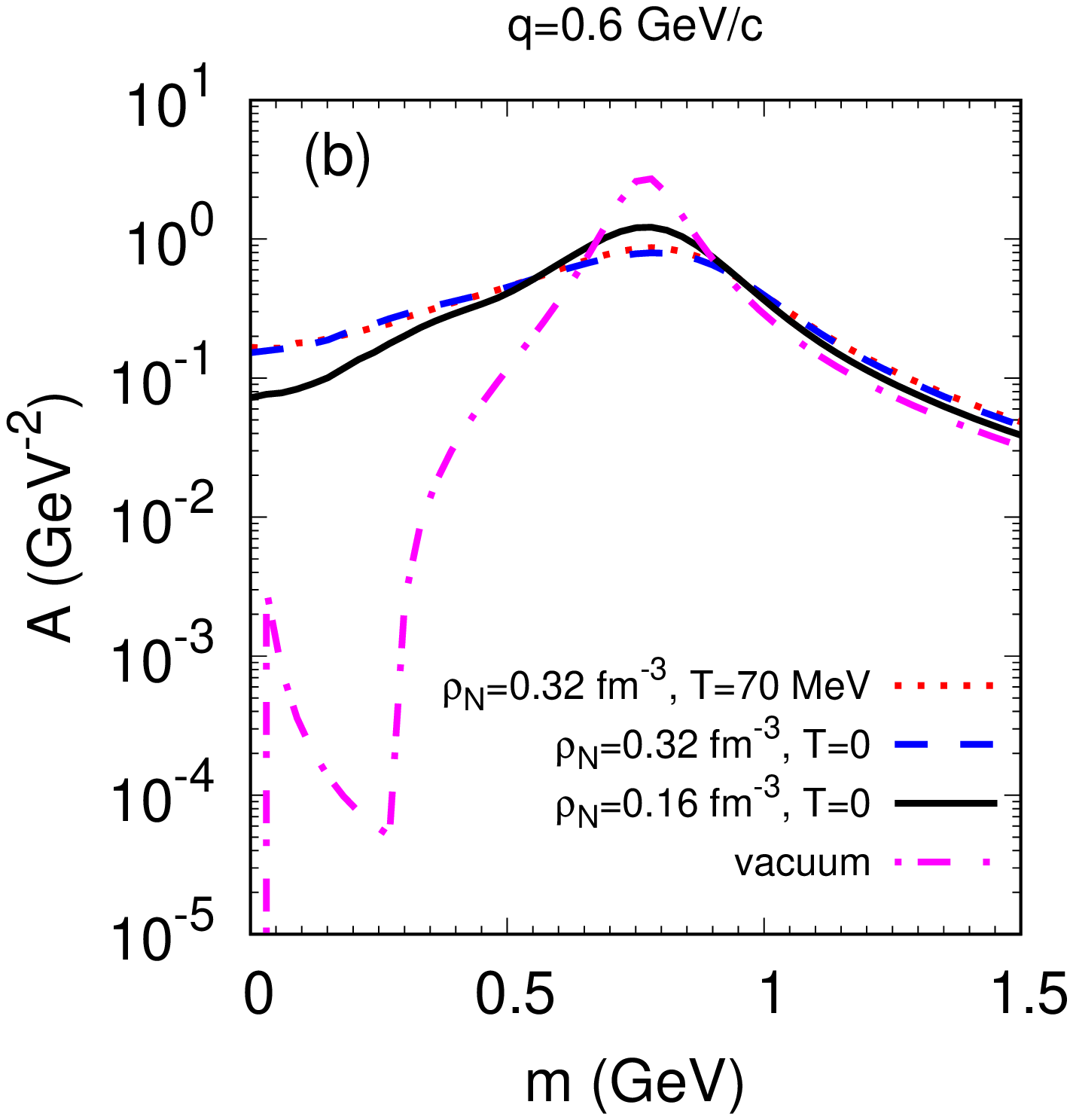}
  \caption{\label{fig:spfun} Spectral function of the $\rho$-meson in nuclear matter, Eq. (\ref{calA}), at different densities and temperatures
    as a function of the meson mass at momentum $q=0$ (a) and $q=0.6$ GeV/c (b).
    Solid (black) line -- $\rho_N=0.16$ fm$^{-3}$, $T=0$; long-dashed (blue) line -- $\rho_N=0.32$ fm$^{-3}$, $T=0$;
    short-dashed (red) line -- $\rho_N=0.32$ fm$^{-3}$, $T=70$ MeV; dot-dashed (magenta) line -- $\rho_N=0$.
    The collisional width is calculated in the resonance + high energy model.}
\end{figure}

The resulting in-medium modifications to the spectral function of the $\rho$-meson are shown in Fig.~\ref{fig:spfun}. For a $\rho$ meson at rest, the effects of the nuclear medium are only
marginal. The spike seen near the $e^+e^-$ threshold is a consequence of the oversimplified description of the $\rho$ width below the $2\pi$ threshold, as discussed earlier in connection with Eq.\ (\ref{Gamma_V2ee}).
Moreover, the behaviour of the free $\rho$ width in this low mass region is missing additional contributions, not included here, from $\rho \to \mu^+ \mu^-$ and, in particular, the  $\rho \to \pi^0 \gamma$ decay channel that has an order of magnitude larger branching ratio as compared to $\rho \to e^+e^-$  \cite{Tanabashi:2018oca}.

In contrast, at finite momentum of the $\rho$, there is a dramatic enhancement of the spectral strength at small invariant masses
due to the additional collisional width. The effect of temperature is quite small and more visible for the meson at rest, in form of a moderate broadening of the spectral strength. 

\begin{figure}
  \includegraphics[scale = 0.50]{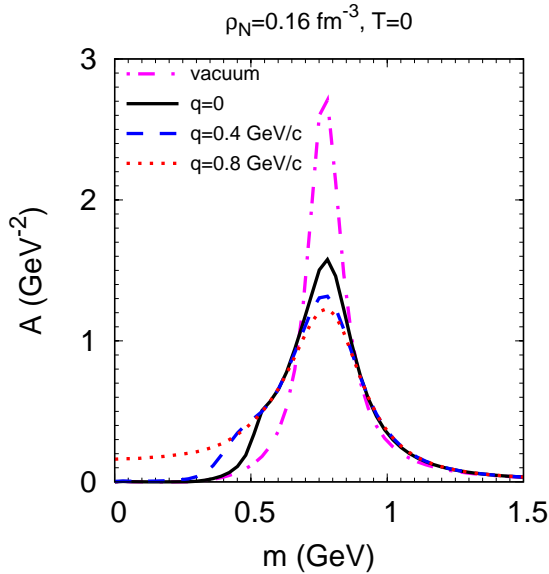}
  \caption{\label{fig:spfun4comp} Spectral function of the $\rho$-meson in ground-state nuclear matter for
    different values of the three momentum $|\bvec{q}|$: 0 -- solid (black) line, 0.4 GeV/c -- long-dashed (blue) line,
    0.8 GeV/c -- short-dashed (red) line. The vacuum spectral function is shown by the dot-dashed (magenta) line.}
\end{figure}
For comparison with Fig.~8 of Ref.~\cite{Peters:1997va}, we present in Fig.~\ref{fig:spfun4comp} the spectral function
calculated at three different values of the $\rho$-meson three momentum. We observe a tendency of a broadening towards
lower invariant masses with increasing $q$. However, our results show a somewhat smaller collisional broadening as compared to the self-energy calculations of Refs.~\cite{Peters:1997va,Post:2003hu}. 
In particular, the double-humped structure in the transverse spectral function obtained there due to the coupling to the $N^*(1520)$ resonance is missing here. 
Given the simplicity of our resonance model for the in-medium $\rho$ spectral function, on the other hand, the overall agreement with the more sophisticated resonance models of Refs.~\cite{Peters:1997va,Post:2003hu} is quite compelling.

\section{Results}
\label{results}

\begin{figure}
  \includegraphics[scale = 0.50]{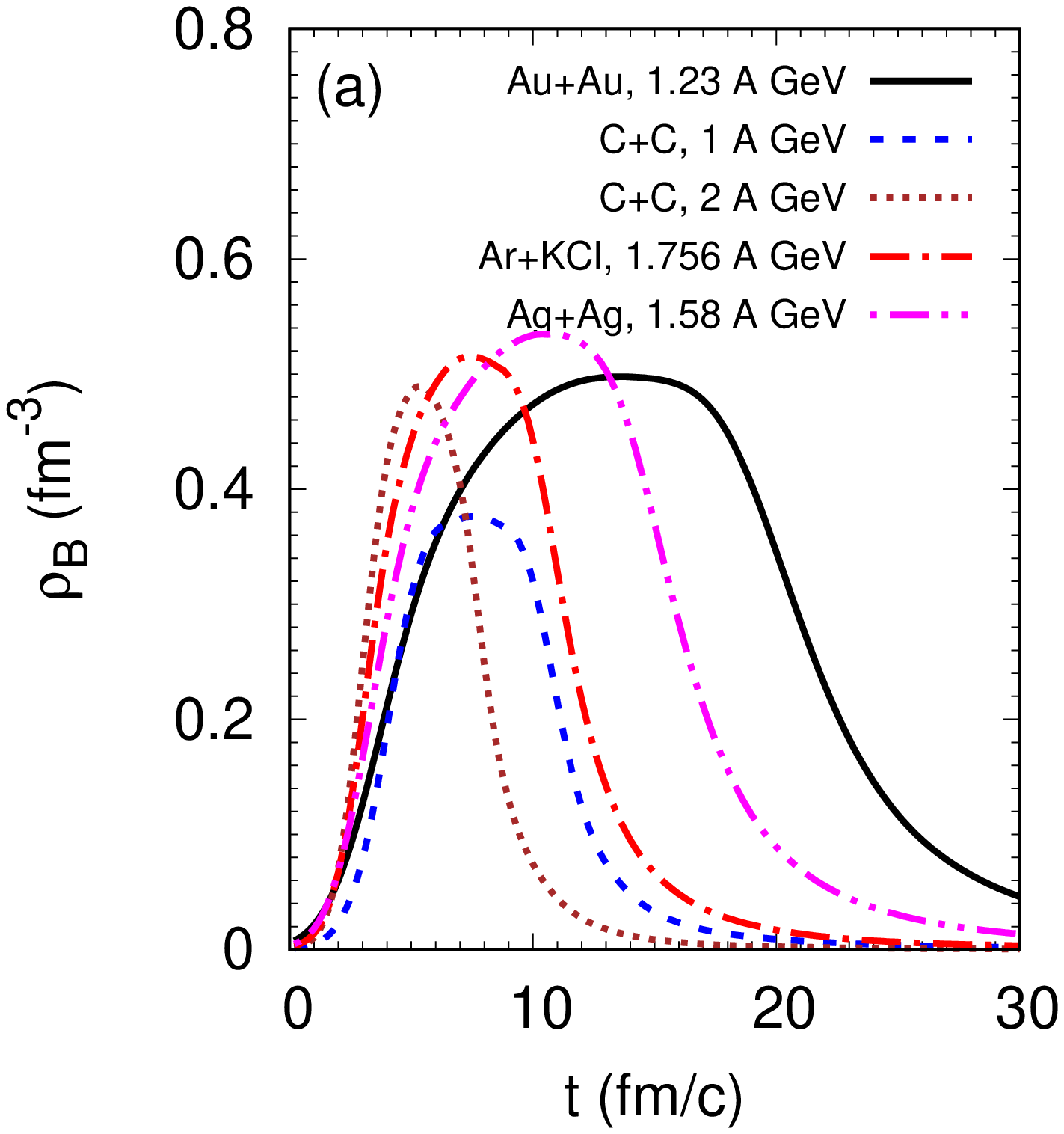}
  \includegraphics[scale = 0.50]{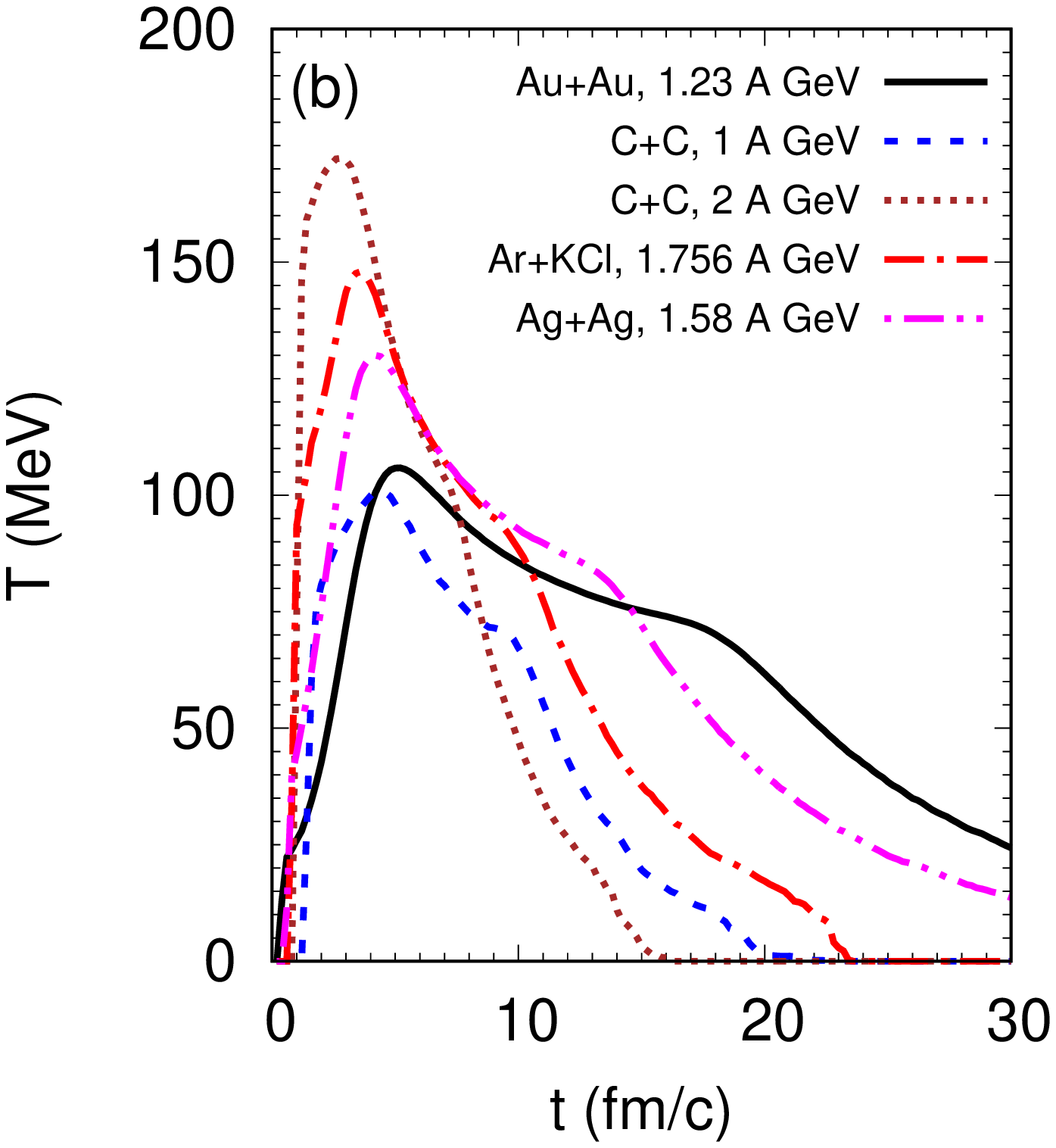}
  \caption{\label{fig:DensTemp}(Color online) The central baryon density (a) and temperature (b) vs time for Au+Au at $1.23 A$~GeV (black solid line),
  C+C at $1 A$~GeV (blue dashed line),  C+C at $2 A$~GeV (brown dotted line), Ar+KCl at $1.756 A$~GeV (red dash-dotted line), and Ag+Ag at $1.58 A$~GeV (magenta dash-double-dotted line) The impact parameter is set to zero for all systems.}
\end{figure}

Before comparing our calculations with experimental data in Sec.~\ref{HIC} below, we first consider the time evolution
of some selected observables. 
Fig.~\ref{fig:DensTemp} shows baryon density and temperature in the position of the center-of-mass
for central collisions of C+C at $1 A$~GeV and $2 A$~GeV, Ar+KCl at $1.756 A$~GeV, Ag+Ag at $1.58 A$~GeV, and Au+Au at $1.23 A$~GeV. 
The temperature has been extracted locally in position space
by fitting $\langle p^2 \rangle$ of the baryons in the local rest frame of nuclear matter using the Fermi distribution, Eq.~(\ref{Fermi}).
Note that this is only an effective equivalent temperature, the colliding systems are not necessarily fully equilibrated at any time.
It is also obvious that there is no thermal equilibrium at the initial inter-penetration stage when the two
counter-streaming flows of nucleons only start to decelerate each other by elastic and inelastic $NN$ collisions. Thus, the extremely high
temperatures at the beginning of the collision must not be considered as real physical ones, but demonstrate that $T$ is just a parameter to fit the non-equilibrium momentum distribution of the baryons by a Fermi distribution having the same $\langle p^2 \rangle$. Earlier studies have in fact  indicated that, at the relatively low bombarding energies considered here, full thermal equilibrium is not achieved during the high-density phase of the collision. \cite{Lang:1991qa,Lang:1992jz}. 

The density evolution looks quite simple and intuitive: it consists of a compression stage followed by a plateau behaviour, and finally the expansion of the system.  
Central baryon densities of up to $2-3 \, \rho_0$ are reached; where $\rho_0 =0.16$ fm$^{-3}$ is the nuclear saturation density.
For the Au+Au system, the temperature evolution pattern shows up a bump at $t \approx 18$ fm/c
correlated with the end of the density plateau. At these times, the calculated temperature
is about $T\approx 80$ MeV. We have checked that the sum of the pion and $\Delta$ multiplicities, as an estimate of the total
inelastic production, saturates at approximately the same time. This implies that during the density plateau stage the temperature drops mainly due to inelastic production.
In contrast, at the expansion stage the temperature drops mostly because fast nucleons leave the central zone faster,
a feature of kinetic free streaming. A similar behavior is observed for other colliding systems.

\begin{figure}

 \vspace*{-.6cm}
  \includegraphics[width=\linewidth]{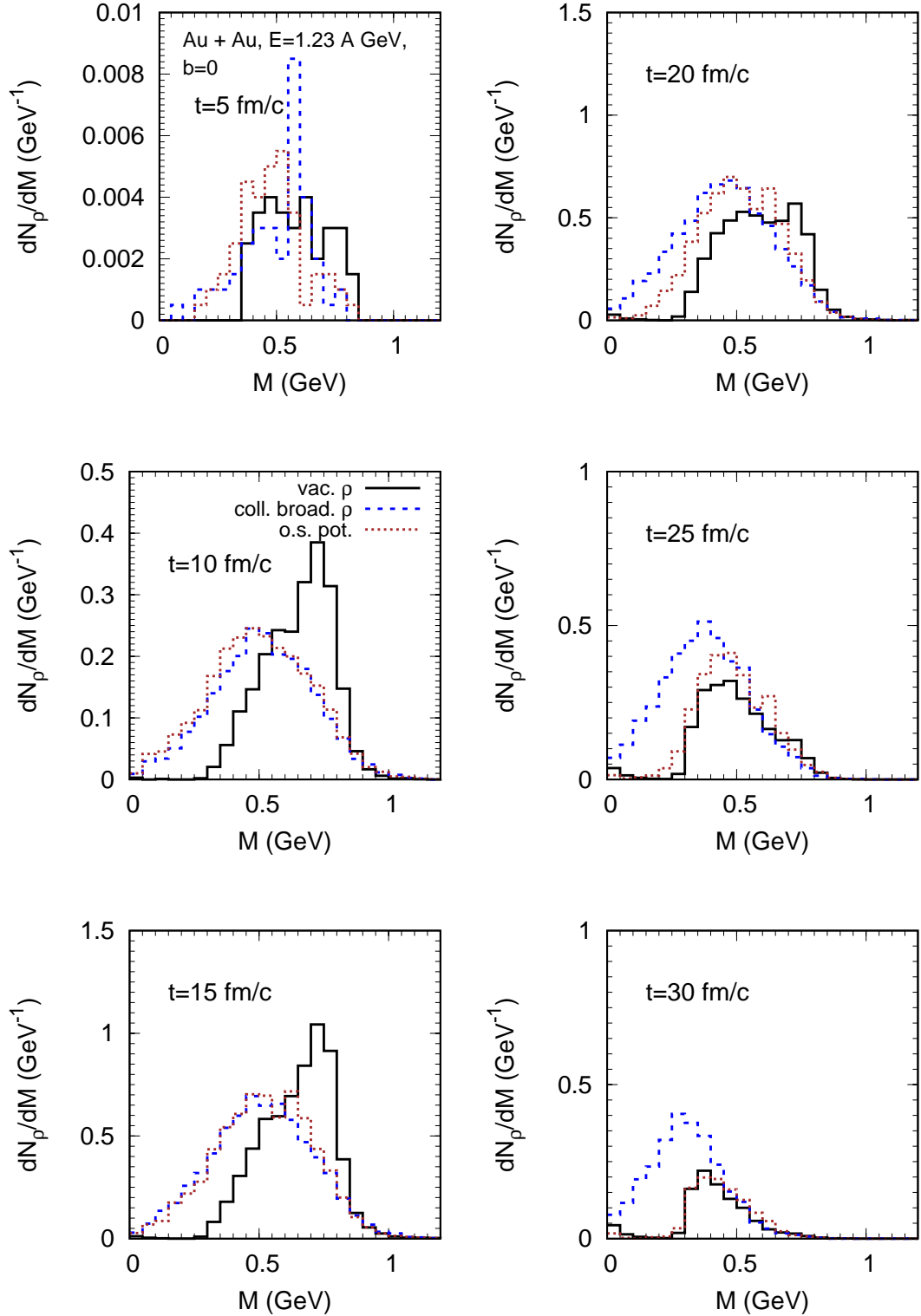}
  
  \vspace{-1.2cm}
  \caption{\label{fig:InvMassDist_rho}(Color online) Invariant mass per-event spectra of $\rho$-mesons produced at different moments in time for central Au+Au ($b=0$~fm) at $1.23 A$~GeV: Black solid line -- vacuum width, blue dashed line -- vacuum and collisional width,
    brown dotted line -- vacuum and collisional width with OSP ansatz.}
\end{figure}

Fig.~\ref{fig:InvMassDist_rho} displays the time evolution of the $\rho$-meson invariant mass distribution.
At the initial stage of a collision, hard first-chance $NN$ collisions and multistep processes allow to produce baryonic resonances in a broad mass range (in particular, the $N^*(1520)$ that dominates the $\rho$ production). Thus, there is not much phase-space limitation for the $\rho$ production here, and a large part of the $\rho$ spectral strength, including the on-shell peak region, is populated in $N^* \to \rho N$ decays.
With increasing time, baryon resonance production becomes governed by soft $\pi N$ collisions which ultimately leads to smaller invariant masses of the produced $\rho$'s. Another effect, which shifts the $\rho$ strength to smaller invariant masses with increasing time is the $\rho \to \pi \pi$ decay, since the $\Gamma_{\rho \to \pi \pi}$ width grows with the invariant mass of the $\rho$, see Eq.~(\ref{Gamma_rho2pipi}). Altogether, this leads to a softening of the $\rho$-invariant-mass spectrum at $t \gtapprox 20$~fm/c, even in calculations with vacuum $\rho$ spectral function.

Including the collisional width in the $\rho$ spectral function leads to a softer $\rho$ invariant-mass spectrum at the early stage, $t \ltapprox 15$ fm/c, due to
the spreading of the $\rho$ spectral strength towards lower invariant masses, see Fig.~\ref{fig:spfun} (b) above.
The collisional width of the  $\rho$-meson has been determined by using the local values of baryon density and temperature calculated on the spatial grid with step size $\approx 0.3-1.1$ fm in each direction (the exact values depend on the colliding system and are chosen suitable to resolve the density gradients).
In the calculation that treats off-shell $\rho$'s as free particles with fixed masses (blue dashed lines in Fig.~\ref{fig:InvMassDist_rho}) 
the excess of the low-mass $\rho$'s ($m \ltapprox 0.3$ GeV) survives until the later times. However, in the calculation using the OSP ansatz (brown dotted lines in Fig.~\ref{fig:InvMassDist_rho}) the low-mass $\rho$'s gradually migrate
closer to the on-shell peak. Therefore, at late times the off-shell transport, through the OSP ansatz of Eqs.\ (\ref{rDotOSP}) -- (\ref{ImPi}), 
produces $\rho$ mass distributions close to those with vacuum $\rho$ width.
These observations are in-line with HSD model calculations \cite{Bratkovskaya:2007jk}.

\begin{figure}

  \vspace*{-2.0cm}
  \includegraphics[width=\linewidth]{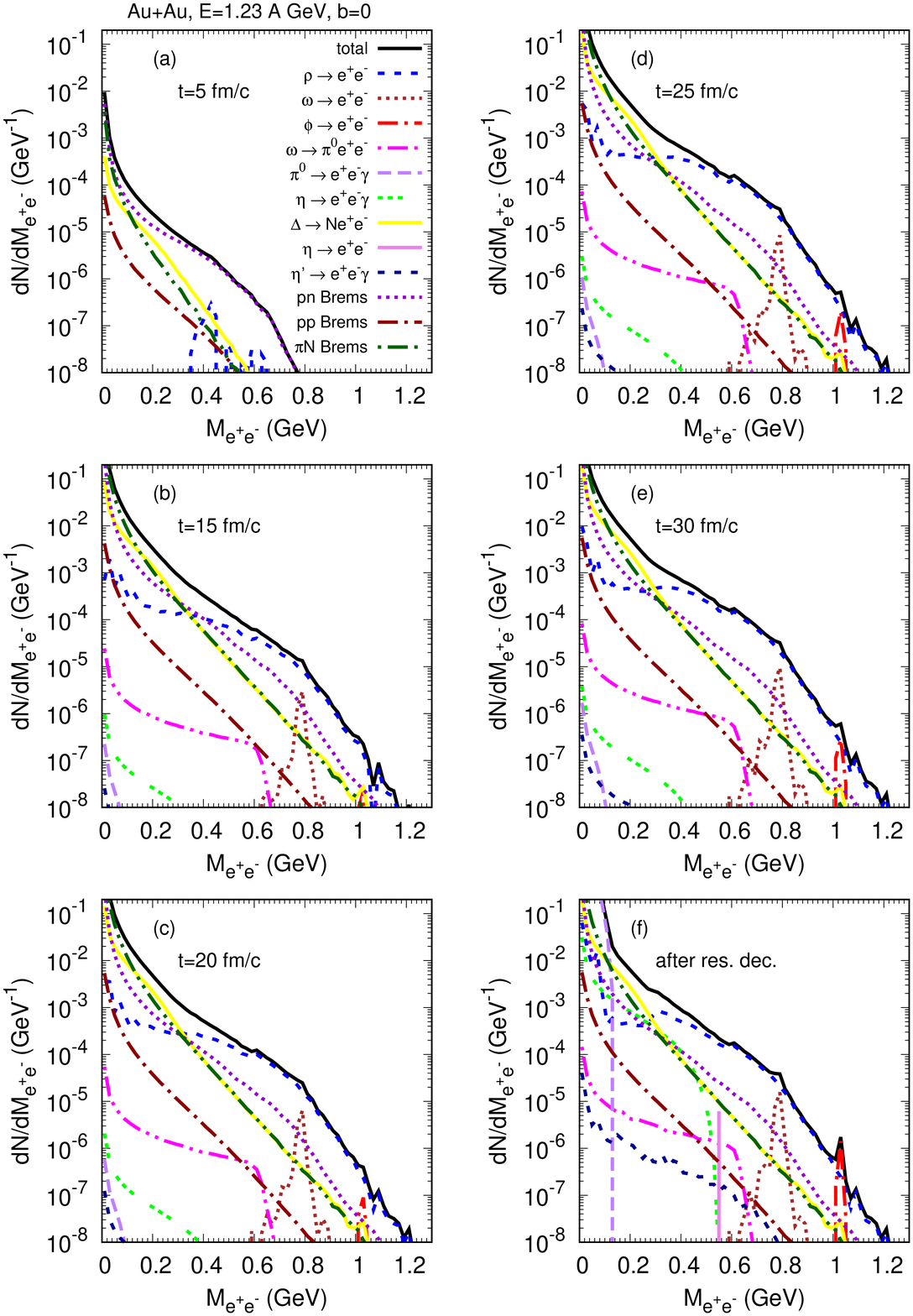}
  
  \vspace{-1.0cm}
  \caption{\label{fig:dNdM_res_woFilt_rmf}(Color online) Invariant mass per-event spectrum of $e^+e^-$ pairs produced in Au+Au at $1.23 A$~GeV,
    $b=0$ fm at different time moments (a)-(e), and at the end of time evolution $t=60$ fm/c after adding up electromagnetic decays of long-lived particles (f).
    Solid black lines show the total spectrum.  
    Other lines show the partial contributions of the different production channels as indicated.
    Calculations include the collisional width of the $\rho$ meson within the OSP ansatz.
    Full acceptance is assumed.}
\end{figure}

Fig.~\ref{fig:dNdM_res_woFilt_rmf} shows the time evolution of the dilepton spectrum in central Au+Au collisions at $1.23 A$~GeV.
At early times ($t \approx 5$ fm/c) the spectrum is practically saturated by the $pn$ bremsstrahlung. Then we observe
a dramatic increase of the $\rho \to e^+ e^-$ component that quickly becomes dominant at $M_{e^+e^-} = 0.5-1$ GeV.
The $\Delta$ Dalitz contribution also develops quite early and dominates at $M_{e^+e^-} \ltapprox 0.3$ GeV.
The $\pi N$ bremsstrahlung contribution is quite close to that of the $\Delta$ Dalitz one.

\begin{figure}

\vspace*{-2.5cm}
  \includegraphics[width=\linewidth]{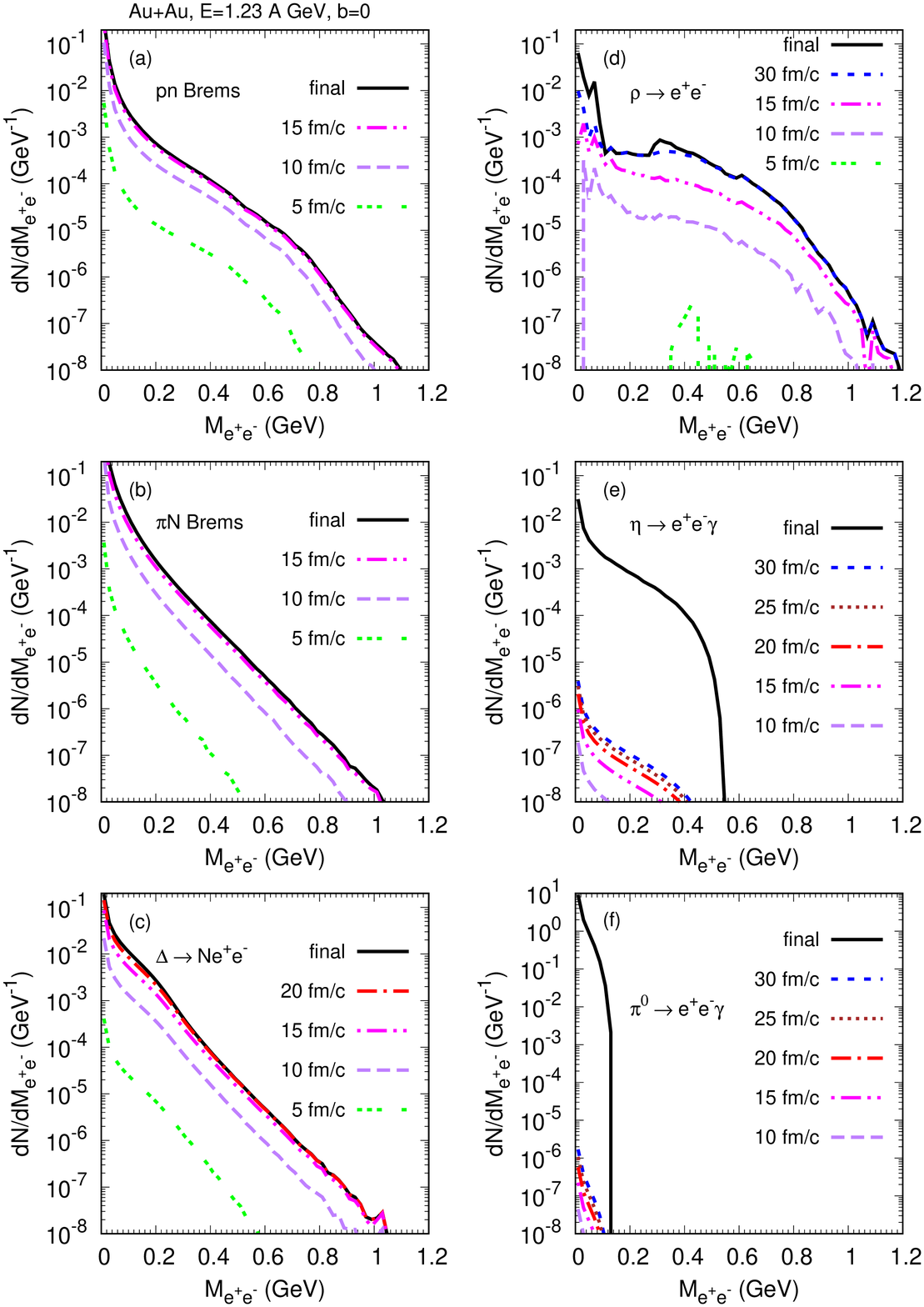}
  
\vspace{-1.2cm}
  \caption{\label{fig:dNdM_res_woFilt_rmf_sum}(Color online) Time evolution of the most important partial components of the $e^+e^-$ invariant mass spectrum
    from Au+Au at $1.23 A$~GeV, $b=0$ fm: $pn$ bremsstrahlung (a), $\pi N$ bremsstrahlung (b), $\Delta(1232)$ Dalitz decay (c), $\rho \to e^+e^-$ decay (d),
    $\eta$ Dalitz decay (e), and $\pi^0$  Dalitz decay (f). Final spectra (solid lines) are obtained at $t=60$ fm/c after adding up electromagnetic decays of long-lived particles.
    Calculations include the collisional width of the $\rho$ meson within the OSP ansatz. Full acceptance is assumed.}
\end{figure}

Individual views of the time evolution of the most important partial components of the dilepton mass spectrum are provided in Fig.~\ref{fig:dNdM_res_woFilt_rmf_sum}.
We observe that the $pn$ and $\pi N$ bremsstrahlung components practically saturate at 15 fm/c when the primary stopping is over and the system reaches the highest compression state
(see Fig.~\ref{fig:DensTemp} above). The $\Delta \to N e^+ e^-$ component at larger dilepton invariant masses, $M_{e^+e^-} \gtapprox 0.4$ GeV, also shows 
a quite early saturation. However, the $\Delta$ Dalitz decays still continue to populate the softer part of dilepton mass spectrum until late times. 

The time evolution of the $\rho \to e^+ e^-$ decay component is much slower which reflects the multistep processes of the $\rho$ production, mostly
mediated by the $N^*(1520)$. The spectrum around the pole mass of the $\rho$ is practically saturated at 30 fm/c since the life-time of the
$\rho$ meson at the pole mass is only 1.3 fm/c. Thus, large-mass $\rho$'s decay very quickly after decoupling from the fireball.
However, the $\rho$'s with masses only slightly above the $2\pi$ threshold are long-lived in vacuum. Thus, their dilepton decays continue until quite
late times, on the order of $\approx$ 60~fm/c. Note that the OSP ansatz leads to 
almost vanishing $\rho$ mass spectrum below $2m_\pi$ at late times,
except for a peak at extremely small invariant masses due to the growing partial width $\Gamma_{\rho \to e^+ e^-}$ towards small invariant masses
(see Fig.~\ref{fig:InvMassDist_rho}). This explains the behaviour of the $\rho \to e^+ e^-$ component in the dilepton mass spectrum
below $2m_\pi$.

\begin{figure}[ht]
  \includegraphics[scale = 0.65]{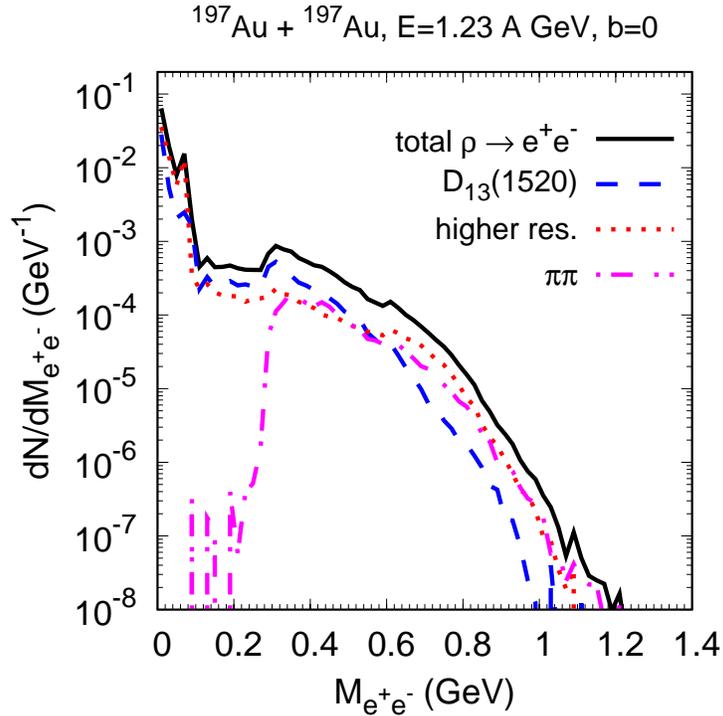}
  \caption{\label{fig:dNdM_res_woFilt_rmf_rho}(Color online) Dilepton invariant mass spectrum produced in $\rho \to e^+ e^-$ decays from Au+Au at $1.23 A$~GeV $b=0$ fm.
    Solid line -- total spectrum. Dashed and dotted lines -- partial contributions from $\rho$'s produced in decays of $N^*(1520)$ resonance and
    higher resonances, respectively. Dot-dashed line -- partial contribution from $\pi \pi \to \rho$ process.
    Calculations are done including the collisional width of $\rho$-meson within the OSP ansatz. Full acceptance is assumed.}
\end{figure}

The Dalitz decays of the $\pi^0$ and $\eta$ mesons have large branching fractions. 
Thus, their contributions are almost entirely dominated by time scales much larger than the GiBUU evolution time. While this is equivalent to the summation of the decays of all produced $\pi^0$'s and $\eta$'s at the end of the GiBUU time evolution, for consistency we also show in Fig.~\ref{fig:dNdM_res_woFilt_rmf_sum} the small contributions of $\pi^0$ and $\eta$ decays that occur during the GiBUU time evolution. 

 The final $\rho \to e^+ e^-$ component of the dilepton invariant mass spectrum is shown together with its sub-components in
  Fig.~\ref{fig:dNdM_res_woFilt_rmf_rho}. 
  Indeed, the decays of the $D_{13}$ $N^*(1520)$ resonance are the main source of $\rho$'s providing the largest contribution both in the total integrated
  spectrum and in the intermediate invariant mass region $0.2-0.4$ GeV. However, the decays of all higher resonances coupled to the $\rho N$ final state
  (cf.\ Fig.~\ref{fig:GamCollres} above) and the $\pi \pi$ collisions provide the dominant contribution at higher invariant masses $\gtapprox 0.6$ GeV.
  It is interesting that the $\pi \pi$ sub-component does not vanish below $2m_\pi$. This is related to the off-shell dynamics of the $\rho$ meson that changes its invariant mass. As discussed after Eq.~(\ref{ImPi}), the OSP ansatz has the effect that a $\rho$ meson moving towards higher density regions tends to shift away from the mass shell (and vice versa).

\subsection{Comparison with HADES data}
\label{HIC}

The HADES collaboration has measured inclusive dilepton spectra at SIS18 energies for the following
systems: $p+p$ collisions at beam energies of 1.25~GeV \cite{Agakishiev:2009yf},
2.2~GeV \cite{Agakishiev:2012tc}, and 3.5~GeV \cite{HADES:2011ab}, $d+p$ collisions at beam energy of $1.25 A$~GeV \cite{Agakishiev:2009yf},
C+C at $1 A$~GeV \cite{Agakishiev:2007ts} and $2 A$~GeV \cite{Agakichiev:2006tg},
Ar+KCl at $1.76 A$~GeV \cite{Agakishiev:2011vf}, and Au+Au at $1.23 A$~GeV \cite{Adamczewski-Musch:2019byl}.
Recently the measurements have also been performed for Ag+Ag at $1.58 A$~GeV although the data have not being published yet.  

Below, if not specially mentioned, the calculated spectra are smeared according to the HADES detector resolution and
filtered through the HADES acceptance filter. After that, the proper angular and momentum cuts are taken into account.
For the Au+Au system at $1.23 A$~GeV the dedicated acceptance filter does not exist yet. Thus, we have applied for that system the filter for $d+p$ at $1.25 A$~GeV where the magnetic field setting is similar \cite{Galatyuk_priv}.
After filtering the opening angle cut $\Theta_{e^+e^-} > 9\degree$ and restrictions on the $e^-$ and $e^+$ momenta $0.1~\mbox{GeV} < p_{e^\pm} < 1.1~\mbox{GeV}$ have been applied for Au+Au.

In heavy-ion collision simulations, the GiBUU evolution time was set to 30 fm/c. This is long enough such that practically only mean field potential interactions,
elastic rescattering of produced particles and resonance decays may occur later on.
These, however, do practically not influence hadron multiplicities and the spectra of produced dileptons. Note that particle multiplicities and spectra are calculated
taking into account the decays of resonances which survived until the end of the GiBUU time evolution.

The HADES data for heavy-ion collisions are given relative to the 
``total $\pi^0$ multiplicity'' (we will refer to this as the pseudo neutral-pion multiplicity in App.\ \ref{hadronnumbers}), $N_{\pi^0}$, experimentally defined as\
\begin{equation}
    N_{\pi^0} \equiv (N_{\pi^+}+N_{\pi^-})/2~,          \label{N_pi0}
\end{equation}
with the charged pion multiplicities $N_{\pi^\pm}$ obtained by extrapolation
to the full solid angle. In the following comparisons we do not use this normalization but instead compare with the dilepton data themselves since these are measured only in a limited acceptance window. A more detailed discussion of pion numbers is given in App.\ \ref{hadronnumbers}.

\subsubsection{$pp$ and $dp$ collisions}
\label{elem}

\begin{figure}
  \includegraphics[scale = 0.50]{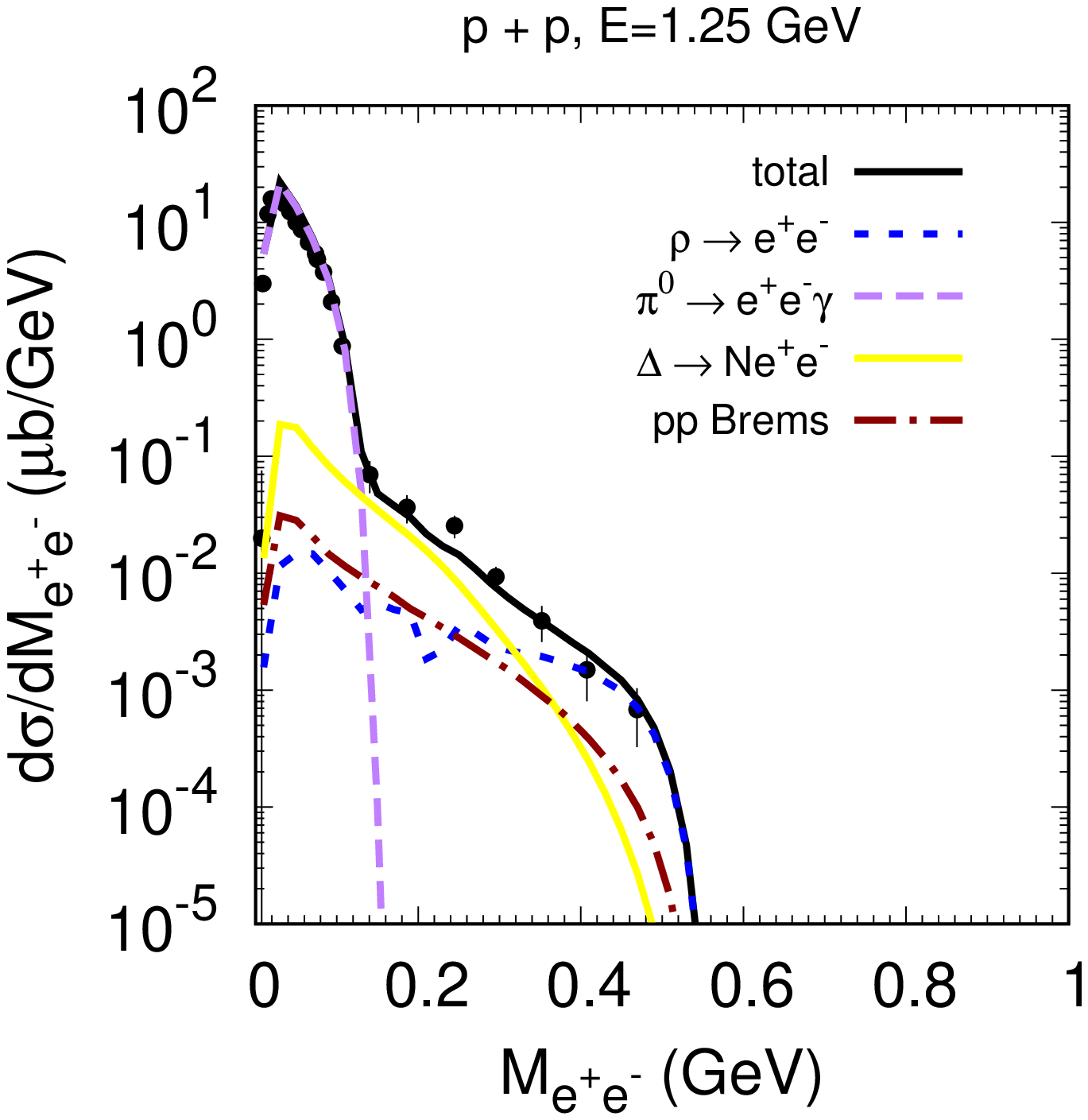}
  \includegraphics[scale = 0.50]{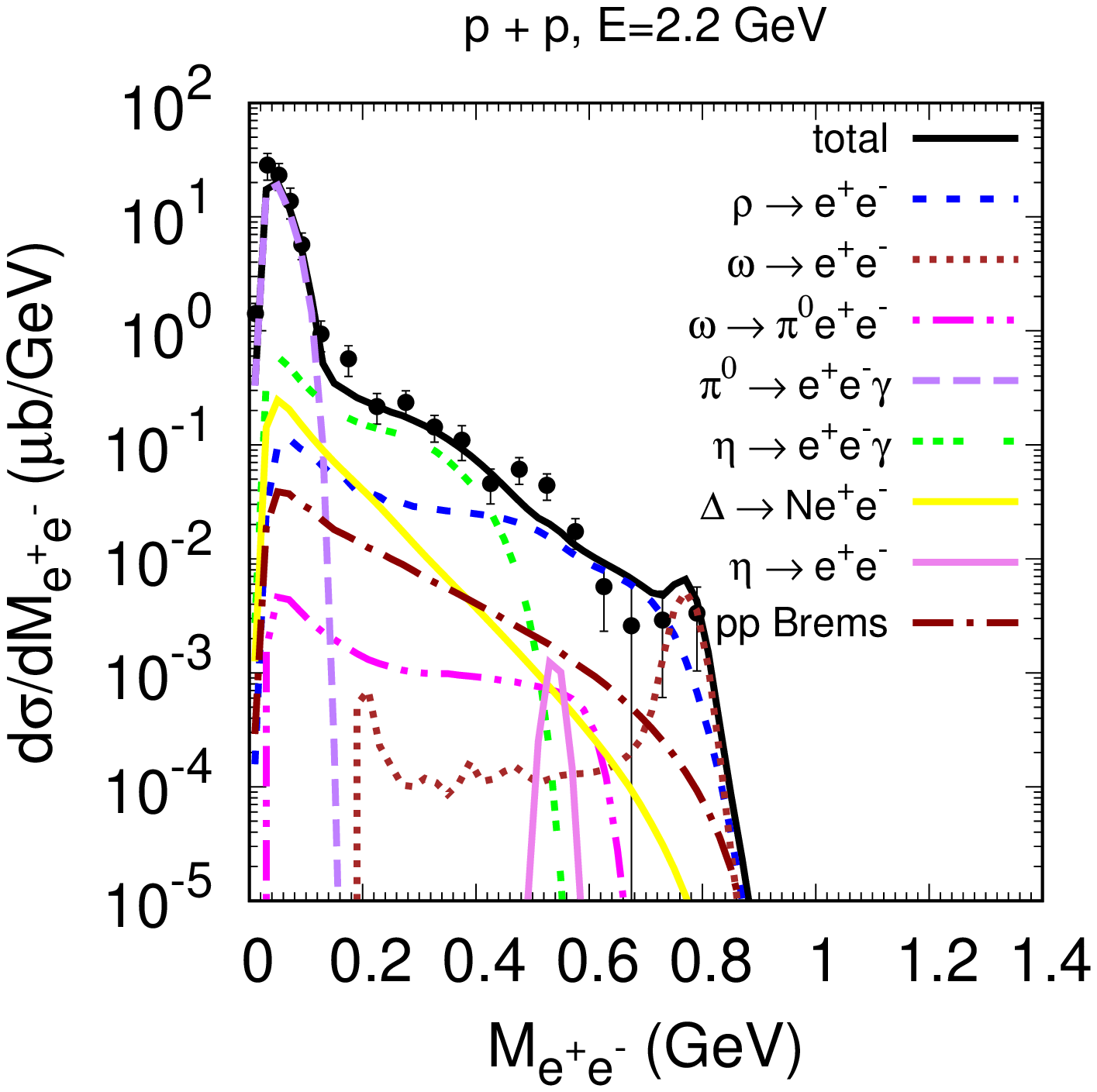}
  \includegraphics[scale = 0.50]{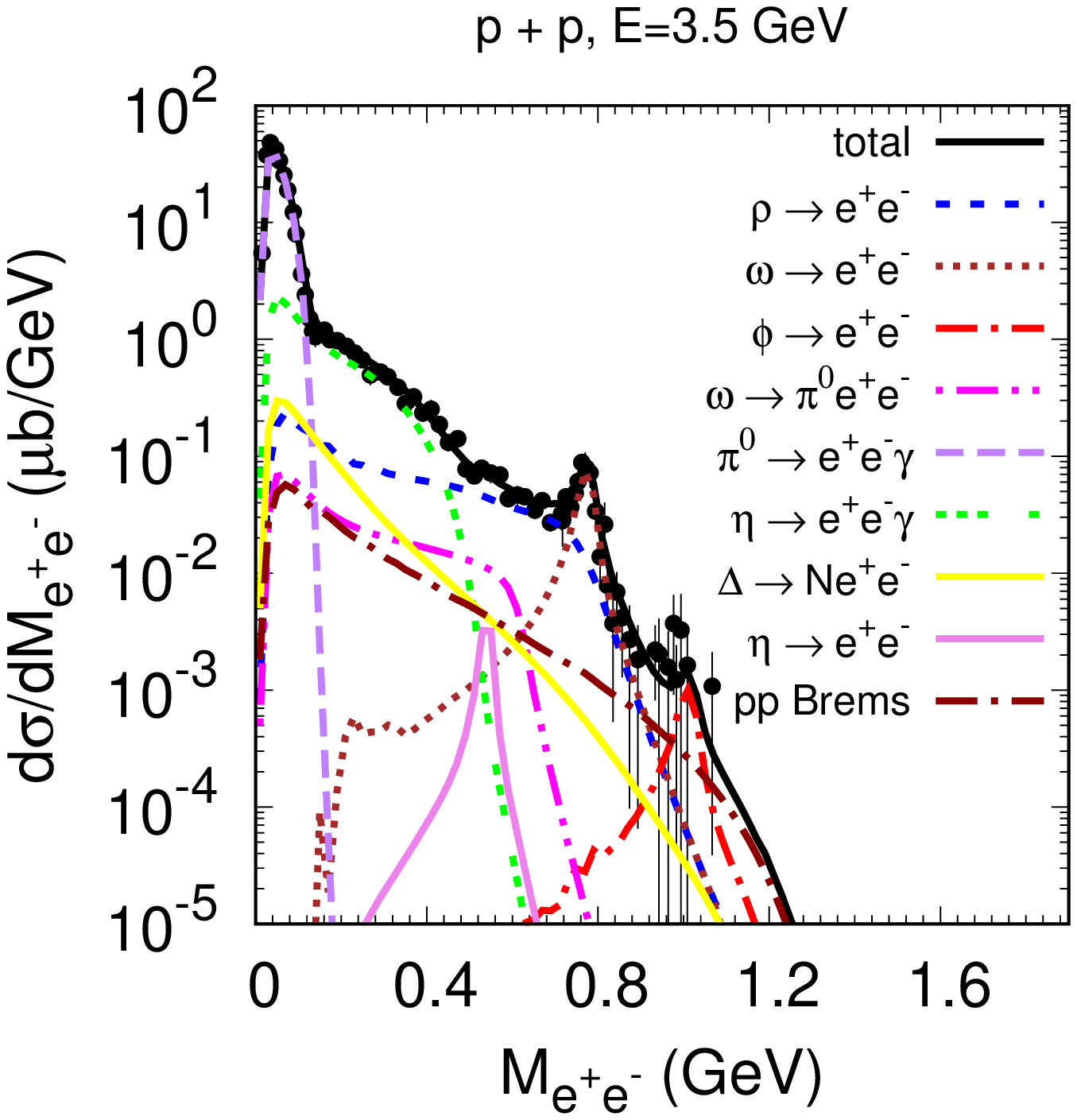}  
  \includegraphics[scale = 0.50]{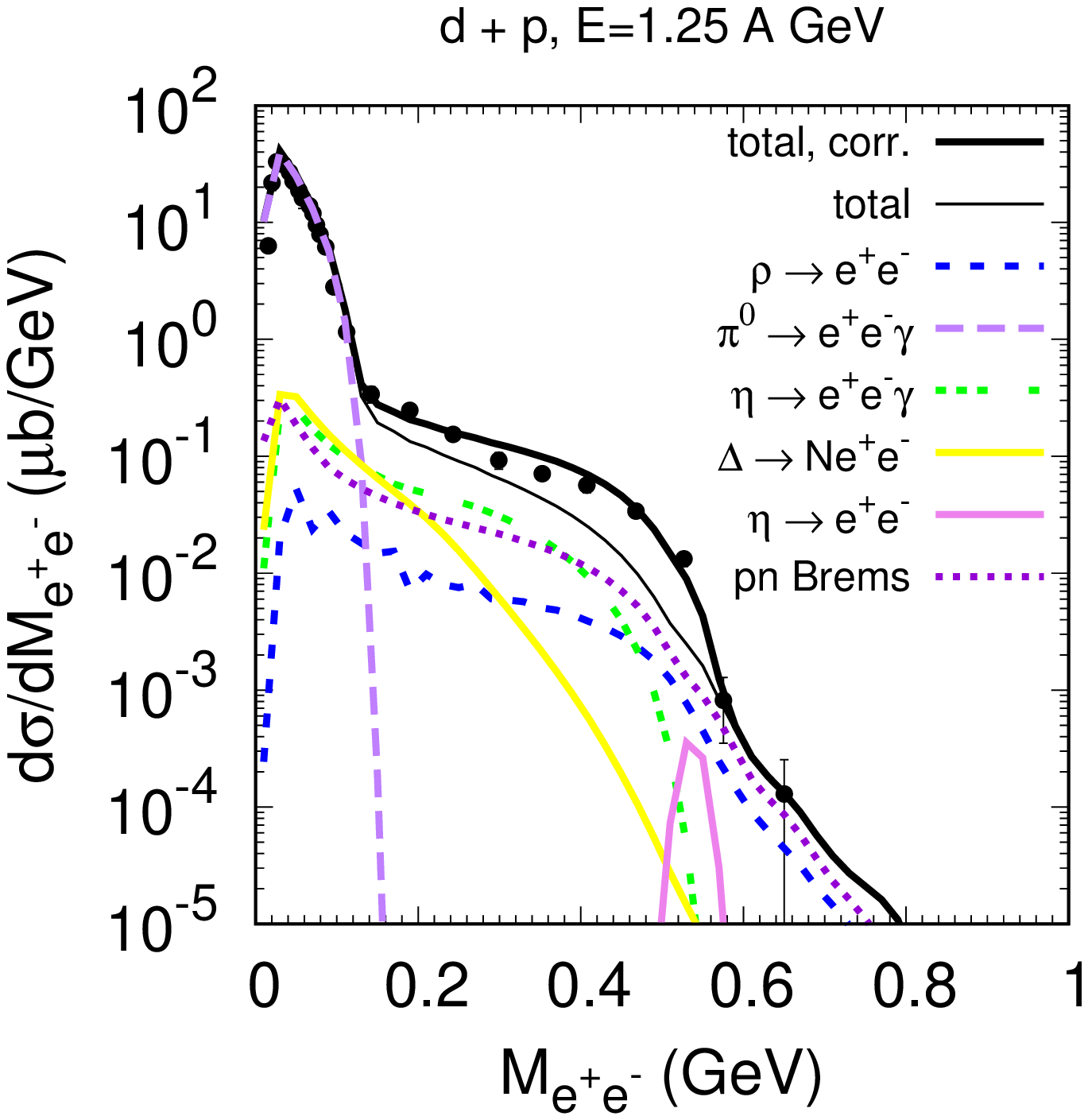}
  \caption{\label{fig:dNdM_res_pp&dp1.25gev}(Color online) Invariant-mass differential cross section of dilepton production in $p+p$ collisions
    at the beam energies 1.25, 2.2, and 3.5~GeV, and in $d+p$ collisions at beam energy $1.25 A$~GeV.
    Thick solid (black) lines show the total calculated cross sections. Other lines show the partial contributions
    of the different production channels as indicated. For the $d+p$ reaction, the cross section of the $pn$ bremsstrahlung
    component (see Eq.~(\ref{fM})) was thereby corrected in order for the total (thick black line) cross section to agree with experiment. The total cross without this correction is shown as the thin solid (black) line for comparison.
    Experimental data are from Refs.~\cite{Agakishiev:2009yf,Agakishiev:2012tc,HADES:2011ab,Agakishiev:2009yf}.
    In the $d+p$ reaction, the $n+p$ collisions were exclusively selected by detecting the fast forward spectator proton.} 
\end{figure}
Fig.~\ref{fig:dNdM_res_pp&dp1.25gev} shows the invariant mass $e^+e^-$ spectra from $p+p$ collisions at 1.25~GeV, 2.2~GeV, and 3.5~GeV,
as well as $d+p$ collisions at $1.25 A$~GeV. The experimental spectra from $p+p$ collisions are very well described by our GiBUU transport model simulations.

In the case of $d+p$ collisions, following Ref.~\cite{Agakishiev:2009yf} only $n+p$ collisions were taken into account in our calculations.
The neutron momentum spread in the deuteron has been taken into account by using the wave function of the full Bonn model \cite{Machleidt:1987hj}. 
Since $\eta$ production is below threshold in $NN$ collisions at 1.25~GeV, the $\eta$ Dalitz component for $d+p$ collisions at $1.25 A$~GeV
is predominantly due to neutron Fermi motion.
While the introduction of a coupling of the virtual photon to an exchanged charged pion led to a significant increase of the mass spectrum around $M_{e^+e^-}\approx 500$ MeV \cite{Shyam:2010vr} this is still not  sufficient to describe the experimental dilepton yield in this region.

There are at least two possible reasons for this remaining discrepancy. First, it might be due to the very simple OBE model used to describe elementary $NN$ scattering in Ref.~\cite{Shyam:2010vr}. 
Secondly, at the invariant masses near the quasi-free threshold $M_{e^+e^-}^{\rm max}=\sqrt{s_{NN}}-2m_N=0.545$ GeV the high-momentum part of the deuteron wave function which is subject to light-cone corrections \cite{Frankfurt:1977vc,Frankfurt:1981mk} might become relevant, which is not taken into account here.
In the present work, we decided to tune the elementary $p+n$ cross section to the experimental $d+p$ data at $1.25 A$~GeV by multiplying the $pn$ bremsstrahlung component of the dilepton production cross section by the factor,
\begin{equation}
    f(M)=C\frac{1 + w M^2/b^2}{(\exp[(a-M)/d]+1)(\exp[(M-b)/d]+1)} + 1~,       \label{fM}
\end{equation}
with dilepton invariant mass $M$ in GeV, $C=1.5$, $d=0.01$, $a=0.10$, $b=0.55$, and $w=3.0$. Note that the particular form of Eq.(\ref{fM})  is chosen for reasons of numerical convenience only. For a further discussion of this tuning factor see Appendix B.

In Fig.~\ref{fig:dNdM_res_pp&dp1.25gev} we show the resulting total dilepton mass spectrum, after multiplication of the $pn$ bremsstrahlung component (shown as dotted magenta line in the figure) by the factor $f(M)$ of Eq.~(\ref{fM}). For comparison, the corresponding total spectrum without this rescaling is also
shown as thin black line. Once fixed phenomenologically from the elementary reaction, we then use the same factor $f(M)$ with the same parameters also for all other cross sections, such as momentum distributions, for heavy-ion collisions at beam energies around $1 A$~GeV.

\begin{figure}
  \includegraphics[scale = 0.50]{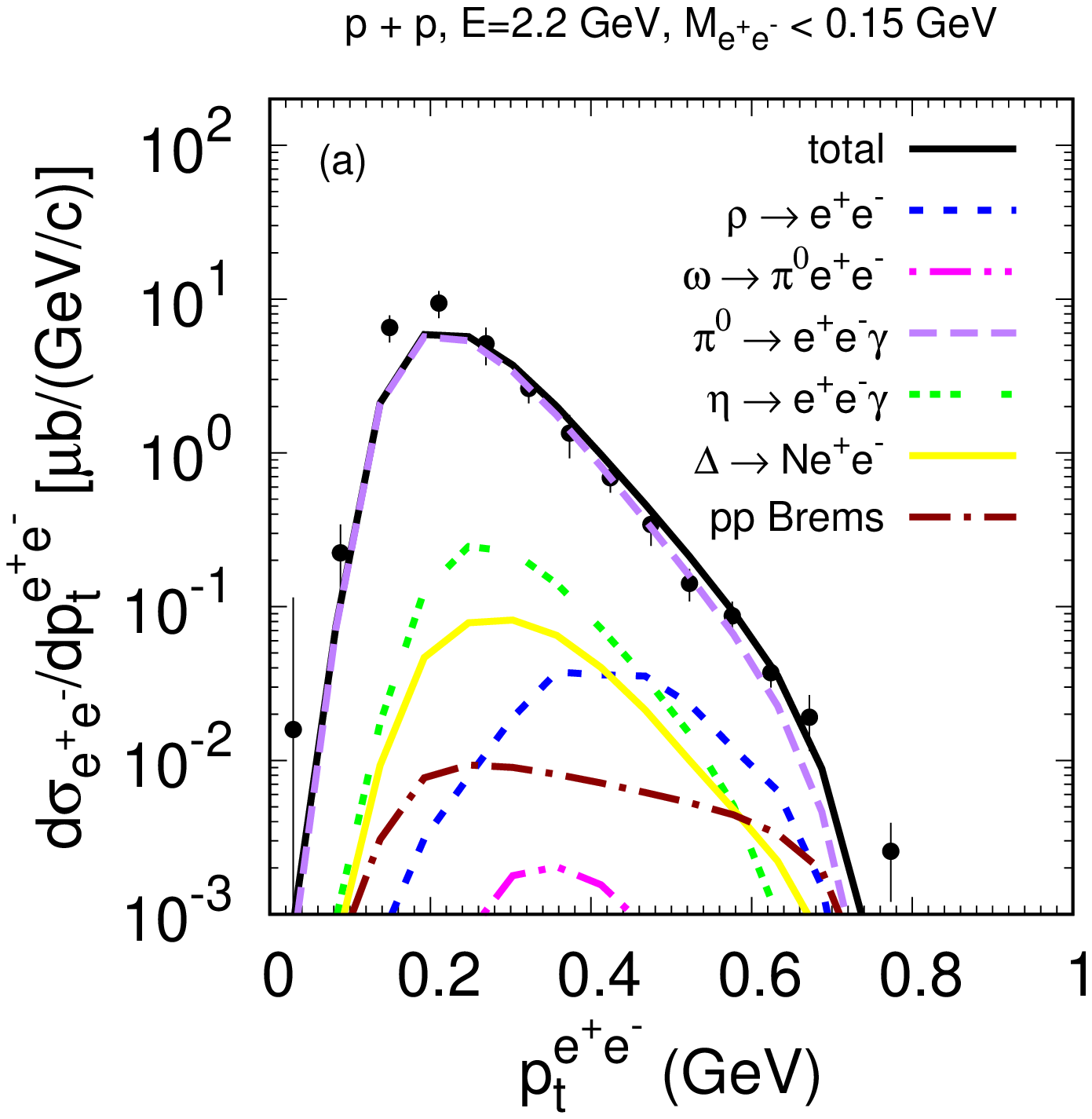}
  \includegraphics[scale = 0.50]{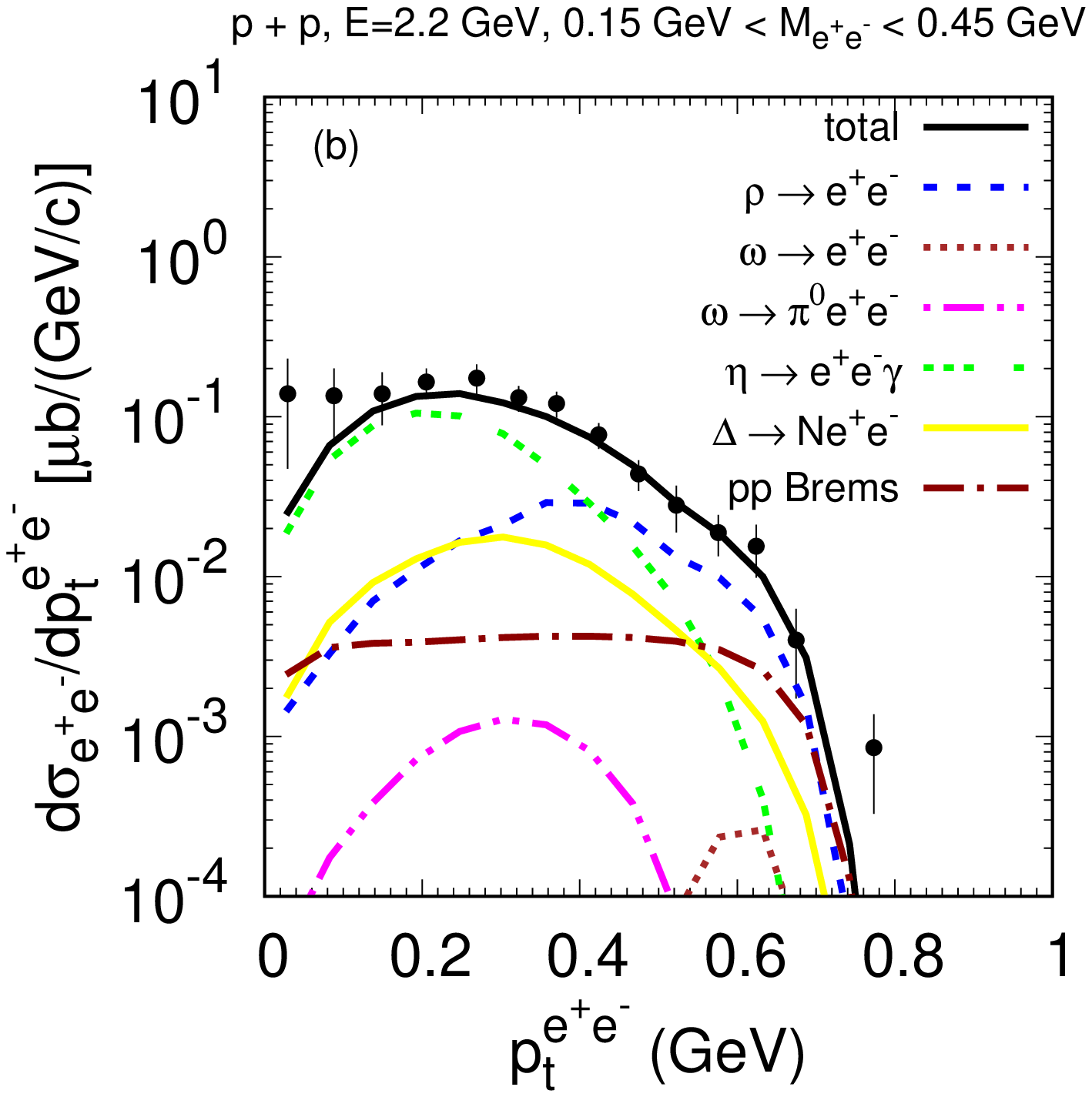}
  \includegraphics[scale = 0.50]{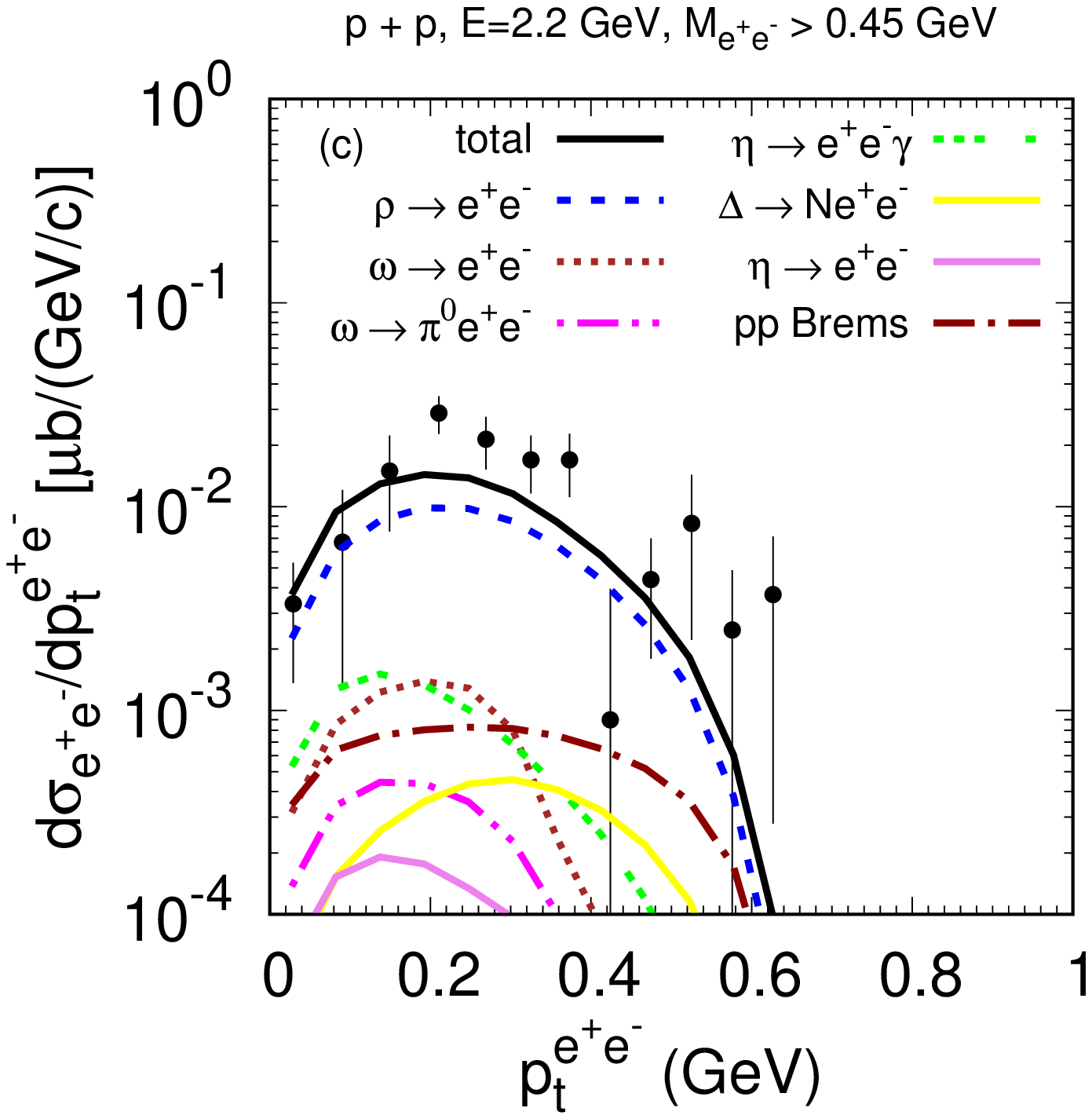}
  \caption{\label{fig:dsigdpt_pp2.2gev}(Color online) Transverse momentum differential cross section of dilepton production in $p+p$ collisions
    at the beam energy 2.2~GeV in the invariant mass intervals $M_{e^+e^-} < 0.15$ GeV (a), $0.15~\mbox{GeV} < M_{e^+e^-} < 0.45$ GeV (b),
    and $M_{e^+e^-} > 0.45$ GeV (c).
    The thick solid (black) line shows the total calculated cross section. The other lines show the partial contributions
    of different production channels as indicated.
    Experimental data are from Ref.~\cite{Agakishiev:2012tc}.}
\end{figure}
Fig.~\ref{fig:dsigdpt_pp2.2gev} displays the transverse momentum differential cross section of $e^+e^-$ production in $pp$ collisions
at 2.2~GeV in different invariant mass windows where the dominant contributions are $\pi^0 \to \gamma e^+e^-$ (a), $\eta \to \gamma e^+e^-$ (b),
and $\rho \to e^+e^-$ (c).
The $\rho \to e^+e^-$ decay and $pp$ bremsstrahlung also provide the two main contributions at large $p_t$'s in the intermediate mass window (b).
There is a good overall agreement of the calculated dilepton $p_t$ spectra with experiment except for the $\rho$-dominated invariant-mass window where the $p_t$-spectrum is slightly underestimated. This might indicate a larger coupling 
of the baryon resonances to the $\rho N$ channel as compared to the one of Ref.~\cite{Manley:1992yb} used in our default calculations (see also Ref.~\cite{Agakishiev:2012tc} for further discussion).

\subsubsection{C + C collisions}

\begin{figure}\
  \includegraphics[scale = 0.60]{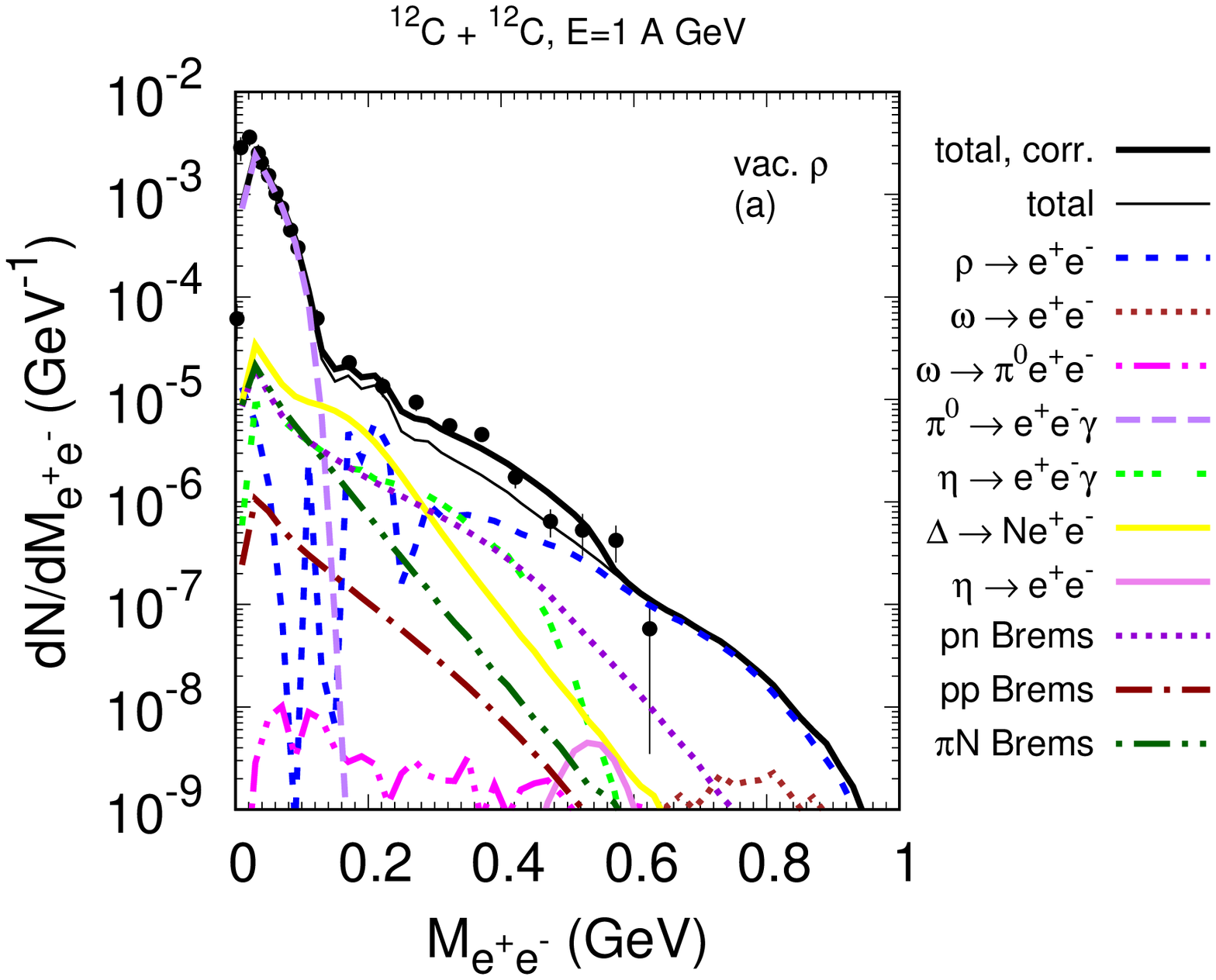}
  \includegraphics[scale = 0.60]{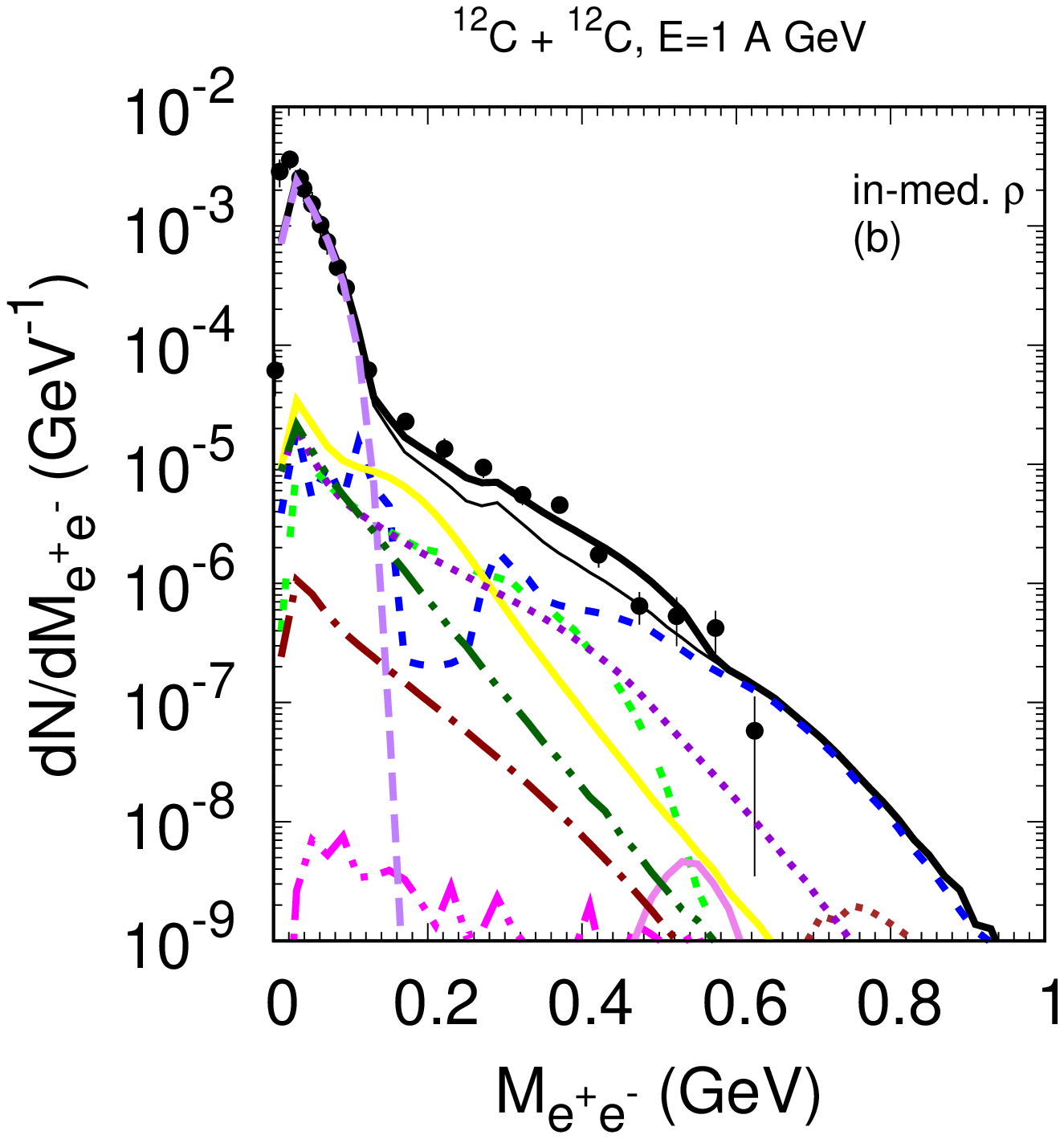}
  \caption{\label{fig:dNdM_CC1}(Color online)
  Invariant mass spectrum of dileptons produced in C+C collisions at $1 A$~GeV calculated
    with vacuum (a) and in-medium (b) $\rho$ spectral function.
    Thin and thick solid lines show the total calculated spectrum before and after correction of the $pn$ bremsstrahlung component (see Eq.(\ref{fM})), respectively.
    Other lines show the partial contributions
    of the different production channels to the total spectrum as indicated.
    Experimental data are from Ref.~\cite{Agakishiev:2007ts}.}
\end{figure}

Fig.~\ref{fig:dNdM_CC1} shows the dilepton invariant mass spectrum from C+C collisions at $1 A$~GeV. 
At small invariant masses the spectrum is saturated by the $\pi^0$ Dalitz decay component.\footnote{The $\pi^0 \to e^+ e^- \gamma$ component extends slightly above $m_\pi$ due to the smearing by the detector resolution.}
At $M_{e^+e^-} \gtapprox 0.4$ GeV the spectrum is dominated by the $\rho \to e^+e^-$ decay. 
At intermediate invariant masses in the range $M_{e^+e^-} \approx 0.2-0.4$ GeV there are four comparable contributions
of the components: $\Delta \to N e^+e^-$, $\eta \to \gamma e^+e^-$, $pn$ bremsstrahlung, and $\rho \to e^+e^-$.
Thus, the intermediate region is quite complex and the disagreement with experiment might be caused by any one of these four components, or an accumulated effect from inaccuracies in several of these four. However, we observe that just using the tuned $pn$ bremsstrahlung component of Eq.(\ref{fM}) solves the problem of missing yield in the intermediate region of invariant mass (thick solid line).

\begin{figure}

\vspace*{-.2cm}
  \includegraphics[scale = 0.60]{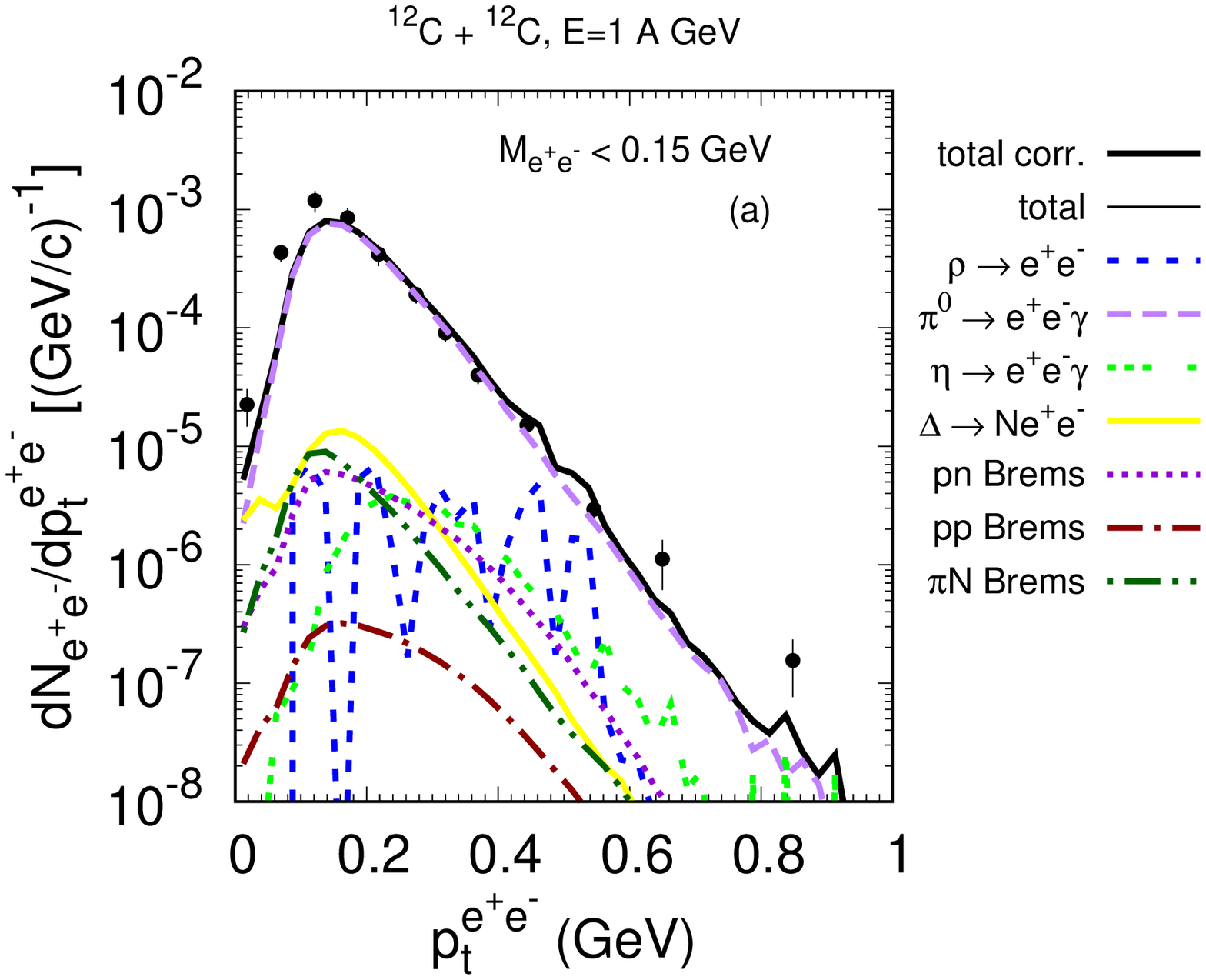}
  \includegraphics[scale = 0.60]{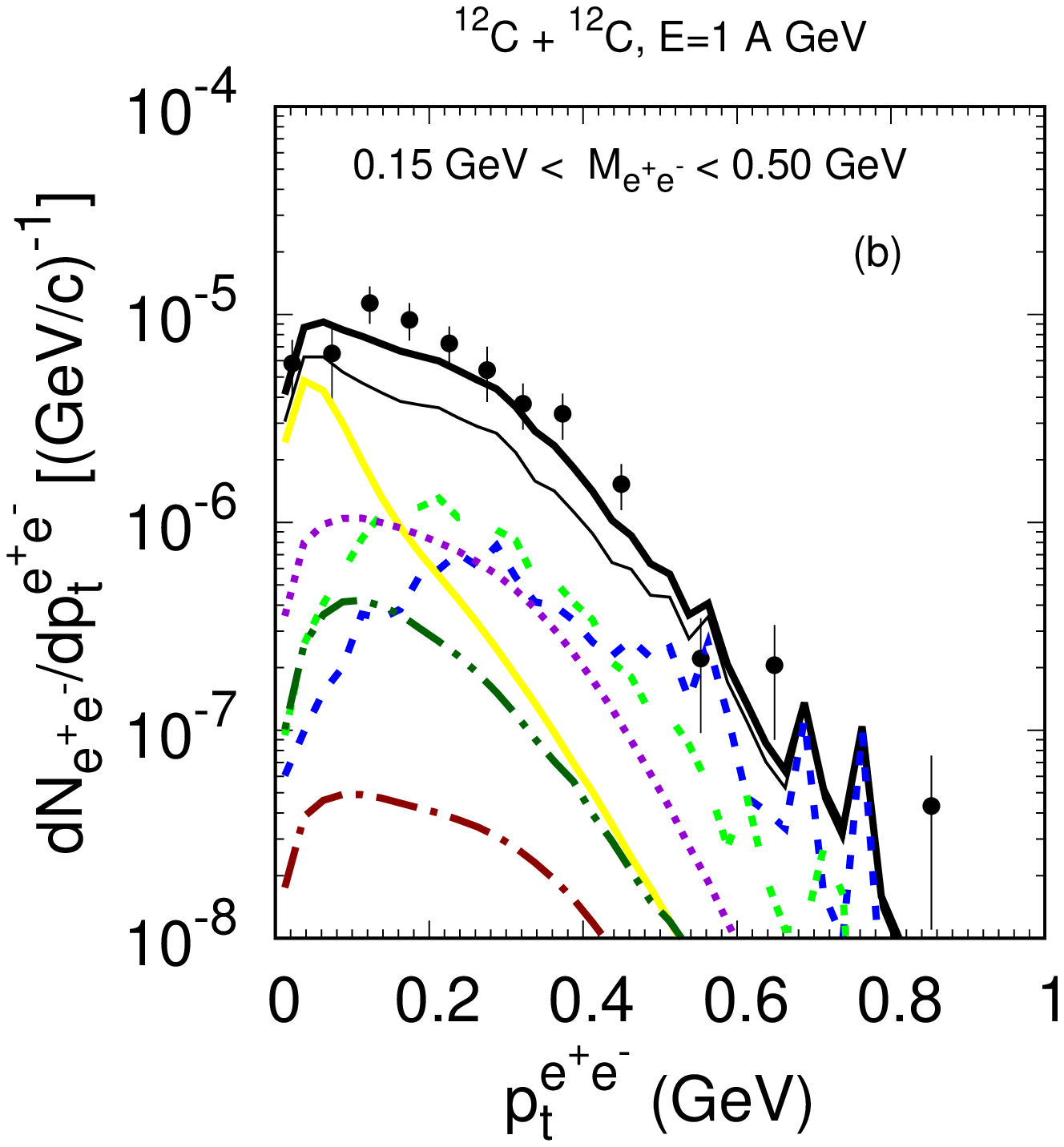}
  
  \vspace{-.2cm}
  \caption{\label{fig:dNdpt_CC1}(Color online)
  Transverse momentum distributions of dileptons produced in C+C collisions at $1 A$~GeV
    in the invariant mass intervals $M_{e^+e^-} < 0.15$ GeV (a), $0.15~\mbox{GeV} < M_{e^+e^-} < 0.50$ GeV (b).
    Calculations were done with in-medium $\rho$ spectral functions.
    Thin and thick solid lines show the total spectrum before and after correction of the $pn$ bremsstrahlung component (see Eq.(\ref{fM})), respectively.
    Other lines show partial contributions
    of the different production channels as indicated.
    Experimental data are from Ref.~\cite{Pachmayer:2008}.}
\end{figure}

\begin{figure}

\vspace*{-.2cm}
  \includegraphics[scale = 0.60]{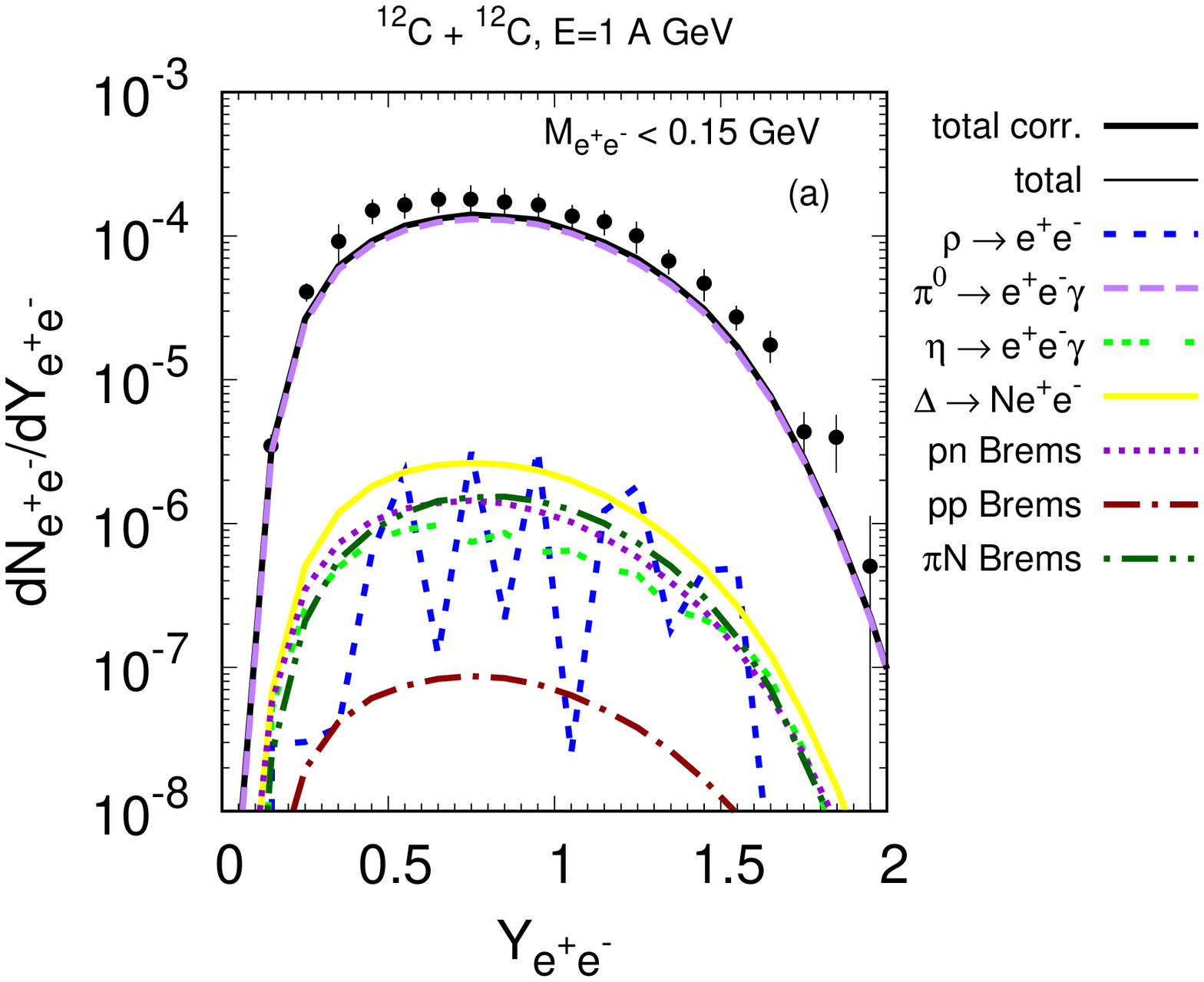}
  \includegraphics[scale = 0.60]{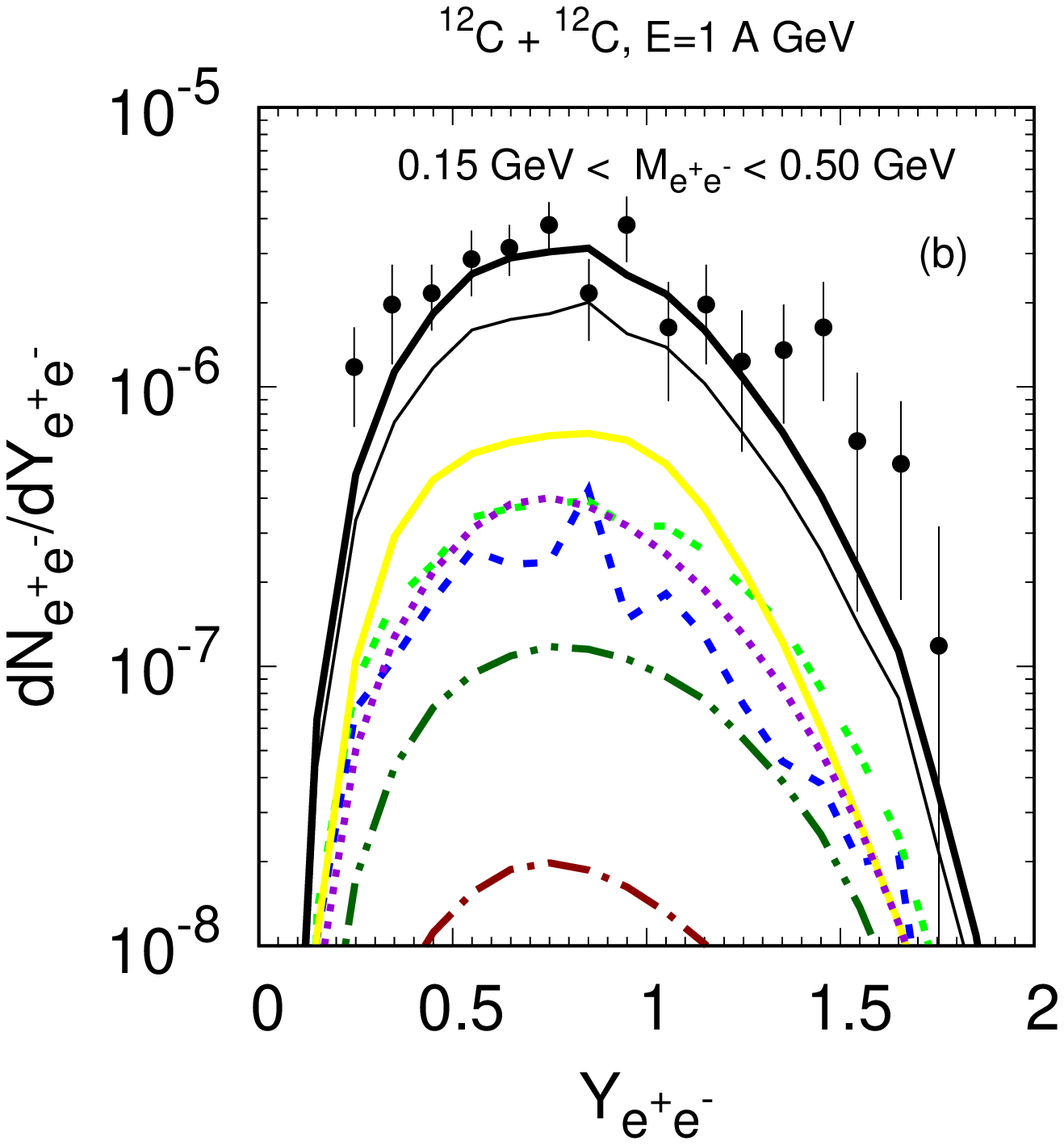}
  
  \vspace{-.2cm}
  \caption{\label{fig:dNdY_CC1}(Color online)
    Rapidity distributions of dileptons produced in C+C collisions at $1 A$~GeV
    in the invariant mass intervals $M_{e^+e^-} < 0.15$ GeV (a), $0.15~\mbox{GeV} < M_{e^+e^-} < 0.50$ GeV (b).
    Calculations were done with in-medium $\rho$ spectral functions.
    The total spectrum before and after correction of the  $pn$ bremsstrahlung
    component (see Eq.(\ref{fM})) is shown by thin and thick solid lines, respectively. Other lines show partial contributions
    of the different production channels as indicated. 
    Experimental data are from Ref.~\cite{Pachmayer:2008}.}
\end{figure}

The physical effect of collisional broadening of the $\rho$ meson on the other hand is still rather weak for this light C+C system.
However, including the collisional width of the $\rho$ meson allows to reduce statistical fluctuations in the $\rho \to e^+ e^-$ component of the spectra at $M_{e^+e^-} < 2 m_\pi$, producing a smoother behaviour of this component.
Nevertheless, significant statistical fluctuations of the $\rho \to e^+ e^-$ component still persist in the transverse momentum and rapidity distributions of dileptons at small  $M_{e^+e^-}$, as seen in Figs.~\ref{fig:dNdpt_CC1} (a), \ref{fig:dNdY_CC1} (a), \ref{fig:dNdpt_CC2} (a), and \ref{fig:dNdY_CC2} (a) below.


The transverse momentum and rapidity distributions of dileptons produced in C+C collisions at $1 A$~GeV are shown
in Fig.~\ref{fig:dNdpt_CC1} and \ref{fig:dNdY_CC1}, respectively.\footnote{Here we show the cases with the in-medium $\rho$ only. Calculations with vacuum $\rho$ produce practically indistinguishable
spectra, except for the somewhat stronger statistical fluctuations in the $\rho \to e^+e^-$ components at small invariant masses.}
In the lowest invariant mass range ($M_{e^+e^-} < 0.15$ GeV), the spectra are saturated by the $\pi^0$ Dalitz decay,
with the other contributions suppressed by more than one order of magnitude.
In the higher invariant mass window ($0.15~\mbox{GeV} < M_{e^+e^-} < 0.50$ GeV), the composition of the dilepton spectra is more complex.
The small-$p_t$ part is governed by the $\Delta$ Dalitz decays while at large $p_t$'s there are comparable contributions
of $\eta$ Dalitz decay, $\rho \to e^+e^-$ decay, and $pn$ bremsstrahlung. We see again that using the tuned $pn$ bremsstrahlung improves the description of the experimental data. This illustrates that the $pn$ bremsstrahlung is an essential component in the spectra and has to be quantitatively brought under control.

\begin{figure}
  \includegraphics[scale = 0.60]{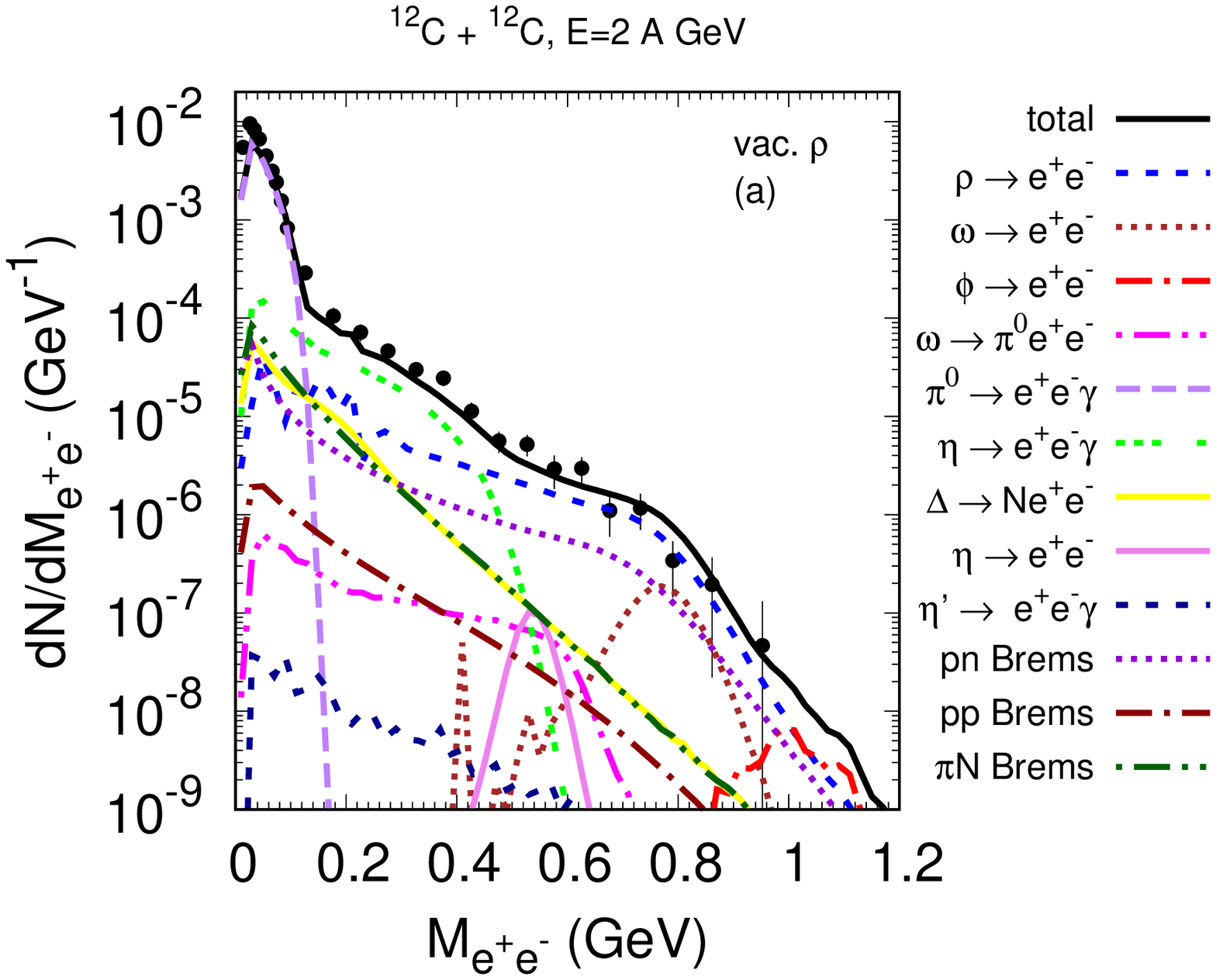}
  \includegraphics[scale = 0.60]{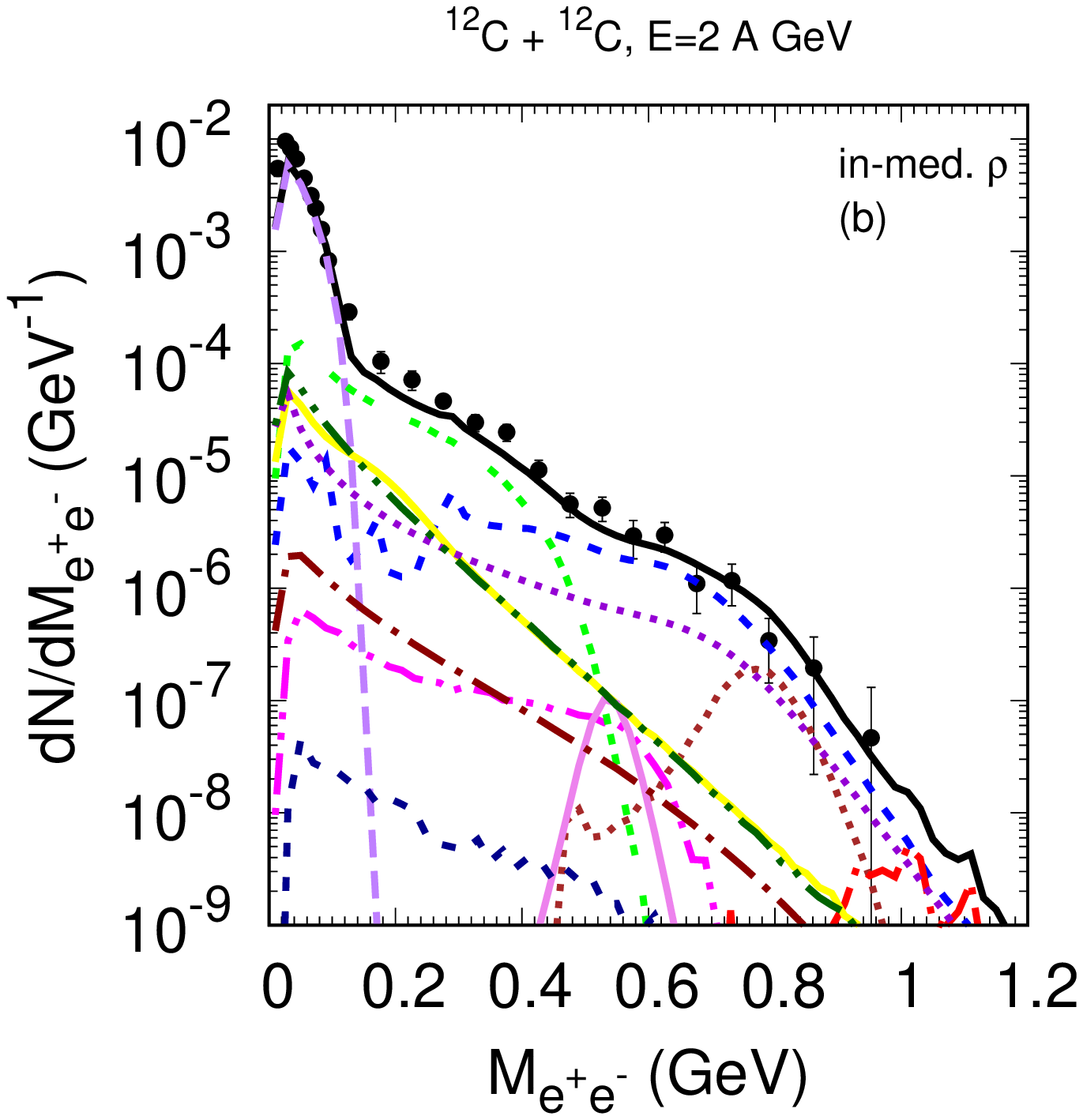}
  \caption{\label{fig:dNdM_CC2}(Color online) Invariant mass spectra of dileptons produced in C+C collisions at $2 A$~GeV
    calculated with vacuum (a) and in-medium (b) $\rho$ spectral function.  The thick solid (black) line shows the total calculated spectrum.
    Other lines show partial contributions
    of the different production channels to the total spectrum as indicated.
    Experimental data are from Ref.~\cite{Agakichiev:2006tg}. }
\end{figure}

The dilepton invariant mass spectrum in C+C collisions at $2 A$~GeV is shown in Fig.~\ref{fig:dNdM_CC2}.
Similar to the case of C+C at $1 A$~GeV, at small and large $M_{e^+e^-}$ the spectrum is dominated by the $\pi^0$ Dalitz
and $\rho \to e^+ e^-$ decays, respectively. However, in contrast to the same system at lower energy, now the intermediate
mass region is almost saturated by the $\eta$ Dalitz decays. There are also rather strong contributions of the $\omega \to e^+ e^-$
and $\phi \to e^+ e^-$ decays in the invariant mass regions near their pole masses. We observe an overall quite perfect agreement between our GiBUU results and the  experimental data.

\begin{figure}
  \includegraphics[scale = 0.55,trim= 0in 0in 2.5in 0in]{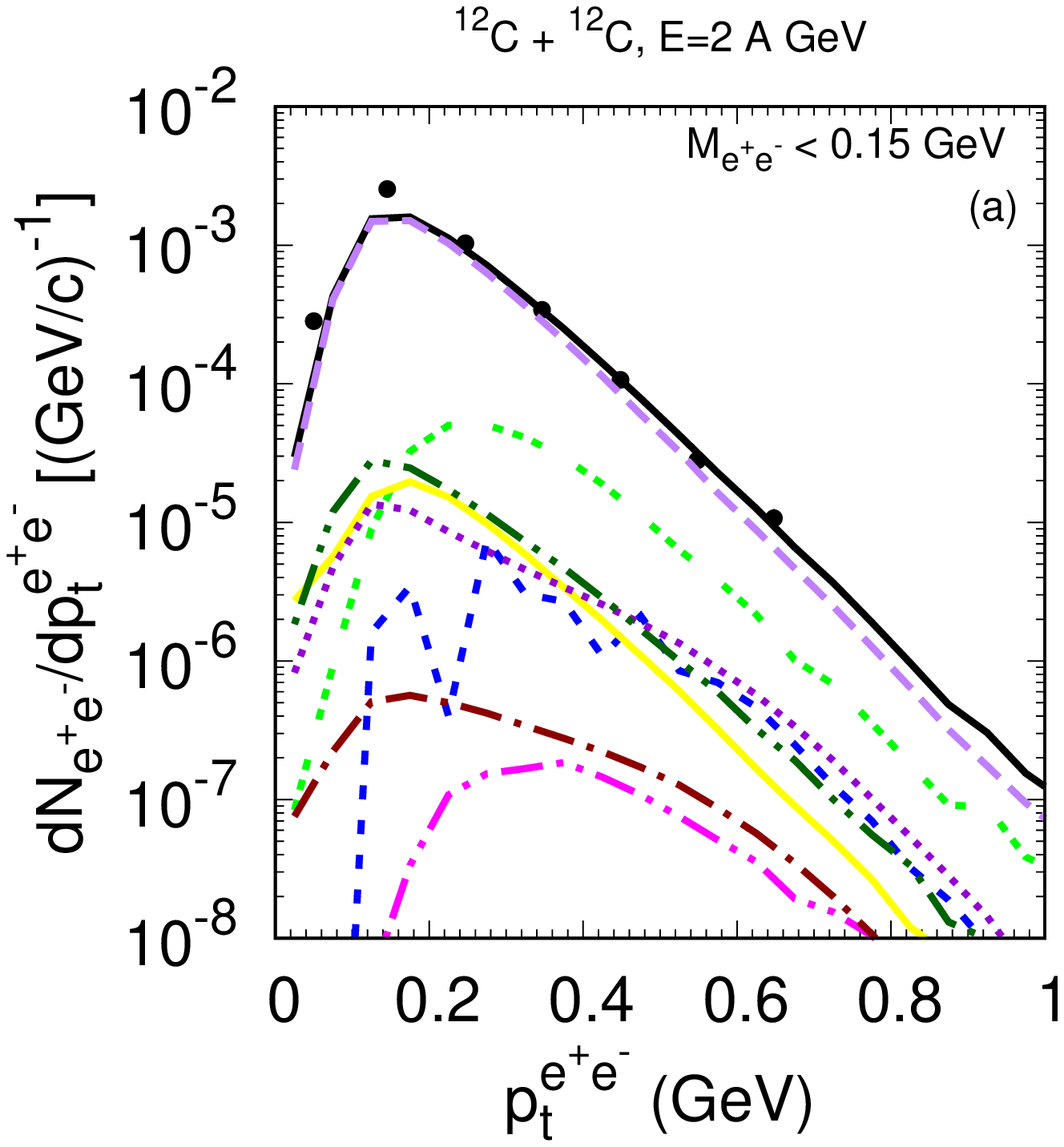}
  \includegraphics[scale = 0.55,trim= 0in 0in 2.5in 0in]{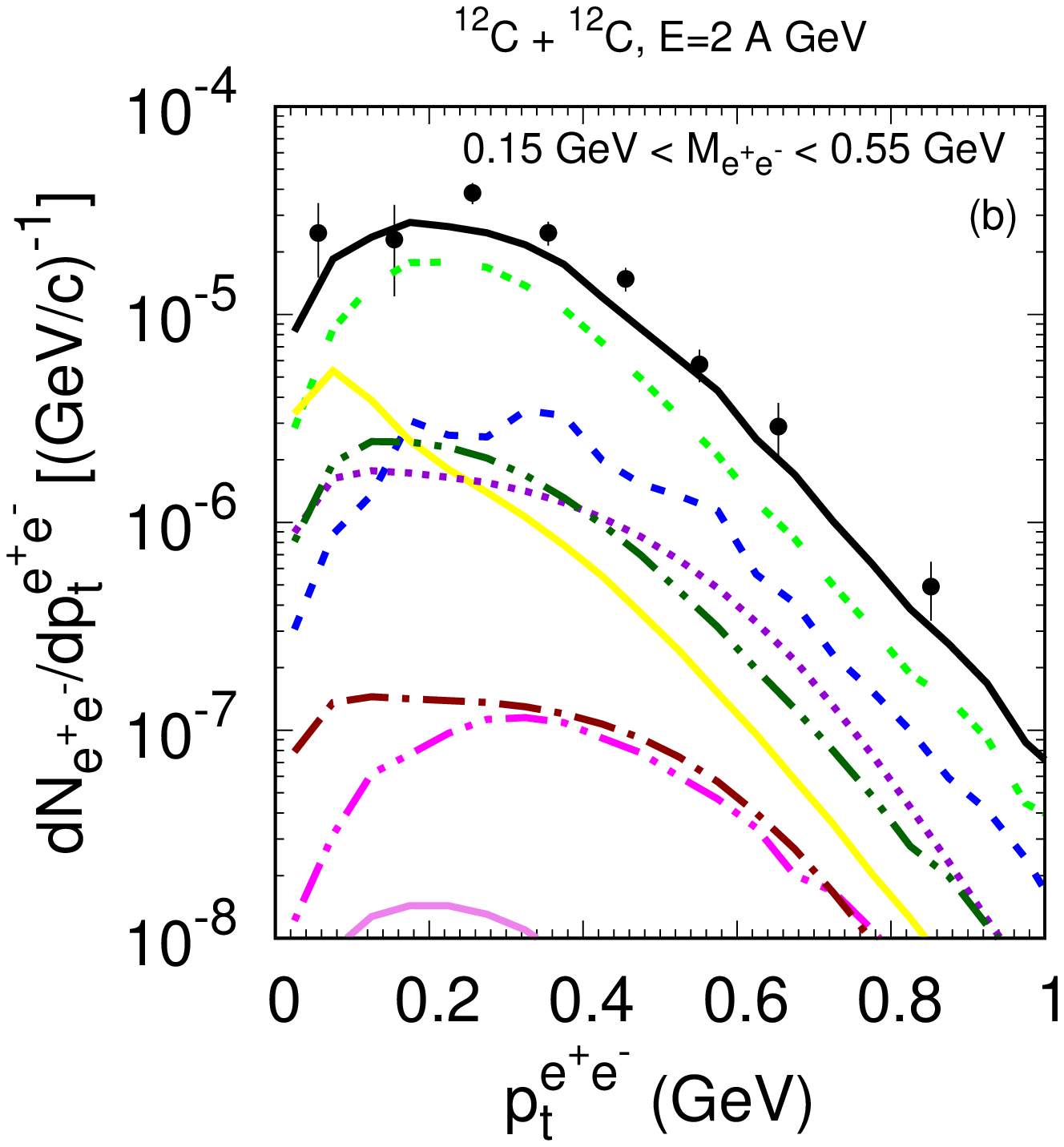}
  \includegraphics[scale = 0.55]{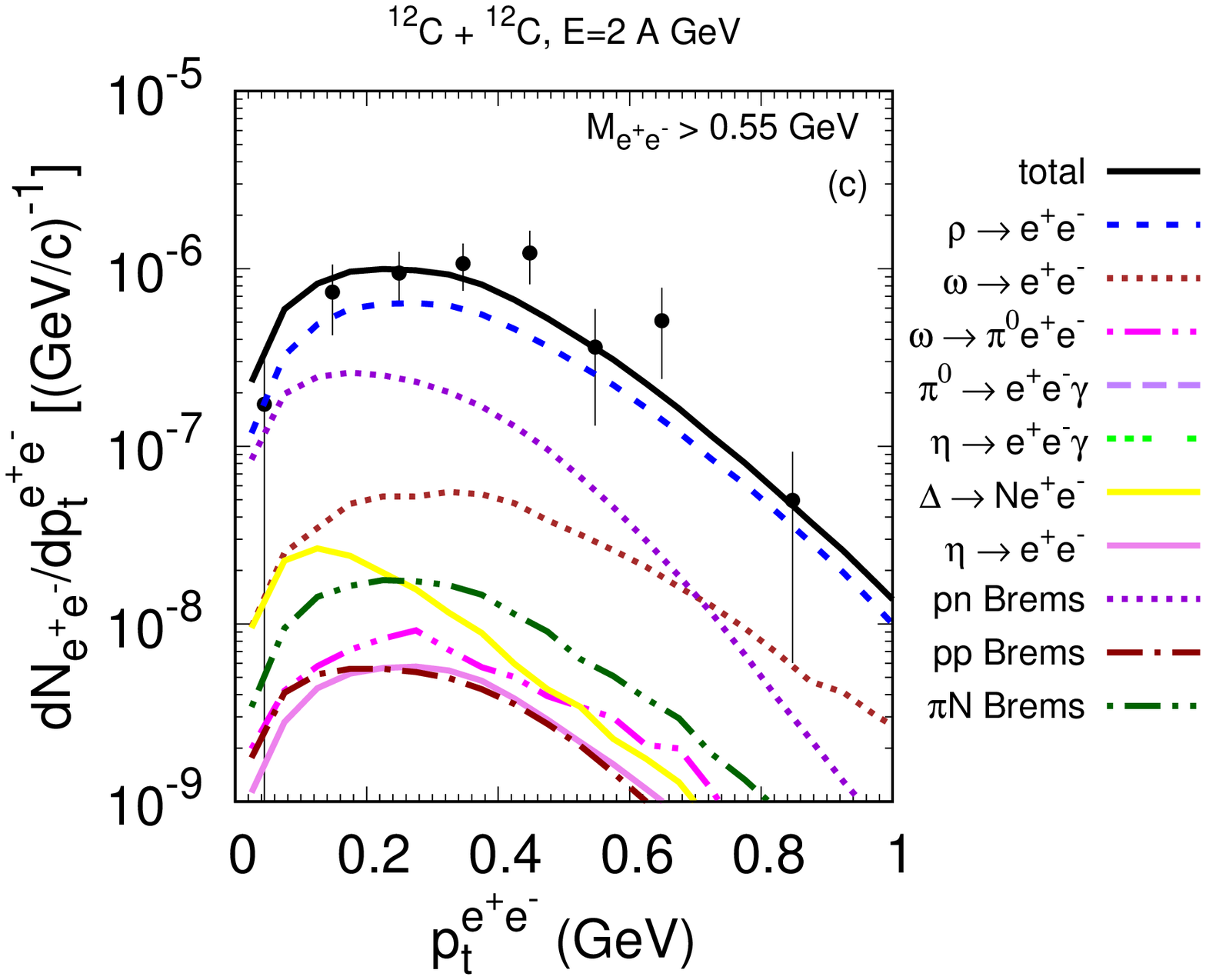}
  \caption{\label{fig:dNdpt_CC2}(Color online)
  Transverse momentum distributions of dileptons produced in C+C collisions at $2 A$~GeV
    in the invariant-mass intervals $M_{e^+e^-} < 0.15$ GeV (a), $0.15~\mbox{GeV} < M_{e^+e^-} < 0.55$~GeV (b),
    and $M_{e^+e^-} > 0.55$ GeV (c).
    Calculations were done with in-medium $\rho$ spectral functions.
    Thick solid (black) lines show the total calculated cross sections, other lines the partial contributions of different production channels as indicated.
    The fluctuations of the $\rho \to e^+e^-$ component in the lowest invariant mass interval (a) are purely statistical.
    Experimental data are from Ref.~\cite{Sudol:2007}.}
\end{figure}
\begin{figure}
  \includegraphics[scale = 0.55,trim= 0in 0in 2.5in 0in]{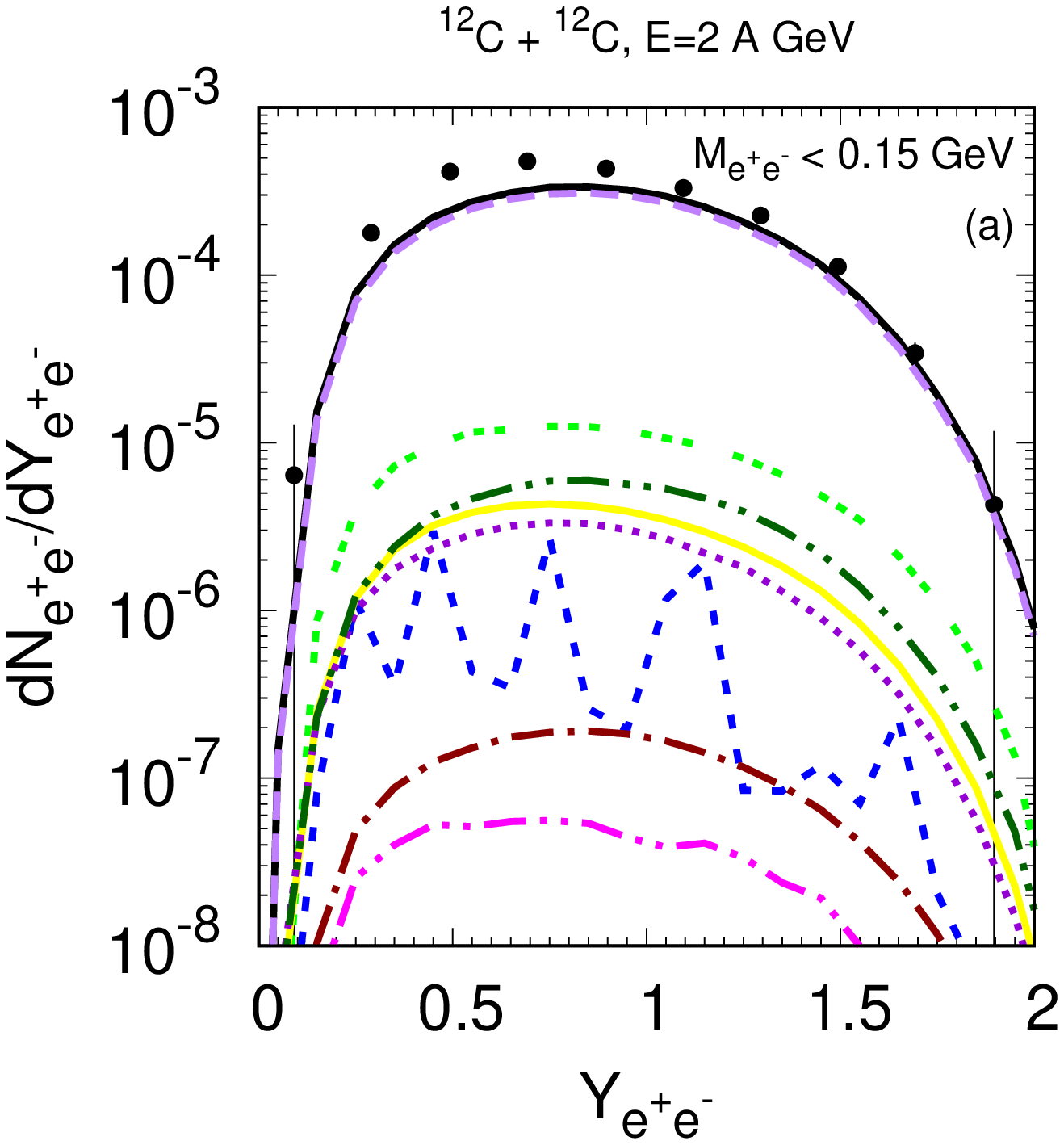}
  \includegraphics[scale = 0.55,trim= 0in 0in 2.5in 0in]{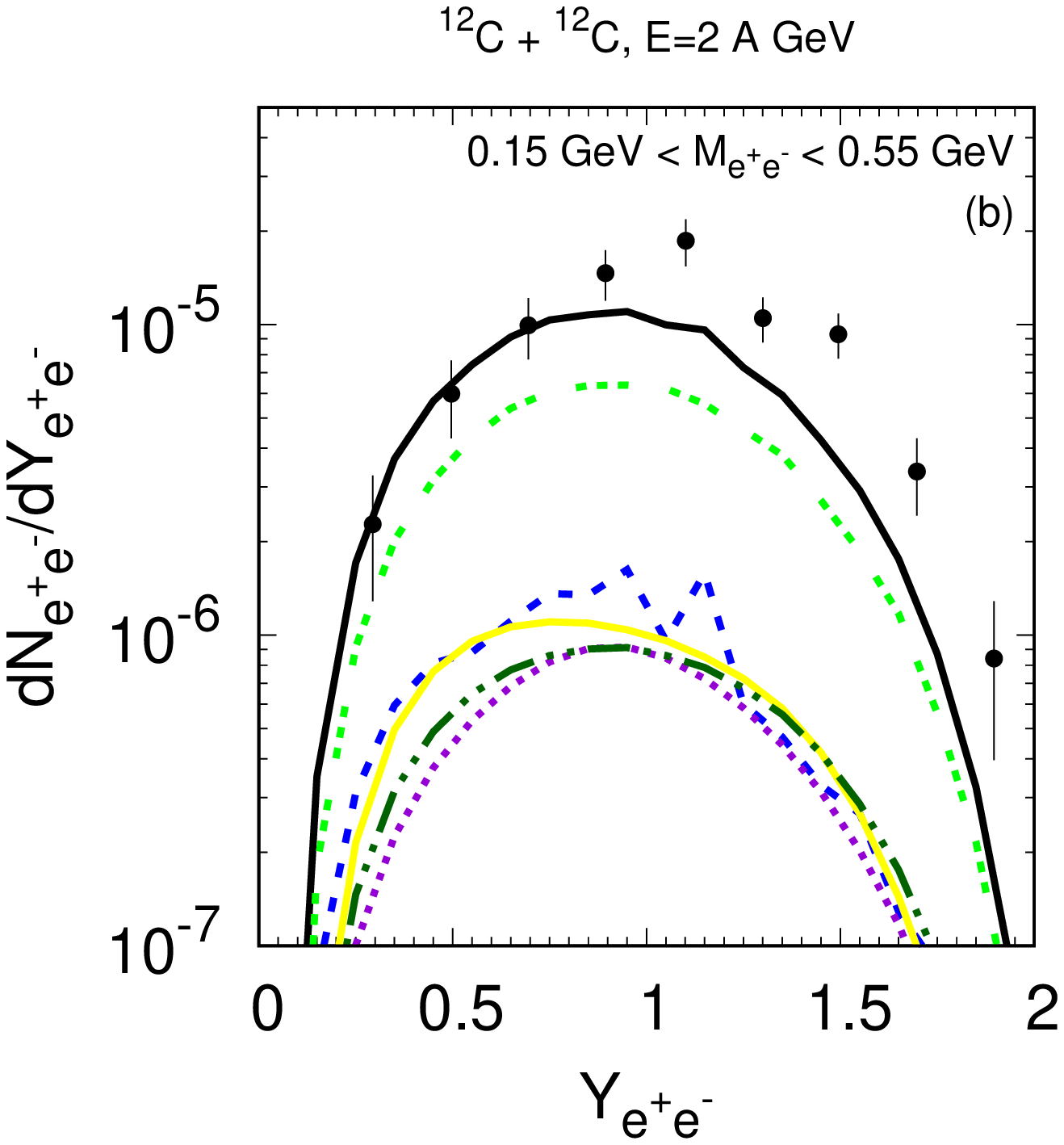}
  \includegraphics[scale = 0.55]{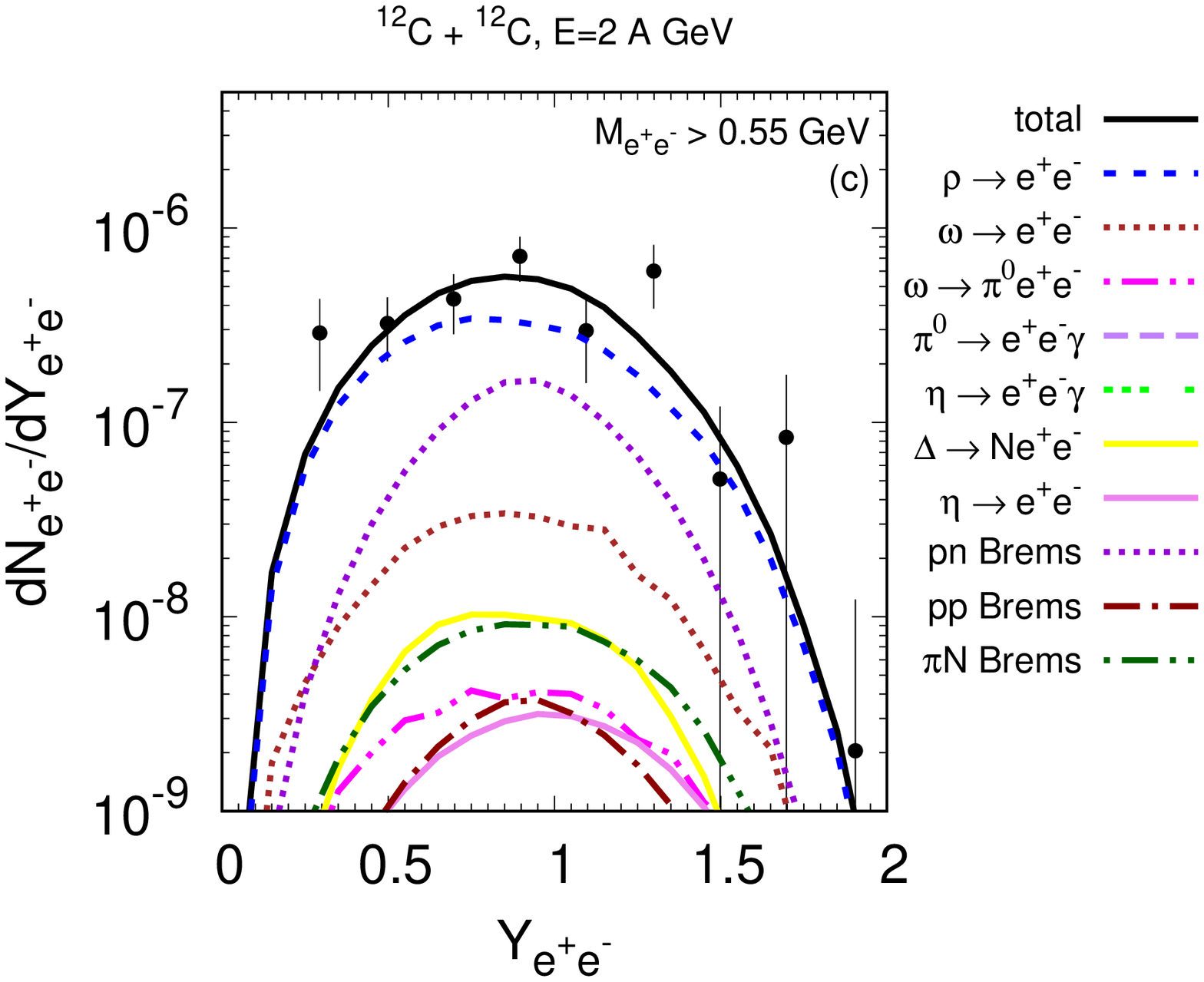}  
  \caption{\label{fig:dNdY_CC2}(Color online)
    Rapidity distributions of dileptons produced in C+C collisions at $2 A$~GeV
    in the invariant-mass intervals $M_{e^+e^-} < 0.15$ GeV (a), $0.15~\mbox{GeV} < M_{e^+e^-} < 0.55$~GeV (b),
    and $M_{e^+e^-} > 0.55$ GeV (c).
    Calculations are done with in-medium $\rho$ spectral function.
    Thick solid (black) lines show the total calculated cross sections, other lines show the partial contributions
    of different production channels as indicated.
    The fluctuations of the $\rho \to e^+e^-$ component in the lowest invariant mass interval (a) are purely statistical.  
    Experimental data are from Ref.~\cite{Sudol:2007}.}
\end{figure}
Fig.~\ref{fig:dNdpt_CC2} shows the transverse momentum distributions of dileptons from C+C collisions at $2 A$~GeV
in the three invariant mass windows with dominant $\pi^0$ Dalitz, $\eta$ Dalitz, and $\rho \to e^+ e^-$ decays, respectively,
in the order of increasing $M_{e^+e^-}$. It is interesting to compare these spectra to those from $p+p$ collisions at 2.2 GeV
in Fig.~\ref{fig:dsigdpt_pp2.2gev}.  The leading components in each mass window experience sharp cutoffs in the $p+p$ case
while they are falling exponentially with $p_t$ in the C+C case. This difference is largely caused by Fermi motion
inside the carbon nuclei.\footnote{The transverse collective flow effect is expected to be small in C+C system.}  
The experimental data in all three invariant mass windows are described very well.

The rapidity distributions of the dileptons from C+C collisions at $2 A$~GeV are shown in Fig.~\ref{fig:dNdY_CC2}. They are not
symmetric around mid-rapidity ($Y=0.9$) due to the experimental acceptance. The experimental data are well described, except for
the intermediate invariant mass region dominated by the $\eta$ Dalitz decay where our calculations underestimate the dilepton yield at forward rapidities. 
A possible reason is the oversimplified description of $\eta$ production
in decays of $N^*(1535)$ which are modeled isotropically in the resonance rest frame.

\subsubsection{Ar + KCl collisions}

\begin{figure}
  \includegraphics[scale=0.6]{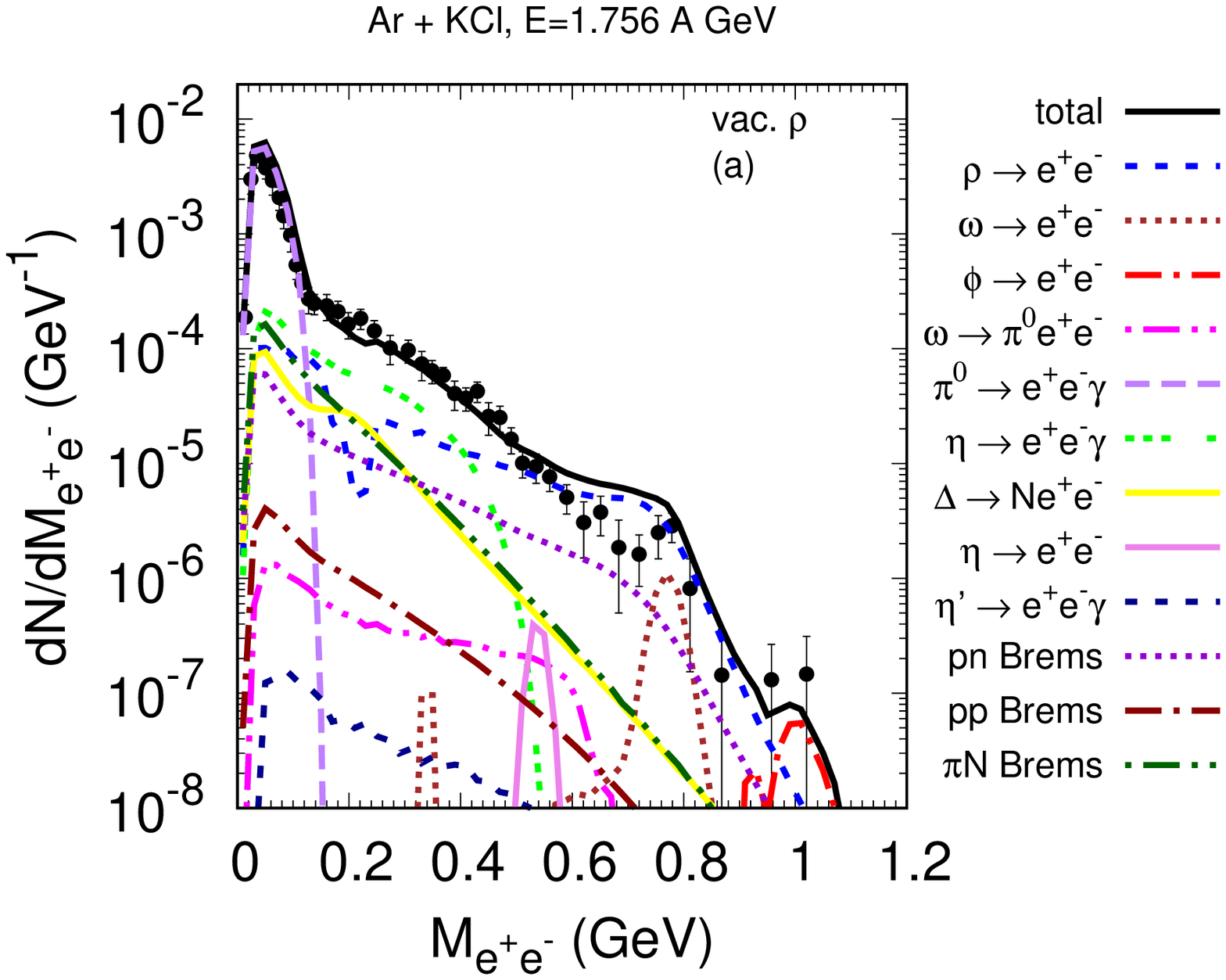}
  \includegraphics[scale=0.6]{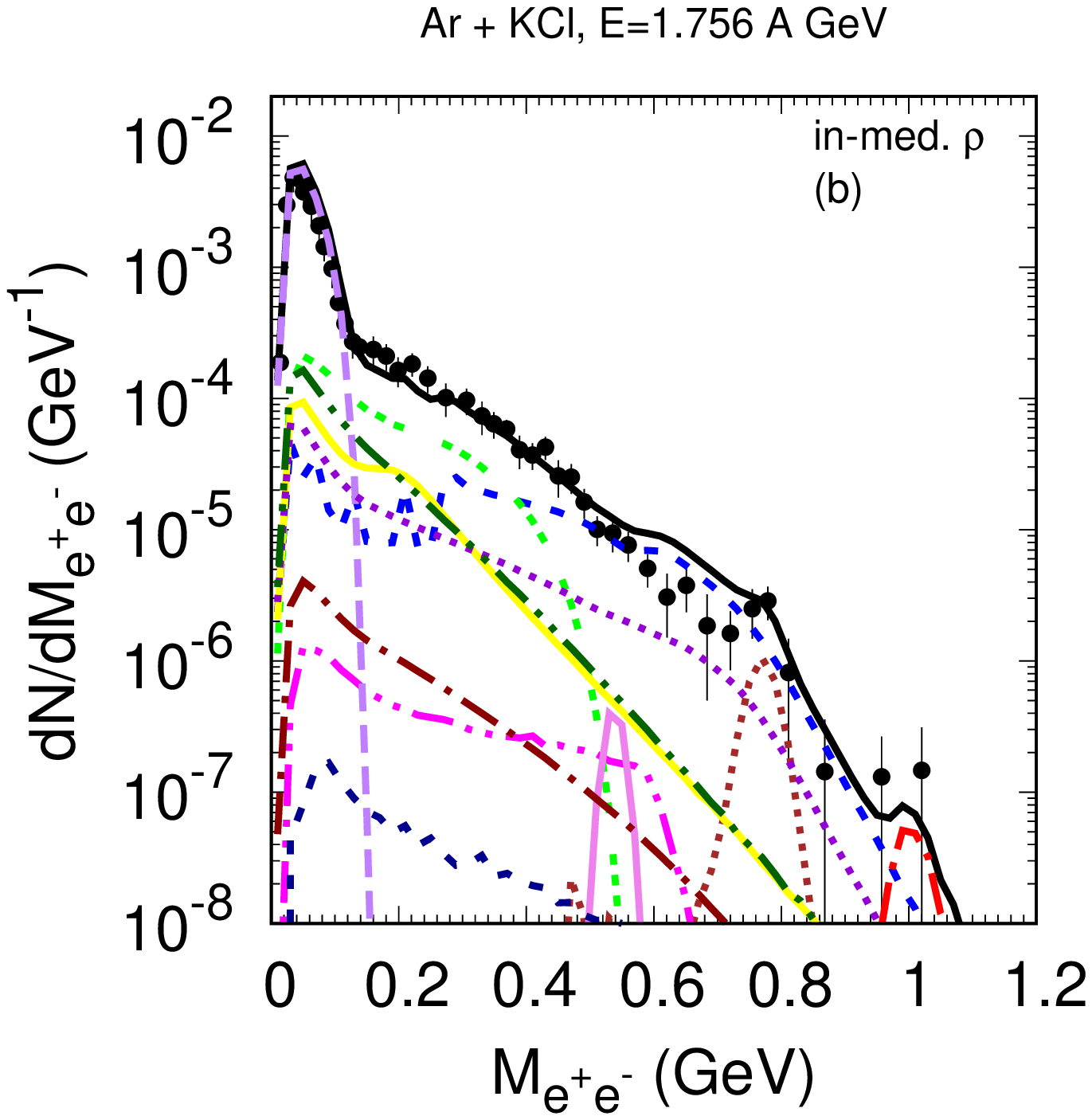}
  \caption{\label{fig:dNdM_ArKCl}(Color online) Invariant mass spectra of dileptons produced in Ar+KCl collisions at $1.756 A$~GeV
    calculated with vacuum (a) and in-medium (b) $\rho$ spectral function.    
    Thick solid (black) line shows the total calculated spectrum.
    Other lines show the partial contributions
    of the different production channels to the total spectrum as indicated.
    Experimental data are from Ref.~\cite{Agakishiev:2011vf}.}
\end{figure}

The dilepton invariant mass spectrum from Ar+KCl collisions at $1.756 A$~GeV 
is shown in Fig.~\ref{fig:dNdM_ArKCl}. 
For this colliding system,\footnote{In the calculations we have replaced KCl by $^{37}$Ar.} similar to C+C at $2 A$~GeV, the spectrum at $M_{e^+e^-} \approx 0.2-0.4$ GeV is dominated by the $\eta$ Dalitz decays.
There are also clearly visible peaks due to the $\omega \to e^+e^-$ and $\phi \to e^+e^-$ decays. In calculations with vacuum $\rho$, we furthermore see a pronounced shoulder at the $\rho$ pole mass which is smoothed by the collisional broadening of the $\rho$. However, a sizable overestimation of the data at the $\rho$ pole mass remains in this case.
\begin{figure}
  \includegraphics[scale = 0.60]{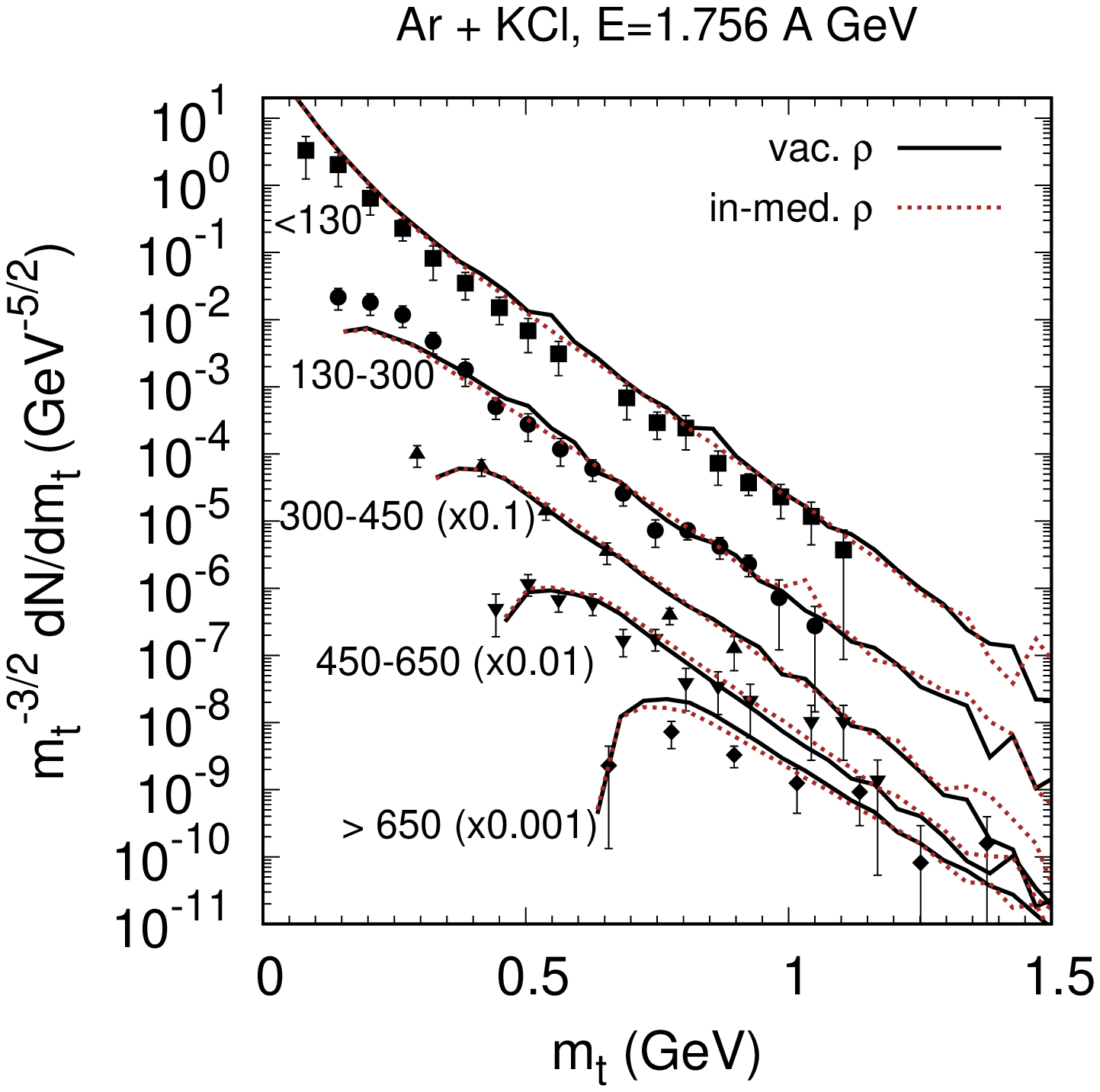}
  \caption{\label{fig:dNee_dmt_ArKCl}(Color online) Transverse mass spectra of dileptons produced in Ar+KCl collisions at $1.756 A$~GeV
    in different invariant mass windows as indicated (in MeV). Calculations with vacuum and in-medium $\rho$ are shown by solid and dotted lines, respectively.
    Experimental data are from Ref.~\cite{Agakishiev:2011vf}. Scaling factors are given in parentheses.}
\end{figure}
Fig.~\ref{fig:dNee_dmt_ArKCl} displays the transverse mass distributions of the dileptons from Ar+KCl collisions at $1.756 A$~GeV.
The calculated $m_t$ spectra agree with experimental data in all invariant mass windows, except for the highest one ($M_{ee} > 0.65$ GeV) where
our calculations produce softer $m_t$ spectra and overestimate the yields, in line with the excess observed in Fig.~\ref{fig:dNdM_ArKCl}.

Similar results (with vacuum $\rho$) are obtained in the SMASH model \cite{Staudenmaier:2017vtq}.
The HSD model \cite{Bratkovskaya:2013vx}
produces a somewhat better description of the dilepton invariant mass yield from Ar+KCl near the $\rho/\omega$ pole masses with a similar qualitative effect
of the collisional broadening of the $\rho$ meson.
In Refs.~\cite{Endres:2015fna,Endres:2015egk,Galatyuk:2015pkq,Staudenmaier:2017vtq,Seck:2017zjr}, the coarse graining approach 
has been applied for the description of heavy-ion collisions at SIS18 energies. This is a hybrid approach based on the assumption of local thermal equilibrium with parameters $T,\mu$ extracted from the microscopic transport calculations. The microscopic transport contribution of the vector-meson decays is then (partly) substituted by the dilepton emission from the thermal system with in-medium spectral functions. 
Coarse-grained transport simulations  describe the entire dilepton invariant mass spectrum for Ar+KCl very well.
Note however that establishing local thermal equilibrium at SIS18 energies can be a delicate issue \cite{Lang:1991qa}, especially for light colliding systems, which needs verification on a case by case basis.   
While in our present calculations we also resort to an equivalent local temperature in evaluating the $\rho$ spectral function for simplicity, to avoid numerically expensive calculations of $\rho$ collision rates from the actual dynamics, in contrast to Refs.~\cite{Endres:2015fna,Endres:2015egk,Galatyuk:2015pkq,Staudenmaier:2017vtq,Seck:2017zjr} on the other hand, we entirely rely on our microscopic transport simulations
in the computation of dilepton emission from the non-equilibrium system.

\subsubsection{Au + Au collisions}
\label{AuAu}

The Au+Au system at $1.23 A$~GeV is currently in the focus of studies by the HADES collaboration
(cf.~\cite{Adamczewski-Musch:2019byl,Adamczewski-Musch:2020vrg} and refs.~therein).
This is the heaviest system measured so far. Therefore, one expects the deviations  due to the various in-medium effects from the superposition of quasi-free $NN$ collisions as reference to be stronger than in the other colliding systems. As we have already seen, inclusive dilepton production is influenced by many reaction processes.
Since the production of mesons $\pi^0, \eta, \rho, \omega$ decaying into dileptons is mediated by baryon resonances, 
it is especially important to have baryon resonance production, absorption and decay in the nuclear medium well under control.
For this purpose the GiBUU model has been extensively applied to various reactions with nuclear targets, such as heavy-ion, (anti-)proton, pion, photon, electron, and neutrino induced (semi-)inclusive production processes \cite{Buss:2011mx}.

\begin{figure}
  \includegraphics[scale = 0.47]{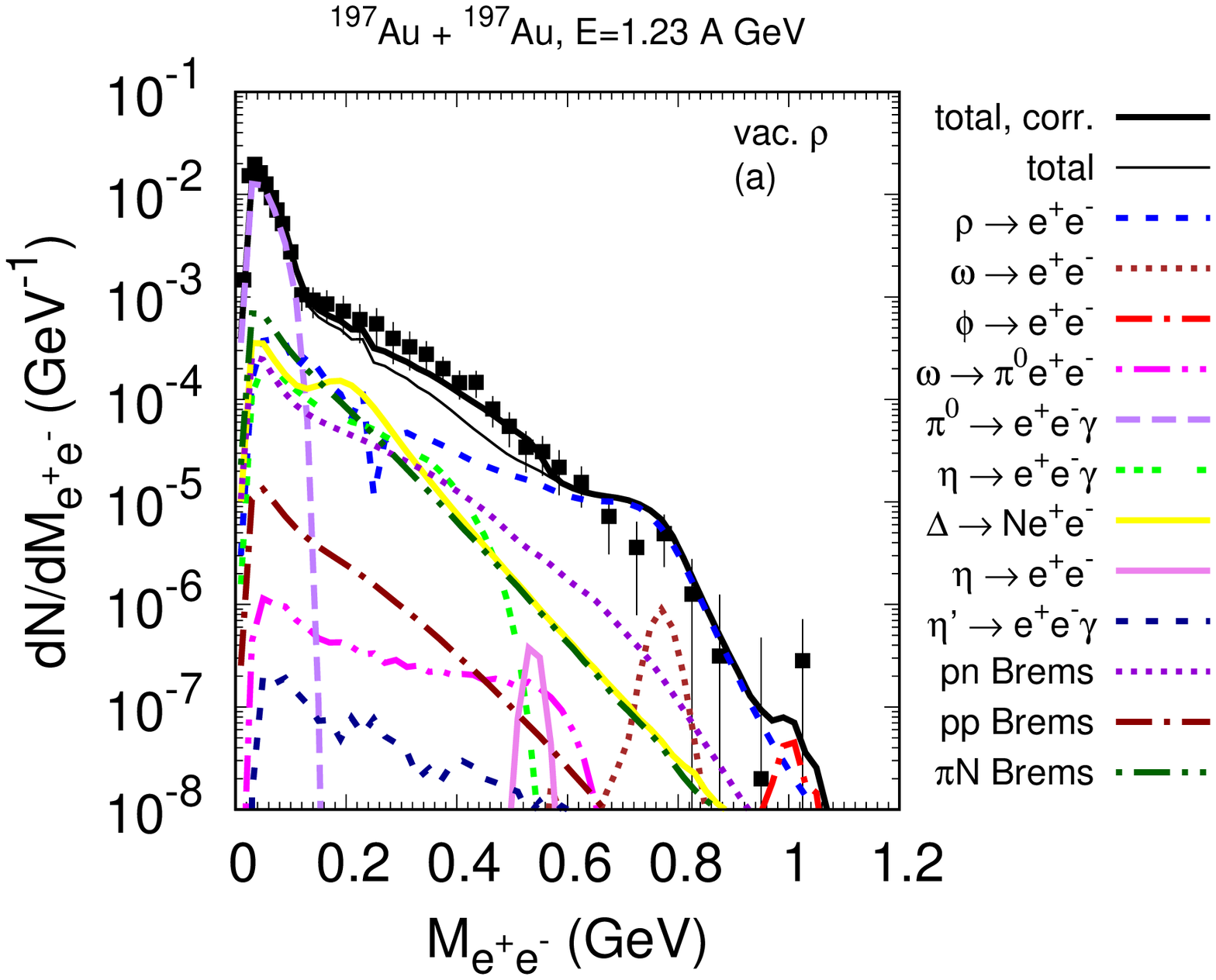}
  \includegraphics[scale = 0.47,trim= 0in 0in 2.5in 0in]{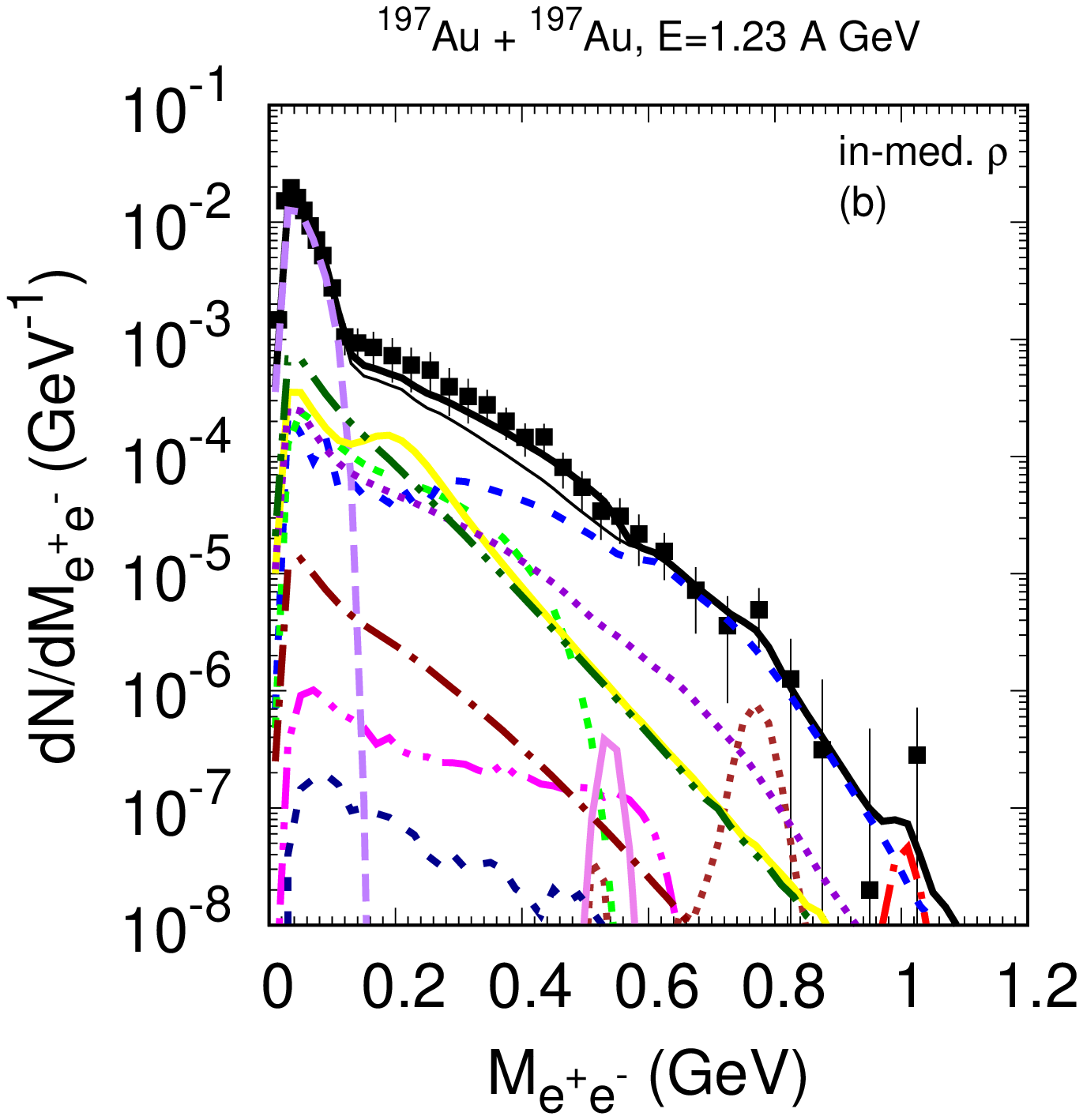}
  \includegraphics[scale = 0.47]{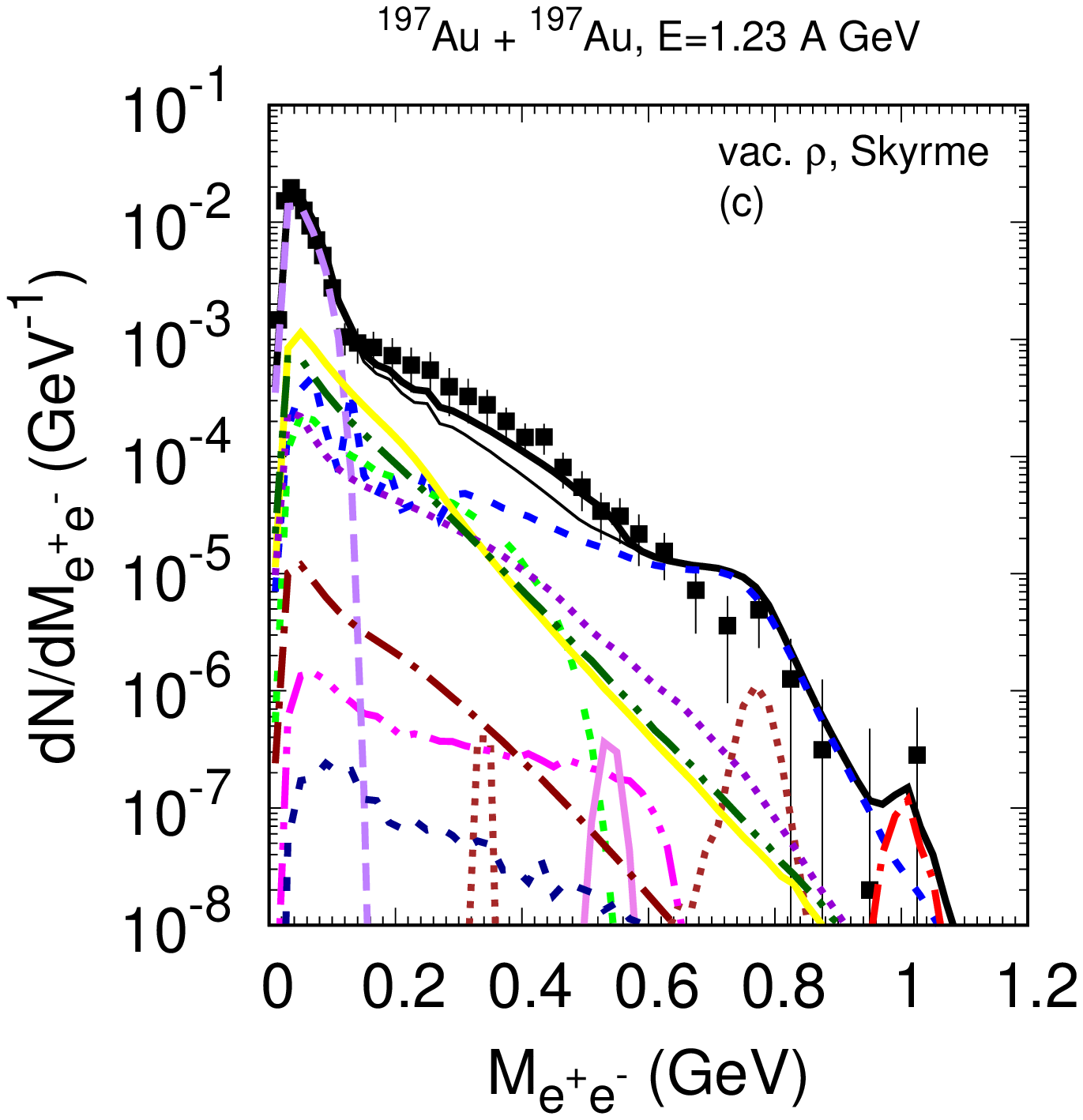}
  \includegraphics[scale = 0.47,trim= 0in 0in 2.5in 0in]{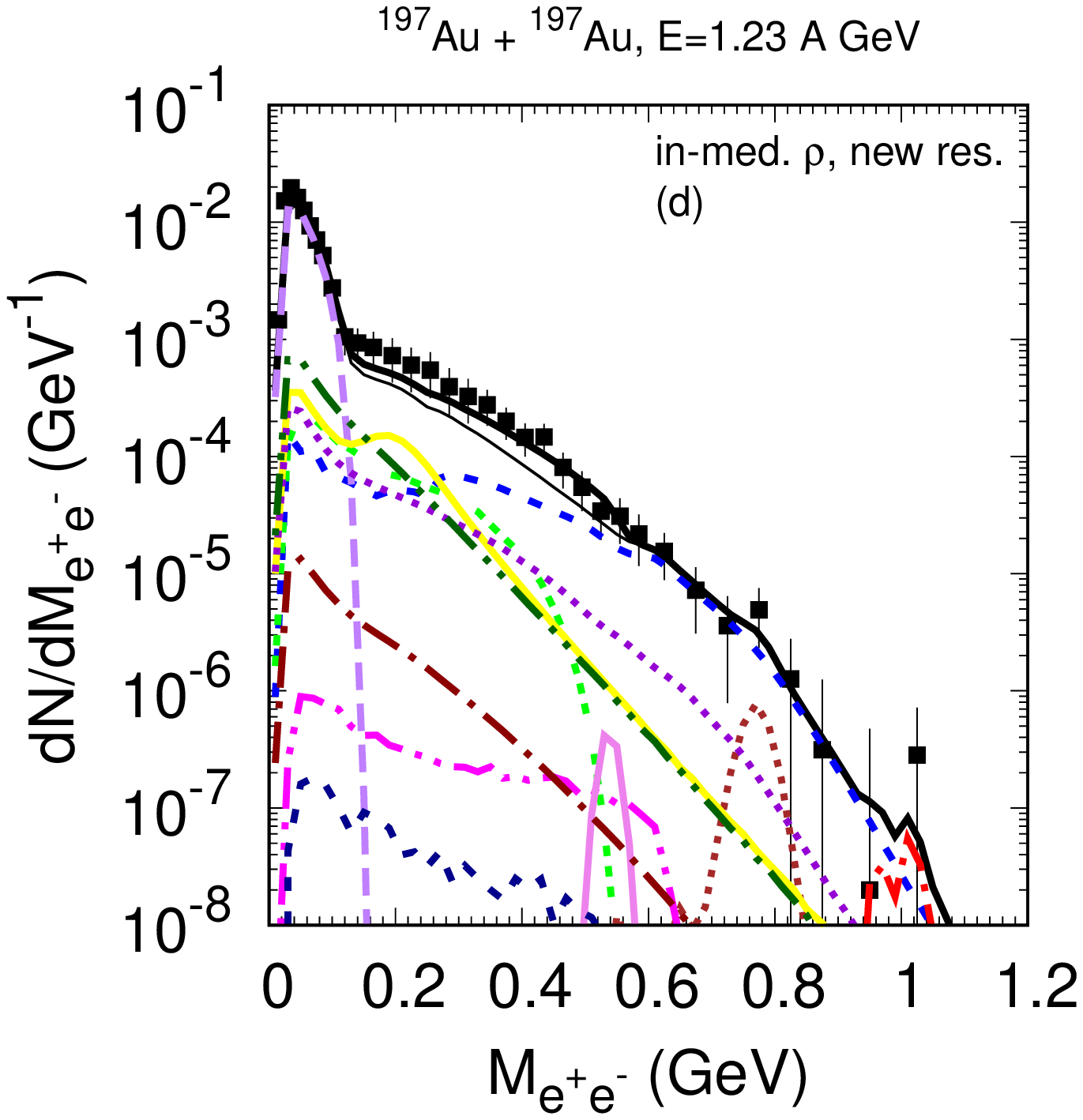}
  \caption{\label{fig:dNdM_AuAu}(Color online) Invariant-mass spectra of dileptons produced in Au+Au collisions at $1.23 A$~GeV
    calculated with vacuum (a), (c) and in-medium (b), (d) $\rho$ spectral function.
    The calculation with the Skyrme-like potential is shown in panel (c). Panel (d) shows the results obtained with updated resonance parameters.
    Thick solid black lines show the total spectra obtained from GiBUU output with multiplying the $pn$ bremsstrahlung component by the factor given in Eq.~(\ref{fM}). Thin solid black lines show the same spectra without this correction. The other lines show the partial contributions
    of different production channels to the total spectra as indicated.
    Experimental data are from Ref.~\cite{Adamczewski-Musch:2019byl}.}
\end{figure}

For comparison, the dilepton invariant-mass spectrum in Au+Au at $1.23 A$~GeV with vacuum $\rho$ spectral function is shown in Fig.~\ref{fig:dNdM_AuAu} (a) with the relativistic and (c) with
the Skyrme-like mean fields (see App.\ \ref{uncert} for detail). The resonance structure in the $\rho \to e^+ e^-$ component is clearly visible in either case.
Compared to that, in Fig.~\ref{fig:dNdM_AuAu} (b) and (d), we see that the collisional broadening of the $\rho$ meson is strong enough in this system, to remove the shoulder in the spectrum near the $\rho$ pole mass and hence, to yield better agreement with data.
As before in this energy range, however, without applying the correction factor of Eq.~(\ref{fM}) to the $pn$ bremsstrahlung component we would again underpredict the data in the intermediate invariant-mass region. This lends further support to the hypothesis that there is indeed some strength missing in the $pn$ bremsstrahlung component at these low energies.

As already mentioned in Sec.~\ref{rhomed}, the set of resonances of Ref.~\cite{Manley:1992yb} is used in GiBUU by default. The nucleon resonance parameters have been recently updated in Ref.~\cite{Hunt:2018wqz}. The influence of these new resonance parameters on the collisional width of the $\rho$-meson has been demonstrated
in Fig.~\ref{fig:GamCollres} above. As shown there, the collisional width of the $\rho$ meson is slightly larger at small masses with the new resonance parameters. Comparing the resulting dilepton invariant-mass spectrum in 
Fig.~\ref{fig:dNdM_AuAu} (d) with that obtained from the analogous calculation with the default resonance parameters in Fig.~\ref{fig:dNdM_AuAu} (b), we observe that the overall effect of the updated resonance parameters is rather small, with an only moderate increase in the $\rho \to e^+ e^-$ component slightly further improving the agreement with experiment  at low invariant masses.

Overall, the agreement of the calculations with the experimental data is very good. In particular, also the $\pi^0$ Dalitz-decay component at small masses is in excellent agreement with the data. This indicates that the number of produced neutral pions is correctly described. It is, therefore, surprising to see that the calculated total number of pions produced is significantly larger than the number of experimentally measured pions, as shown in App.~\ref{hadronnumbers}. The latter discrepancy seems to be fairly model-independent since other generators give roughly the same overestimate (see Fig.~7 in \cite{Adamczewski-Musch:2020vrg}). We will discuss this problem further in App.~\ref{hadronnumbers}.

\subsubsection{Ag + Ag collisions}

In the recent Ref.~\cite{Staudenmaier:2020xqr},  the SMASH (+ coarse graining) model predictions for the dilepton invariant mass spectra from Ag+Ag at the beam energy $1.58 A$~GeV were given, indicating practically full coincidence of the Ag+Ag and Au+Au spectra.
\begin{figure}
  \includegraphics[scale = 0.6]{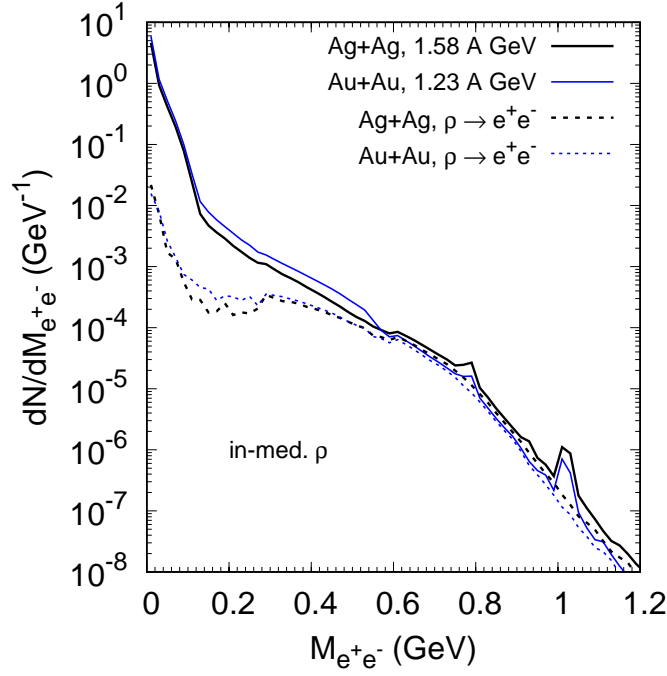}
  \caption{\label{fig:dNdM_AuAu_vs_AgAg}(Color online) Dilepton invariant-mass spectra for Ag+Ag at $1.58 A$~GeV (black lines)
  and Au+Au at $1.23 A$~GeV (blue lines). Solid and dashed line show the total spectra and their partial $\rho \to e^+ e^-$ components, respectively. 
  Calculations were done with the in-medium $\rho$ spectral function. The centrality selection is $0-40\%$ for both systems, and modeled 
  in the sharp cutoff approximation for the impact parameter, with $b < 7.7$~fm for Ag+Ag, and $b < 9.3$ fm for Au+Au.
  The total spectrum for Au+Au is shown with the bremsstrahlung correction of Eq.~(\ref{fM}). All spectra are given in full acceptance.}
\end{figure}
Fig.~\ref{fig:dNdM_AuAu_vs_AgAg} displays our predictions for the total dilepton invariant mass spectrum and its partial $\rho \to e^+e^-$ component for Ag+Ag at $1.58 A$~GeV. The corresponding spectra for Au+Au at $1.23 A$~GeV are also shown for comparison. In the $\pi^0$ Dalitz region, the spectra for Ag+Ag and Au+Au are almost identical since the (true) $\pi^0$ multiplicities per event are quite close:  9.0 for Ag+Ag vs 11.0 for Au+Au. However, at larger invariant masses the spectra from Ag+Ag and Au+Au differ significantly. In the intermediate mass range, $M_{e^+e^-} \approx 0.2-0.6$ GeV, the yield for Ag+Ag is smaller while in the $\rho$-dominated region,  $M_{e^+e^-} \gtapprox 0.6$ GeV, it is larger than the yield for Au+Au. The first effect is due to the bremsstrahlung correction that is here included for Au+Au only. As we have seen above,  such a correction is necessary at the lower energy in Au+Au. Whether and to what extend it is also needed for Ag+Ag at $1.58 A$~GeV remains uncertain. The second effect, i.e.~the enhancement by $\approx 30\%$  for Ag+Ag relative to Au+Au in the $\rho$ mass region, is more robust. We have checked that a similar enhancement presents also in calculation with a vacuum $\rho$ spectral function. A small, although statistically significant, difference between predictions of GiBUU and SMASH models is most probably due to different resonance parameters.

\subsection{Excess radiation}
\label{excess}

In-medium effects, such as multiple scattering, secondary particle interactions and, possibly, modifications of the vacuum spectral properties of the $\rho$ are also being discussed in terms of the excess radiation, that is the dilepton spectrum without $\eta$ and $\omega$ decay contributions, and with the $NN$ reference spectrum subtracted. The latter describes dilepton radiation in the picture of first-chance $NN$ collisions and is defined as follows (see \cite{Agakishiev:2011vf,Adamczewski-Musch:2019byl}):
\begin{equation}
  \frac{dN_{\rm ref}}{dM_{e^+e^-}} =      \left( \frac{c_{pp}}{\sigma_{pp}^{\rm tot}} \frac{d\sigma_{e^+e^-}^{pp}}{dM_{e^+e^-}}
                                         +       \frac{c_{np}}{\sigma_{np}^{\rm tot}} \frac{d\sigma_{e^+e^-}^{np}}{dM_{e^+e^-}}
                                          \right)
                                     A_{\rm part}  ~.           \label{dNrefdMee}
\end{equation}
Here, $c_{pp}$ and $c_{np}$ are the fractions of $pp+nn$ and $np$ collisions, respectively, calculable as follows:
\begin{eqnarray}
   c_{pp} &=& \frac{(P_{pp}+P_{nn}) \sigma_{pp}^{\rm tot}}{(P_{pp}+P_{nn}) \sigma_{pp}^{\rm tot} + P_{np} \sigma_{np}^{\rm tot}}~,   \label{c_pp}\\
   c_{np} &=& \frac{P_{np} \sigma_{np}^{\rm tot}}{(P_{pp}+P_{nn}) \sigma_{pp}^{\rm tot} + P_{np} \sigma_{np}^{\rm tot}}~,   \label{c_np}
\end{eqnarray}
where $P_{pp}=(Z/A)^2$, $P_{nn}=(N/A)^2$, and $P_{np}=2ZN/A^2$ are the probabilities that a randomly chosen $NN$ pair will be, respectively, $pp$, $nn$ or $np$ one
for symmetric $(A,Z)+(A,Z)$ nuclear collisions.
The total $pp$ and $np$ cross sections calculated using GiBUU  at $E_{\rm lab}=1.25$~GeV are  $\sigma_{pp}^{\rm tot} = 48$ mb and $\sigma_{np}^{\rm tot} = 39$ mb, in a good agreement with empirical data \cite{Tanabashi:2018oca}. The effective participant number can be determined as the ratio 
of the $\pi^0$ multiplicities in the studied $AA$ and in the $NN$ collision, i.e.
\begin{equation}
   A_{\rm part} = \frac{N_{\pi^0}^{AA}}{N_{\pi^0}^{NN}}~,      \label{A_part}
\end{equation}
where 
\begin{equation}
    N_{\pi^0}^{NN} = \frac{c_{pp}}{\sigma_{pp}^{\rm tot}} \sigma_{\pi^0}^{pp} + \frac{c_{np}}{\sigma_{np}^{\rm tot}} \sigma_{\pi^0}^{np}~.    \label{N_pi0^NN}
\end{equation}
Note that it is assumed in Eqs.~(\ref{dNrefdMee}) - (\ref{N_pi0^NN}) that the total cross sections, the dilepton and the $\pi^0$ production cross sections 
in $pp$ and $nn$ collisions are the same. For dileptons this assumption is obviously quite rough. It is needed here, however,  because Eq.~(\ref{dNrefdMee}) is used in the experimental analysis as well \cite{Agakishiev:2011vf,Adamczewski-Musch:2019byl}. For internal consistency, we apply Eq.~(\ref{dNrefdMee}) 
with all quantities calculated from GiBUU. To avoid some possible misunderstanding, in this subsection the $\pi^0$ multiplicity refers 
properly to the charge neutral pions, and not to the average charged pion multiplicity of Eq.~(\ref{N_pi0}). 

\begin{figure}
  \includegraphics[scale = 0.60]{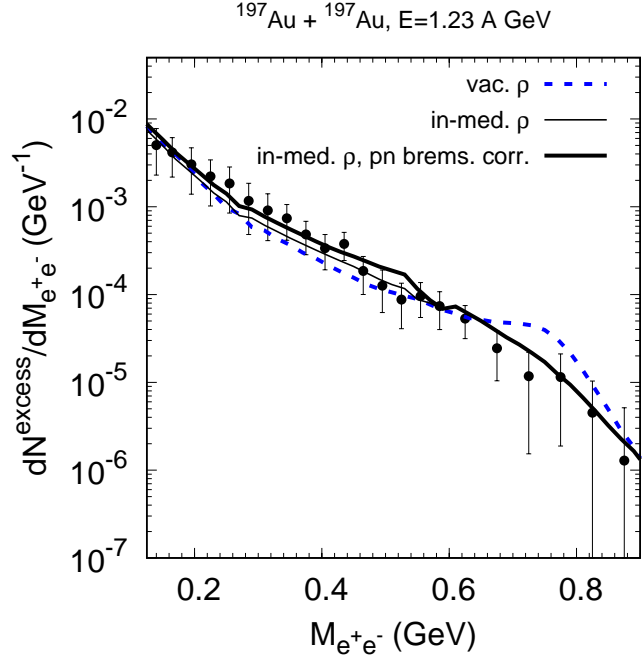}
  \caption{\label{fig:dNdM_AuAu_exc}(Color online) The invariant mass spectrum of excess dileptons produced in Au+Au collisions at $1.23 A$~GeV
    calculated according to Eq.~(\ref{dNexcessdMee}).
    The dashed (blue) versus the thick and thin solid (black) lines depict calculations with vacuum $\rho$ width versus in-medium $\rho$ width with and without correction of the $pn$ bremsstrahlung according to Eq.~(\ref{fM}), respectively. Calculations were done without filtering, i.e.~in full acceptance. 
    Experimental data are from Ref.~\cite{Adamczewski-Musch:2019byl}.}
\end{figure}

Fig.~\ref{fig:dNdM_AuAu_exc} shows the excess dilepton radiation spectrum
\begin{equation}
    \frac{dN^{\rm excess}}{dM_{e^+e^-}} =  \frac{dN^{AA}}{dM_{e^+e^-}} - \frac{dN_{\rm ref}}{dM_{e^+e^-}}  \label{dNexcessdMee}
\end{equation}
calculated in full acceptance. Here $dN^{AA}/dM_{e^+e^-}$ is the invariant mass spectrum in $AA$ collisions with the $\eta$ and $\omega$ decay components removed.
In contrast to the smoothly dropping experimental spectrum with dilepton invariant mass,
the calculation with vacuum $\rho$ shows up a bump at the $\rho$ pole mass and a valley in the intermediate mass region. The collisional broadening of
the $\rho$ meson improves the agreement with experiment, although the deviation still remains. The correction of the $pn$ bremsstrahlung by Eq.~(\ref{fM})
further improves the agreement in the intermediate mass region.

The system mass dependence of the excess radiation can be studied by using the yield ratio \cite{Agakishiev:2011vf,Adamczewski-Musch:2019byl}
\begin{equation}
  R_{AA} = \frac{dN^{AA}}{dM_{e^+e^-}} \left(\frac{dN_{\rm ref}}{dM_{e^+e^-}}\right)^{-1}~.   \label{R_AA}
\end{equation}
The $\eta$ contribution is subtracted in both, the $AA$ and the reference, spectra. 
At low $e^+e^-$ invariant masses, where the dilepton spectrum is saturated by the $\pi^0$ Dalitz decays, one obtains $R_{AA}=1$
in full acceptance. Taking into account the HADES acceptance in the calculation of dilepton spectra while keeping the full $\pi^0$ multiplicities
in the definition of the participant number, Eq.~(\ref{A_part}), results in deviations from unity of $R_{AA}$ in the $\pi^0$ Dalitz region. 
\begin{figure}[ht]
  \includegraphics[scale = 0.50]{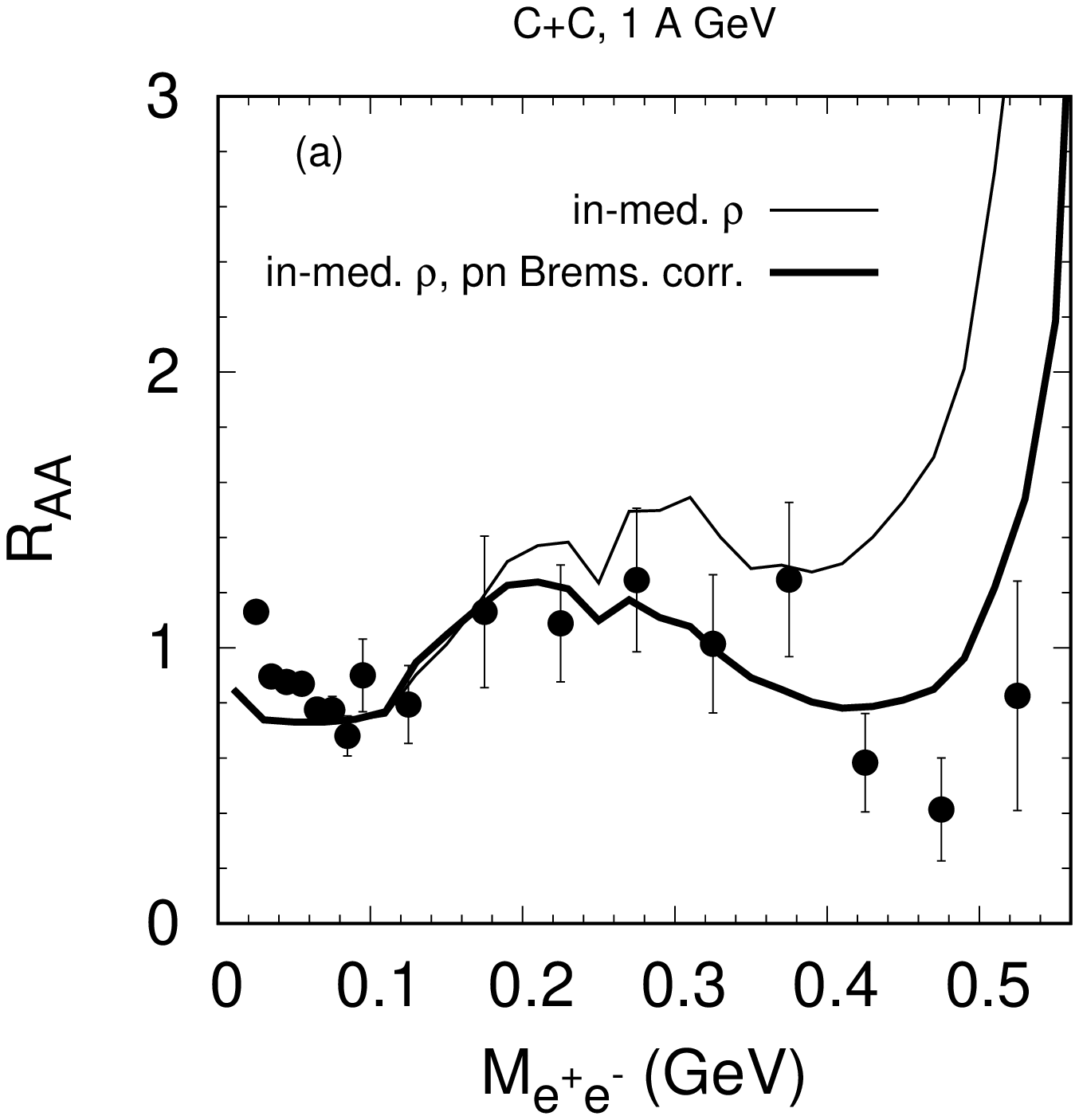}
  \includegraphics[scale = 0.50]{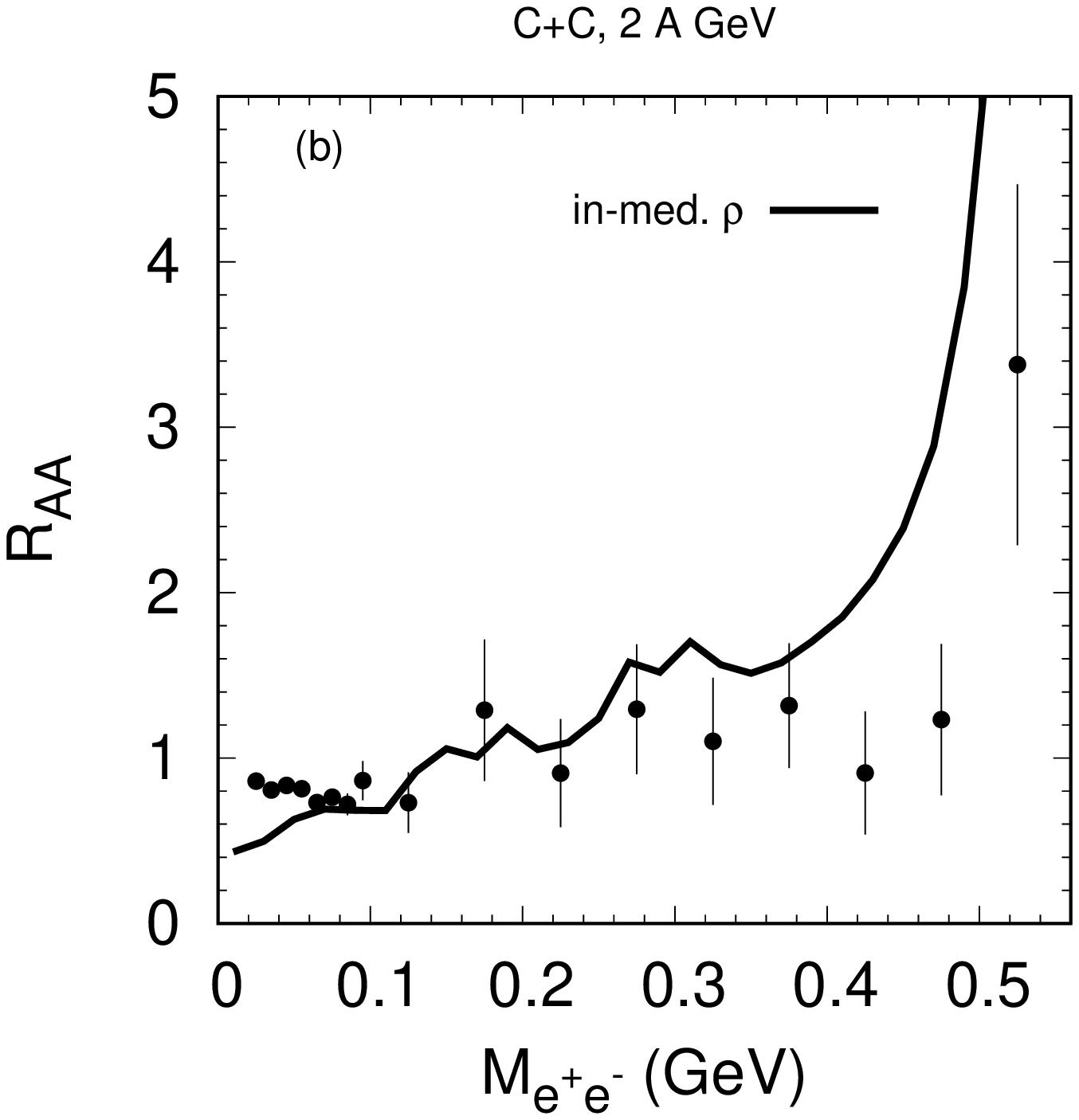}
  \includegraphics[scale = 0.50]{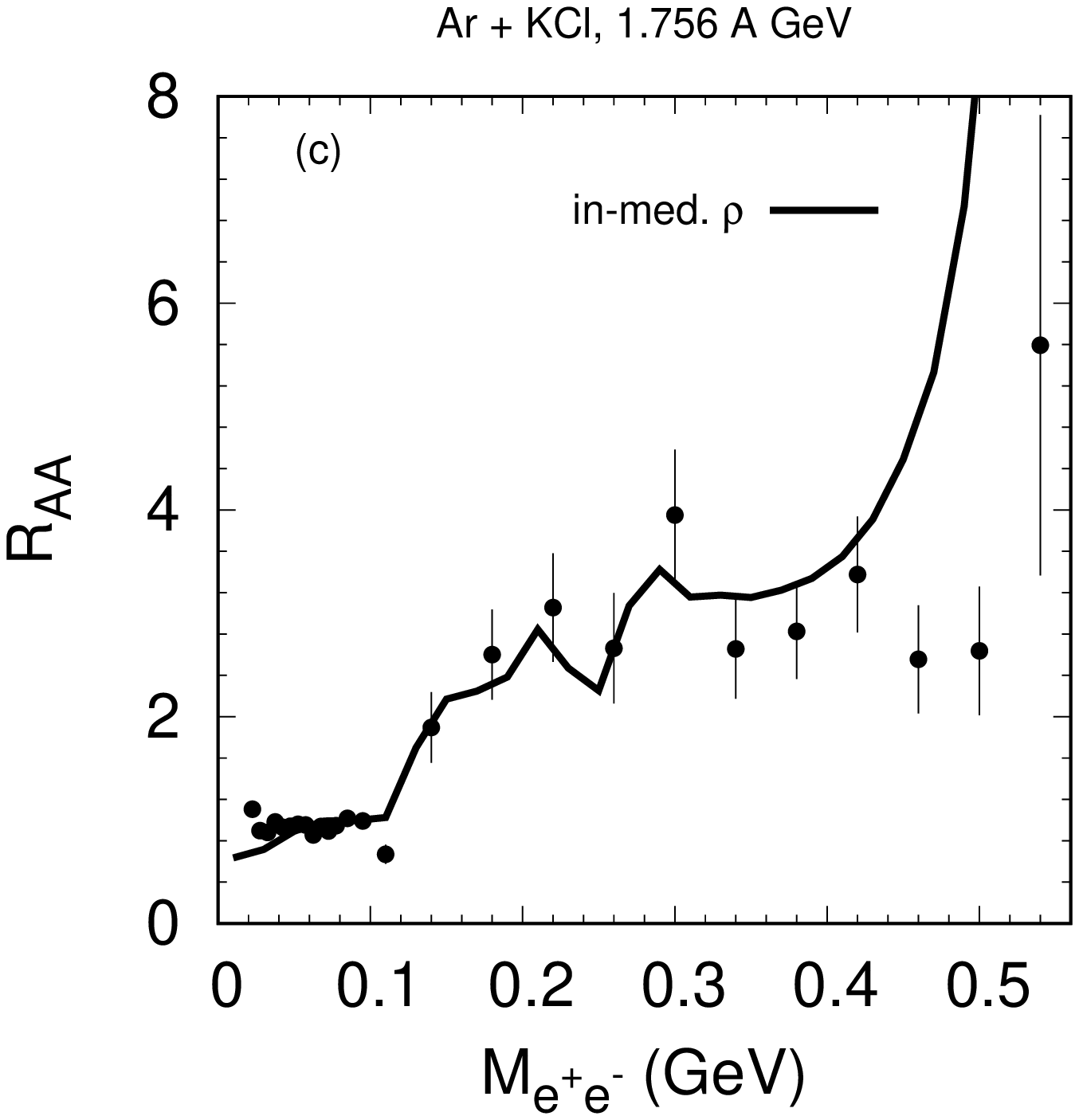}
  \includegraphics[scale = 0.50]{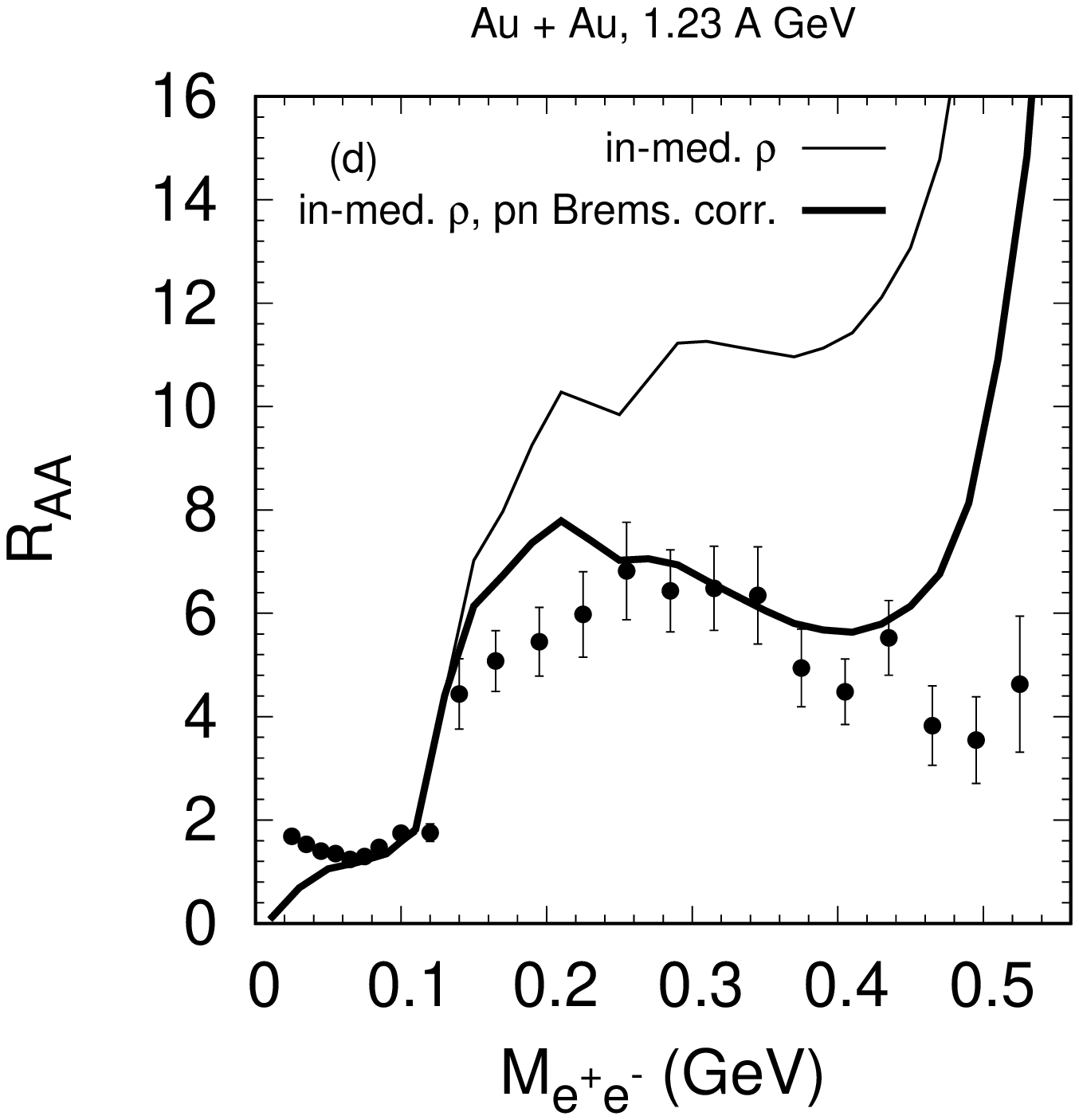}
  \caption{\label{fig:R_AA} The yield ratio of Eq.~(\ref{R_AA}) normalized at the experimental value for  $M_{e^+e^-} \approx 0.07$ GeV 
    as a function of the dilepton invariant mass. Panels (a), (b), (c), and (d) correspond to  C+C at $1 A$~GeV, C+C at $2 A$~GeV, Ar+KCl at $1.756 A$~GeV, and Au+Au at $1.23 A$~GeV.
    The calculations include collisional broadening of the $\rho$ meson.
    For C+C at $1 A$~GeV and Au+Au at $1.23 A$~GeV the thick and thin solid lines show, respectively, the result with and without
    the $pn$ bremsstrahlung correction factor in Eq.~(\ref{fM}). 
    Experimental data are from Refs.~\cite{Agakishiev:2011vf,Adamczewski-Musch:2019byl}.}
\end{figure}

Fig.~\ref{fig:R_AA} shows the yield ratio for the different colliding systems.
For C+C at 1 and $2 A$~GeV, following Ref.~\cite{Agakishiev:2011vf},  the calculations have been performed within Ar+KCl acceptance, both for the spectrum 
from heavy-ion collision and for the reference spectrum.
For the Au+Au system, the $d+p$ acceptance filter at $1.25 A$~GeV has been applied, while the $pp$ and $np$ components of the reference spectrum have been 
calculated with $pp$ and $dp$ acceptance filters at $1.25 A$~GeV, respectively.
Our calculations correctly reproduce
the main trend present in the HADES data, i.e.~the enhanced dilepton yields at intermediate invariant masses with respect to the yield at small invariant masses for the heavy colliding systems, Ar+KCl and Au+Au.
Again, the correction of the $pn$ bremsstrahlung improves the agreement with the C+C data at $1 A$~GeV and the Au+Au data at $1.23 A$~GeV. The figure also shows that in the mass-region between about 0.2 and 0.4 GeV not only in-medium effects contribute, but that there is also a strong sensitivity to the $pn$-bremsstrahlung.
We reiterate, however, that the Ar+KCl acceptance filter was used for C+C and its reference spectrum, while the $d+p$ acceptance filter was used for Au+Au. These differences blur the discussion of physical effects.

\section{Summary and conclusions}
\label{summary}

We have performed microscopic transport calculations of dilepton ($e^+e^-$) production in heavy-ion collisions at $E_{\rm beam}=1-2$~A~GeV.  
The calculations were based on the GiBUU model which has been already successfully applied to describe dilepton production in $pA$ reactions \cite{Weil:2012ji}. The main model inputs such as resonance parameters, elementary cross sections, and dilepton production channels are the same as in 
Ref.~\cite{Weil:2012ji}.
As compared to Ref.~\cite{Weil:2012ji}, the present calculations include the $pp$ and $np$ bremsstrahlung described in the
framework of the boson-exchange model \cite{Shyam:2010vr} that provides a somewhat better description of the $d+p$ data (cf.~our Fig.~\ref{fig:dNdM_res_pp&dp1.25gev}
and Fig.~3 of Ref.~\cite{Weil:2012ji} where the $np$ bremsstrahlung is calculated in the soft-photon approximation).

The most important novel feature of the present work is the self-consistent description of the $\rho$-meson spectral function in the nuclear medium. Self-consistency here implies
that the collisional width of the $\rho$-meson that enters the spectral function is calculated from the collision term of the transport equation, i.e.~it includes the contribution
of the $\rho N$ collisions.\footnote{In the previous microscopic transport studies of dilepton production \cite{Bratkovskaya:2007jk,Weil:2012ji,Bratkovskaya:2013vx}
the collisional broadening of the $\rho$-meson has been included in the linear density approximation.}
The non-equilibrium momentum distribution of the baryons at every space-time point of the colliding system is thereby approximated by an equivalent equilibrium distribution which is then
used to calculate the collisional width of the $\rho$-meson. The applied procedure is based on using the density and $\langle p^2 \rangle$ of the baryons in the local rest frame of the baryonic matter. Such a procedure is expected
to be accurate enough for the highly-compressed state of the baryonic matter formed in the central zone of a heavy-ion collision.

In order to describe the propagation of the off-shell $\rho$-meson in the presence of collisional broadening, we have applied the relativistic off-shell potential ansatz where the deviation of the actual off-shell particle mass squared from the pole mass squared is proportional to the total width of the particle, with the constant of proportionality defined at the production time
of the particle. This allows to recover the vacuum spectral function of the particle when it is emitted to the vacuum outside the baryonic matter.
  
We compared the results of GiBUU calculations with HADES data on the dilepton invariant mass spectra for $p+p$ at $E_{\rm beam}=1.25$, 2.2, and 3.5~GeV, $d+p$ at $1.25 A$~GeV,
C+C at 1 and $2 A$~GeV, Ar+KCl at $1.756 A$~GeV, and Au+Au at $1.23 A$~GeV, and also provided our predictions for Ag+Ag at $1.58 A$~GeV.
The model calculations agree with all HADES data at $E_{\rm beam} \approx 2 A$~GeV and above.
The data for $p+p$ at $1.25 A$~GeV are also described quite well. However, for $d+p$ at $1.25 A$~GeV,  C+C at $1 A$~GeV, and  Au+Au at $1.23 A$~GeV
there is a systematic underprediction of the dilepton yield in the intermediate invariant-mass range, $M_{e^+e^-} \approx 0.2-0.5$ GeV, which deserves further study in the future.

Including the collisional broadening of the $\rho$-meson smears out the peak in the dilepton invariant-mass spectrum near the $\rho$ pole mass and increases the dilepton yield in the intermediate invariant-mass range. The effect is most clearly  visible in the heaviest system Au+Au, while   
it does practically not influence the dilepton spectra for C+C and only weakly changes that for Ar+KCl. The overall strength of the $\rho$ collisional broadening alone, however, is not sufficient to completely account for the missing strength in the intermediate invariant mass region in Au+Au at $1.23 A$~GeV.

This motivated us to adjust the $n+p$ bremsstrahlung cross section
at $E_{\rm beam}=1.25$~GeV by a dilepton invariant-mass dependent factor so as to describe the inclusive dilepton data for $d+p$ at $1.25 A$~GeV. Multiplying the $n+p$ bremsstrahlung component in C+C at $1 A$~GeV and Au+Au at $1.23 A$~GeV by the same factor, without further adjustments, then leads to good agreement with the data on the inclusive dilepton invariant-mass spectra in these heavier systems as well.
Further circumstantial evidence for the increased $n+p$ bremsstrahlung cross section in vicinity of $1 A$~GeV beam energy is also provided by an improved agreement of the excess dilepton yield (see Fig.~\ref{fig:dNdM_AuAu_exc}) and yield ratios (see Figs.~\ref{fig:R_AA} (a), (d)) with the experimental data. 

While the suggested enhancement of the $n+p$
bremsstrahlung is thus effective in improving the intermediate mass dilepton yields of all systems in the range of $1 - 1.5 A$~GeV beam energy, this does not exclude other possibilities. In particular, 
a related alternative problem at these low beam energies might be the $\eta$ production near threshold
which is extremely difficult to constrain from experimental data in $p+p$ and $p+n$ collisions with high precision.
This thus introduces an additional uncertainty in the input cross sections. Therefore, the missing dilepton yield in
$d+p$ collisions at $1.25 A$~GeV might be also at least partly attributed to missing strength in the $\eta$ Dalitz decay component.
The detailed theoretical analysis of the exclusive $dp \to e^+ e^- n p p_{\rm fast}$ cross sections \cite{Adamczewski-Musch:2017oij}, which is beyond the scope of our present work, would be needed to further test the suggested enhancement of the $n+p$ bremsstrahlung component. The availability of precisely determined exclusive $dp \to \gamma e^+ e^- n p p_{\rm fast}$ cross sections would also be useful for better constraining the different partial components of the dilepton invariant-mass spectra. 

Last but not least, the processes involving the deuteron, which would be sub-threshold for the corresponding $p+p$ or $n+p$ collisions on a free target proton at the same beam energy per nucleon, depend on the deuteron wave function at high momenta. The latter is subject to significant
relativistic corrections, as follows from the light-cone description of the deuteron \cite{Frankfurt:1977vc,Frankfurt:1981mk}.
This possibility was not considered in the present work and remains to be studied in future.

In summary, the present calculations, based on non-equilibrium transport theory, give a good description of a wealth of data on dilepton production in heavy-ion collisions, obtained in the HADES experiment. While there are many detailed theoretical uncertainties to be investigated, which we have specified both in the main text and in Appendix B, they are not large enough to affect the overall agreement of the GiBUU calculations with experiment significantly. The main message then is that no additional postulates of thermal equilibrium and of chiral symmetry restoration are necessary to understand these data.

\begin{acknowledgments}
  We thank Tetyana Galatyuk, Mark Strikman, and Janus Weil for stimulating discussions and interest in this work.
  We are especially grateful to Volker Metag for many constructive suggestions concerning the comparison of our calculations with experimental data. 
  The support by the Frankfurt Center for Scientific Computing is gratefully acknowledged.
  This work was financially supported by the German Federal Ministry of Education and
  Research (BMBF), Grant No. 05P18RGFCA.
\end{acknowledgments}

\newpage

\appendix

\begin{appendices}

\section{Hadron Numbers}
\label{hadronnumbers}

\begin{subappendices}

\renewcommand{\thesubsection}{A\arabic{subsection}}

\subsection{Pion numbers}
\label{pionnumbers}

At the end of Sect.\ \ref{AuAu} we have briefly mentioned a problem connected with the measured vs.~calculated pion numbers. We, therefore, now list in Table~\ref{tab:pions}  the calculated pseudo neutral-pion multiplicities, Eq.~(\ref{N_pi0}), together with the corresponding HADES data.
\begin{table}[htb]
  \caption{\label{tab:pions}
    Pseudo neutral $\pi^0$ multiplicities $N_{\pi^0}$ as defined by Eq.~(\ref{N_pi0}) from GiBUU for various colliding systems. The results for Au+Au and Ag+Ag are geometrically weighted in the impact parameter ranges that corresponds to $0-40\%$ centrality which are 
    $b < 9.3$~fm and $b < 7.7$~fm, respectively. For the other systems the impact parameter distributions reproducing the HADES trigger were used.}
  \begin{center}
    \begin{tabular}{lllll}
    \hline
    \hline
    system  \hspace{3cm}   &  $N_{\pi^0}$ \hspace{1cm}    &  exp.  \hspace{2cm}   & Ref. \\
    \hline
    Au+Au, $1.23 A$~GeV      &  12.8                      &  $8.65\pm0.52$        & \cite{Adamczewski-Musch:2019byl}\\
    C+C, $1 A$~GeV           &  0.53                      &  $0.52\pm0.08$        & \cite{Agakishiev:2007ts}\\
    C+C, $2 A$~GeV           &  1.00                      &  $1.16\pm0.16$        & \cite{Agakishiev:2009zv}\\
    Ar+KCl, $1.76 A$~GeV     &  4.1                       &  $3.50\pm0.25$        &  \cite{Agakishiev:2011vf}\\
    Ag+Ag, $1.58 A$~GeV      &  9.8                       &                       &                          \\ 
    \hline
    \hline
    \end{tabular}
  \end{center}
\end{table}

While the calculated pseudo neutral-pion multiplicities for C+C at $1 A$~GeV and $2 A$~GeV agree with the experimental numbers very well, and those for Ar+KCl still reasonably well, the theoretical values for the Au+Au system lie about $50\%$ above the experimentally determined values.  
Also the charged pion multiplicities listed in Table~\ref{tab:mult} below overestimate HADES measurements by about $50\%$, i.e.\ just by the same factor as for the pseudo neutral pions for Au+Au at $0-40\%$ centrality.\footnote{The charged pion multiplicities for GiBUU reported in Ref.~\cite{Adamczewski-Musch:2020vrg} are slightly different
because of the default mode of GiBUU using Skyrme-like baryonic mean fields.}

In our calculations, the numbers of true neutral pions, 17.4 at $0-10\%$ and 11.0 at  $0-40\%$ centrality, are only about $12\%$ lower than the numbers of pseudo neutral pions, i.e. 20.3 at $0-10\%$ and 12.8
at $0-40\%$ centrality.  We can thus exclude this difference as the main source of the discrepancy between the dilepton yield in the $\pi^0$ Dalitz region
and the pseudo neutral-pion multiplicity. 

The fact that the virtual photons from $\pi^0$ Dalitz decays are in agreement with the experimentally measured dileptons in the corresponding invariant-mass range, while the total pion yields are not, thus remains a puzzle.

 Any simple mechanism to reduce the pion yields would inevitably also reduce the $\pi^0$ Dalitz contribution to the dilepton invariant-mass spectra, which, however, agrees very well with the data, see Fig.~\ref{fig:dNdM_AuAu}. Our calculations using a medium-dependent suppression of $NN \leftrightarrow N\Delta$ cross sections from \cite{Song:2015hua} have indeed shown that effect. Such a medium-dependent suppression therefore then requires an additional explanation of missing dilepton yield in the $\pi^0$ Dalitz region. One possibility could be a shortcoming of the acceptance filter used in our calculations. While for C+C, where good agreement for the pion numbers is obtained, a filter specific for that system exists, this is not the case for Au+Au. A designated HADES acceptance filter for Au+Au would certainly help to close in on such a possibility in the future.

\subsection{Hadron multiplicities}
\begin{figure}
  \includegraphics[scale = 0.50]{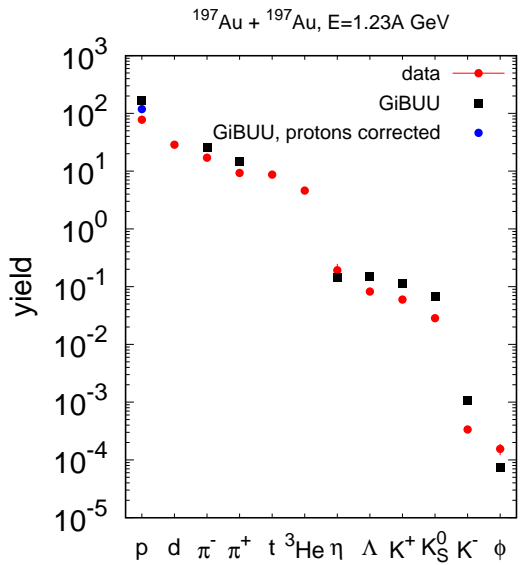}
  \caption{\label{fig:yields}(Color online) Particle multiplicities per event in Au+Au collisions at $1.23 A$~GeV ($\sqrt{s_{NN}}=2.4$ GeV).
    Multiplicities calculated by using GiBUU are shown by black solid boxes. The blue solid circle shows the calculated proton multiplicity
    corrected with a help of Eq.(\ref{N_p^corr}). Experimental data (see Table~\ref{tab:mult}) are displayed by red solid circles.}
\end{figure}

As a further benchmark test, Fig.~\ref{fig:yields} shows particle multiplicities in central collisions of Au+Au at $1.23 A$~GeV.
The calculated multiplicities are weighted with the impact parameter in the range $b=0-4.7 (6.6)$~fm for the $0-10 (20)\%$ most central events
\cite{Adamczewski-Musch:2017sdk} 
and include all particles present in the system after the decays of unstable resonances.

The experimental multiplicities of protons and $\Lambda$'s are overestimated since some of these particles are bound in nuclear clusters.
In particular, for protons the corrected multiplicity due to clustering can be estimated by subtracting the total number of protons in clusters using the experimental cluster multiplicities from the multiplicity of protons calculated within GiBUU:
\begin{equation}
   N_p^{\rm corr.} = N_p^{\rm GiBUU} - N_d^{\rm exp} - N_t^{\rm exp} - 2N_{^3\mbox{He}}^{\rm exp} - 2N_{^4\mbox{He}}^{\rm est}    = 118.4, \label{N_p^corr}
\end{equation}
where $N_p^{\rm GiBUU}=168.7$ is the calculated proton multiplicity; $N_d^{\rm exp}=28.7 \pm 0.8$,
$N_t^{\rm exp}=8.7 \pm 1.1$, and $N_{^3\mbox{He}}^{\rm exp}= 4.6 \pm 0.3$ are the multiplicities of deuterons,
tritons and $^3\mbox{He}$ measured experimentally \cite{Harabasz:2020sei}. The estimated multiplicity of $\alpha$ particles, 
$N_{^4\mbox{He}}^{\rm est} \approx 0.4 N_{^3\mbox{He}}^{\rm exp}$, according to EOS data for central Au+Au collisions at $1 A$~GeV (see Ref.~\cite{Neubert:1999sv} 
and refs.~therein).  
Table~\ref{tab:mult} summarizes calculated and measured particle multiplicities. The corrected proton multiplicity still overestimates the experimental value by $\approx 50\%$. This can be partly explained
by missing heavier cluster contributions in Eq.~(\ref{N_p^corr}). The detailed analysis of cluster production is obviously a difficult problem that is not in the focus of this work.

The calculated $\eta$ multiplicity agrees very well with the experimental value. This implies that the $\eta \to e^+ e^- \gamma$ decay component of the dilepton invariant
spectra is described correctly.

\begin{table}[htb]
  \caption{\label{tab:mult}
    Particle multiplicities per event for Au+Au at $E_{\rm beam}=1.23 A$~GeV. For protons the corrected multiplicity, Eq.~(\ref{N_p^corr}), is given in parentheses.}
  \begin{center}
    \begin{tabular}{lllll}
    \hline
    \hline
    particle \hspace{1cm}  & $N^{\rm GiBUU}$                      &  $N^{\rm exp}$    \hspace{1cm}   & centrality  & ref. \\
    \hline
    $p$       & 168.7 (118.4) \hspace{1cm  }&  $77.6 \pm 2.4$               & $0-10\%$    & \cite{Harabasz:2020sei} \\
    $\pi^-$   & 25.6                      &  $17.1\pm1.2$                 & $0-10\%$    & \cite{Adamczewski-Musch:2020vrg} \\
    $\pi^+$   & 14.9                      &  $9.3\pm0.6$                  & $0-10\%$    & \cite{Adamczewski-Musch:2020vrg} \\
    $\eta$    & 0.14                    &  $0.192\pm0.056$              & $0-20\%$    & \cite{Behnke:2016} \\
    $\Lambda$ & 0.15                    &  $(8.22\pm0.74)\cdot10^{-2}$ \hspace{1cm}& $0-10\%$    & \cite{Adamczewski-Musch:2018xwg} \\ 
    $K^+$     & 0.11                    &  $(5.98\pm0.679)\cdot10^{-2}$ & $0-10\%$     &  \cite{Adamczewski-Musch:2017rtf} \\
    $K^0_S$   & 0.07                    &  $(2.84\pm0.26)\cdot10^{-2}$  & $0-10\%$     &  \cite{Adamczewski-Musch:2018xwg} \\
    $K^-$     & ($1.06\pm0.05)\cdot10^{-3}$        &  $(3.36\pm0.39)\cdot10^{-4}$  & $0-20\%$     &  \cite{Adamczewski-Musch:2017rtf}\\
    $\phi$    & ($7.3\pm1.4)\cdot10^{-5}$        &  $(1.55\pm0.34)\cdot10^{-4}$ & $0-20\%$     &    \cite{Adamczewski-Musch:2017rtf}\\
    \hline
    \hline
    \end{tabular}
  \end{center}
\end{table}

\end{subappendices}

\section{Residual Uncertainties}
\label{uncert}

\begin{subappendices}

\renewcommand{\thesubsection}{B\arabic{subsection}}

\subsection{Mean-field potentials}

One important aspect of the GiBUU transport model is the self-consistent baryonic mean-field potential that can be provided
either by a Skyrme-like energy density functional or by the relativistic mean-field (RMF) Lagrangian of the non-linear Walecka model. In order to estimate the uncertainty caused by the specific choice of mean-field model, we have also performed calculations
using the soft momentum-dependent (SM) Skyrme-like potential \cite{Welke:1988zz} with an incompressibility or bulk modulus
of nuclear matter at normal density, $K=215$~MeV. This is quite close to the value $K=210$~MeV for the NL2 version \cite{Lang:1991qa} of the RMF model used in the majority of calculations presented in this work. 

While the equation of state is therefore almost the same for SM and RMF NL2,
the energy dependence of the optical potential is quite different (see Figs.~2 and 3 of Ref.~\cite{Buss:2011mx}), however. This is a well-known effect of the original RMF model that leads to too repulsive Schroedinger-equivalent potentials at high momenta \cite{Blaettel:1993uz}. Note, however, that around a beam energy of 1 GeV both potentials are close to the phenomenological one. Thus, we think that the influence of different momentum dependencies in SM and RMF NL2 on heavy-ion collisions at
$\approx 1 A$~GeV beam energy is rather weak, although a dedicated study is definitely needed to clarify this issue.

As a matter of fact, due to the different computing prescriptions (see App. D.4.2 of Ref.~\cite{Buss:2011mx} for detail), the so-called 'free' invariant energy, 
$\sqrt{s}_{\rm free}$, used in the calculations of the baryon-baryon cross sections also differ between the Skyrme-like and RMF modes of calculation. In particular, we see from Fig.~D69 of Ref.~\cite{Buss:2011mx} that the $NN$ cross sections are dialed at larger invariant energies in the RMF mode than in the SM mode (we have also checked this by direct comparison of the $\sqrt{s}_{\rm free}$ distributions
in the baryon-baryon collisions for Au+Au at $1.23 A$~GeV). One therefore also expects differences in the $\Delta$ resonance production between the two modes of calculation. Dilepton invariant-mass spectra calculated with the RMF versus the SM mean fields are compared in Figs.~\ref{fig:dNdM_AuAu} (a) and (c).
Indeed, the $\Delta \to N e^+ e^-$ components differ somewhat between the two modes: in the RMF mode (Fig.~\ref{fig:dNdM_AuAu} (a)) there is a bump near $M_{e^+e^-}=0.2$ GeV, while in the Skyrme-like mode (Fig.~\ref{fig:dNdM_AuAu} (c)) the $\Delta$ Dalitz component monotonically drops with $M_{e^+e^-}$. This agrees with our expectation of effectively more energetic $NN \to N\Delta$ collisions in the RMF mode. Overall, however, the difference between results obtained with the RMF and the Skyrme potential is not relevant.

Also note that the $\Delta$ potential in the GiBUU default Skyrme-like mode is scaled by a factor of 2/3 relative to the nucleon potential, corresponding to results from the $\Delta$-hole model \cite{Ericson:1988gk}. In contrast, the RMF calculation was run with the scalar and vector potentials acting on the $\Delta$ assumed to be the same as the nucleonic ones. We have checked, however, that changing the scaling factor of the $\Delta$ potential from $2/3 \to 1$ in the Skyrme-like mode does not lead to any visible changes in the $\Delta$ Dalitz component of the dilepton spectrum.

\subsection{In-medium cross sections and widths}

GiBUU, like any other microscopic transport model, relies on various elementary cross sections which are typically taken from or adjusted to experimental data.  
Consistent calculations of possible in-medium change of all these cross sections are not available.
However, in some cases there are available models which can be relatively easily
included in transport simulations.
One example is the in-medium suppression of the $NN \leftrightarrow N\Delta$
cross sections mentioned in the end of Appendix \ref{pionnumbers}.

Another example is the collisional broadening of  the $\Delta$ resonance in the nuclear medium.
This effect can be included by means of the potential model of $\Delta$ spreading of Ref.~\cite{Oset:1987re}
that is optionally implemented in GiBUU (see Appendix B of Ref. \cite{Buss:2011mx} for details).
In GiBUU, the cross sections of the $\Delta$ quasielastic scattering $\Delta N \to \Delta N$, two-body absorption 
$\Delta N \to N N$, and three-body absorption $\Delta N N \to N N N$ are modified such that the corresponding collision rates
reproduce those of Ref. \cite{Oset:1987re} in cold nuclear matter of a given density. In order to ensure the relation to Ref. \cite{Oset:1987re}, the kinematics of the incoming pion is determined  from the $\pi N \to \Delta$ process on the nucleon from the Fermi sea for the $\Delta$ of a given invariant mass. In contrast to these processes the $\Delta$ production $N N \to N \Delta$ is kept
unmodified and the production $N N N \to N N \Delta$ is not included. As a result, the pion absorption gets effectively increased, but the pion production does not. Accidentally, the pseudo neutral-pion multiplicity calculated with the $\Delta$ spreading potential in Au+Au at $1.23 A$~GeV, $N_{\pi^0}=7.8$, becomes close to the experiment (see Table~\ref{tab:pions}). However, for the light system C+C at $1 A$~GeV we have now $N_{\pi^0}=0.37$ that underestimates the data by 30\%.
This leads to the underestimation of the dilepton yields in the $\pi^0$ Dalitz region\footnote{As discussed in Appendix \ref{pionnumbers} any in-medium change of resonance production cross sections would also affect the $\pi^0$ Dalitz decay contribution in the dilepton spectrum.}.
Another effect of the in-medium $\Delta$ width is the washing out
of the bump at $M_{e^+e^-} \approx 0.2$ GeV in the $\Delta$ Dalitz decay component that is present in calculations employing RMF. 
The $\rho$ direct component at $M_{e^+e^-} \gtapprox 0.4$ GeV gets also somewhat reduced in calculation with $\Delta$ spreading potential.
This, however, does not much influences the total invariant-mass spectrum of dileptons that remains practically unchanged, except reduction of the $\pi^0$ Dalitz peak mentioned above.

There is also a possibility that the collisional width of $\rho$ influences various
$\rho$-mediated dilepton production processes according to VDM.
In particular, this in-medium effect may modify the $pn$ bremsstrahlung cross section.  
In Sect.\ \ref{elem} we have introduced a tuning factor, Eq.(\ref{fM}), to bring the elementary $pn$ bremsstrahlung cross section into agreement with experiment. 
We expect that including the in-medium broadening of the $\rho$ meson into the exchange current contribution of the $pn$ cross section (Fig.~1c of \cite{Shyam:2010vr}) would somewhat shift the strength towards lower masses. The detailed evaluation of this effect
is, however, clearly beyond the scope of the present investigation.

\subsection{Centrality selection}

Finally, there also remains an uncertainty related to the centrality selection. For the dilepton spectra calculations in the Au+Au system at $1.23 A$~GeV  we have applied a sharp cutoff approximation,
i.e.~the spectra are geometrically weighted in the impact parameter range $b < 9.3$~fm which corresponds to $0-40\%$ centrality \cite{Adamczewski-Musch:2017sdk}. The Glauber 
Monte-Carlo model on the other hand produces an impact-parameter distribution smeared by $\approx 1$~fm at the upper bound instead of the sharp cutoff (see e.g.~the upper Fig.~8 in \cite{Adamczewski-Musch:2017sdk}).
We have therefore varied the upper limit of the impact parameter by $\pm 1$~fm to estimate that such variations can at most lead to overall changes of about $\mp 15\%$ in the dilepton yields for Au+Au at $1.23 A$~GeV.

\end{subappendices}

\end{appendices}

\bibliography{dilepHIC}

\begin{thebibliography}{94}%
\makeatletter
\providecommand \@ifxundefined [1]{%
 \@ifx{#1\undefined}
}%
\providecommand \@ifnum [1]{%
 \ifnum #1\expandafter \@firstoftwo
 \else \expandafter \@secondoftwo
 \fi
}%
\providecommand \@ifx [1]{%
 \ifx #1\expandafter \@firstoftwo
 \else \expandafter \@secondoftwo
 \fi
}%
\providecommand \natexlab [1]{#1}%
\providecommand \enquote  [1]{``#1''}%
\providecommand \bibnamefont  [1]{#1}%
\providecommand \bibfnamefont [1]{#1}%
\providecommand \citenamefont [1]{#1}%
\providecommand \href@noop [0]{\@secondoftwo}%
\providecommand \href [0]{\begingroup \@sanitize@url \@href}%
\providecommand \@href[1]{\@@startlink{#1}\@@href}%
\providecommand \@@href[1]{\endgroup#1\@@endlink}%
\providecommand \@sanitize@url [0]{\catcode `\\12\catcode `\$12\catcode
  `\&12\catcode `\#12\catcode `\^12\catcode `\_12\catcode `\%12\relax}%
\providecommand \@@startlink[1]{}%
\providecommand \@@endlink[0]{}%
\providecommand \url  [0]{\begingroup\@sanitize@url \@url }%
\providecommand \@url [1]{\endgroup\@href {#1}{\urlprefix }}%
\providecommand \urlprefix  [0]{URL }%
\providecommand \Eprint [0]{\href }%
\providecommand \doibase [0]{http://dx.doi.org/}%
\providecommand \selectlanguage [0]{\@gobble}%
\providecommand \bibinfo  [0]{\@secondoftwo}%
\providecommand \bibfield  [0]{\@secondoftwo}%
\providecommand \translation [1]{[#1]}%
\providecommand \BibitemOpen [0]{}%
\providecommand \bibitemStop [0]{}%
\providecommand \bibitemNoStop [0]{.\EOS\space}%
\providecommand \EOS [0]{\spacefactor3000\relax}%
\providecommand \BibitemShut  [1]{\csname bibitem#1\endcsname}%
\let\auto@bib@innerbib\@empty
\bibitem [{\citenamefont {Hatsuda}\ and\ \citenamefont
  {Lee}(1992)}]{Hatsuda:1991ez}%
  \BibitemOpen
  \bibfield  {author} {\bibinfo {author} {\bibfnamefont {T.}~\bibnamefont
  {Hatsuda}}\ and\ \bibinfo {author} {\bibfnamefont {S.~H.}\ \bibnamefont
  {Lee}},\ }\href {\doibase 10.1103/PhysRevC.46.R34} {\bibfield  {journal}
  {\bibinfo  {journal} {Phys. Rev. C}\ }\textbf {\bibinfo {volume} {46}},\
  \bibinfo {pages} {34} (\bibinfo {year} {1992})}\BibitemShut {NoStop}%
\bibitem [{\citenamefont {Brown}\ and\ \citenamefont
  {Rho}(1991)}]{Brown:1991kk}%
  \BibitemOpen
  \bibfield  {author} {\bibinfo {author} {\bibfnamefont {G.}~\bibnamefont
  {Brown}}\ and\ \bibinfo {author} {\bibfnamefont {M.}~\bibnamefont {Rho}},\
  }\href {\doibase 10.1103/PhysRevLett.66.2720} {\bibfield  {journal} {\bibinfo
   {journal} {Phys. Rev. Lett.}\ }\textbf {\bibinfo {volume} {66}},\ \bibinfo
  {pages} {2720} (\bibinfo {year} {1991})}\BibitemShut {NoStop}%
\bibitem [{\citenamefont {Leupold}\ \emph {et~al.}(1998)\citenamefont
  {Leupold}, \citenamefont {Peters},\ and\ \citenamefont
  {Mosel}}]{Leupold:1997dg}%
  \BibitemOpen
  \bibfield  {author} {\bibinfo {author} {\bibfnamefont {S.}~\bibnamefont
  {Leupold}}, \bibinfo {author} {\bibfnamefont {W.}~\bibnamefont {Peters}}, \
  and\ \bibinfo {author} {\bibfnamefont {U.}~\bibnamefont {Mosel}},\ }\href
  {\doibase 10.1016/S0375-9474(97)00634-9} {\bibfield  {journal} {\bibinfo
  {journal} {Nucl. Phys. A}\ }\textbf {\bibinfo {volume} {628}},\ \bibinfo
  {pages} {311} (\bibinfo {year} {1998})},\ \Eprint
  {http://arxiv.org/abs/nucl-th/9708016} {arXiv:nucl-th/9708016} \BibitemShut
  {NoStop}%
\bibitem [{\citenamefont {Leupold}\ and\ \citenamefont
  {Mosel}(1998)}]{Leupold:1998bt}%
  \BibitemOpen
  \bibfield  {author} {\bibinfo {author} {\bibfnamefont {S.}~\bibnamefont
  {Leupold}}\ and\ \bibinfo {author} {\bibfnamefont {U.}~\bibnamefont
  {Mosel}},\ }\href {\doibase 10.1103/PhysRevC.58.2939} {\bibfield  {journal}
  {\bibinfo  {journal} {Phys. Rev. C}\ }\textbf {\bibinfo {volume} {58}},\
  \bibinfo {pages} {2939} (\bibinfo {year} {1998})},\ \Eprint
  {http://arxiv.org/abs/nucl-th/9805024} {arXiv:nucl-th/9805024} \BibitemShut
  {NoStop}%
\bibitem [{\citenamefont {Bernard}\ and\ \citenamefont
  {Meissner}(1988)}]{Bernard:1988db}%
  \BibitemOpen
  \bibfield  {author} {\bibinfo {author} {\bibfnamefont {V.}~\bibnamefont
  {Bernard}}\ and\ \bibinfo {author} {\bibfnamefont {U.~G.}\ \bibnamefont
  {Meissner}},\ }\href {\doibase 10.1016/0375-9474(88)90114-5} {\bibfield
  {journal} {\bibinfo  {journal} {Nucl. Phys. A}\ }\textbf {\bibinfo {volume}
  {489}},\ \bibinfo {pages} {647} (\bibinfo {year} {1988})}\BibitemShut
  {NoStop}%
\bibitem [{\citenamefont {Adamova}\ \emph {et~al.}(2003)\citenamefont {Adamova}
  \emph {et~al.}}]{Adamova:2002kf}%
  \BibitemOpen
  \bibfield  {author} {\bibinfo {author} {\bibfnamefont {D.}~\bibnamefont
  {Adamova}} \emph {et~al.} (\bibinfo {collaboration} {CERES/NA45}),\ }\href
  {\doibase 10.1103/PhysRevLett.91.042301} {\bibfield  {journal} {\bibinfo
  {journal} {Phys. Rev. Lett.}\ }\textbf {\bibinfo {volume} {91}},\ \bibinfo
  {pages} {042301} (\bibinfo {year} {2003})},\ \Eprint
  {http://arxiv.org/abs/nucl-ex/0209024} {arXiv:nucl-ex/0209024} \BibitemShut
  {NoStop}%
\bibitem [{\citenamefont {Agakichiev}\ \emph {et~al.}(2005)\citenamefont
  {Agakichiev} \emph {et~al.}}]{Agakichiev:2005ai}%
  \BibitemOpen
  \bibfield  {author} {\bibinfo {author} {\bibfnamefont {G.}~\bibnamefont
  {Agakichiev}} \emph {et~al.} (\bibinfo {collaboration} {CERES}),\ }\href
  {\doibase 10.1140/epjc/s2005-02272-3} {\bibfield  {journal} {\bibinfo
  {journal} {Eur. Phys. J. C}\ }\textbf {\bibinfo {volume} {41}},\ \bibinfo
  {pages} {475} (\bibinfo {year} {2005})},\ \Eprint
  {http://arxiv.org/abs/nucl-ex/0506002} {arXiv:nucl-ex/0506002} \BibitemShut
  {NoStop}%
\bibitem [{\citenamefont {Arnaldi}\ \emph {et~al.}(2006)\citenamefont {Arnaldi}
  \emph {et~al.}}]{Arnaldi:2006jq}%
  \BibitemOpen
  \bibfield  {author} {\bibinfo {author} {\bibfnamefont {R.}~\bibnamefont
  {Arnaldi}} \emph {et~al.} (\bibinfo {collaboration} {NA60}),\ }\href
  {\doibase 10.1103/PhysRevLett.96.162302} {\bibfield  {journal} {\bibinfo
  {journal} {Phys. Rev. Lett.}\ }\textbf {\bibinfo {volume} {96}},\ \bibinfo
  {pages} {162302} (\bibinfo {year} {2006})},\ \Eprint
  {http://arxiv.org/abs/nucl-ex/0605007} {arXiv:nucl-ex/0605007} \BibitemShut
  {NoStop}%
\bibitem [{\citenamefont {Arnaldi}\ \emph
  {et~al.}(2009{\natexlab{a}})\citenamefont {Arnaldi} \emph
  {et~al.}}]{Arnaldi:2008fw}%
  \BibitemOpen
  \bibfield  {author} {\bibinfo {author} {\bibfnamefont {R.}~\bibnamefont
  {Arnaldi}} \emph {et~al.} (\bibinfo {collaboration} {NA60}),\ }\href
  {\doibase 10.1140/epjc/s10052-009-0878-5} {\bibfield  {journal} {\bibinfo
  {journal} {Eur. Phys. J. C}\ }\textbf {\bibinfo {volume} {61}},\ \bibinfo
  {pages} {711} (\bibinfo {year} {2009}{\natexlab{a}})},\ \Eprint
  {http://arxiv.org/abs/0812.3053} {arXiv:0812.3053 [nucl-ex]} \BibitemShut
  {NoStop}%
\bibitem [{\citenamefont {van Hees}\ and\ \citenamefont
  {Rapp}(2006)}]{vanHees:2006ng}%
  \BibitemOpen
  \bibfield  {author} {\bibinfo {author} {\bibfnamefont {H.}~\bibnamefont {van
  Hees}}\ and\ \bibinfo {author} {\bibfnamefont {R.}~\bibnamefont {Rapp}},\
  }\href {\doibase 10.1103/PhysRevLett.97.102301} {\bibfield  {journal}
  {\bibinfo  {journal} {Phys. Rev. Lett.}\ }\textbf {\bibinfo {volume} {97}},\
  \bibinfo {pages} {102301} (\bibinfo {year} {2006})},\ \Eprint
  {http://arxiv.org/abs/hep-ph/0603084} {arXiv:hep-ph/0603084} \BibitemShut
  {NoStop}%
\bibitem [{\citenamefont {Rapp}\ \emph {et~al.}(1997)\citenamefont {Rapp},
  \citenamefont {Chanfray},\ and\ \citenamefont {Wambach}}]{Rapp:1997fs}%
  \BibitemOpen
  \bibfield  {author} {\bibinfo {author} {\bibfnamefont {R.}~\bibnamefont
  {Rapp}}, \bibinfo {author} {\bibfnamefont {G.}~\bibnamefont {Chanfray}}, \
  and\ \bibinfo {author} {\bibfnamefont {J.}~\bibnamefont {Wambach}},\ }\href
  {\doibase 10.1016/S0375-9474(97)00137-1} {\bibfield  {journal} {\bibinfo
  {journal} {Nucl. Phys.}\ }\textbf {\bibinfo {volume} {A617}},\ \bibinfo
  {pages} {472} (\bibinfo {year} {1997})},\ \Eprint
  {http://arxiv.org/abs/hep-ph/9702210} {arXiv:hep-ph/9702210 [hep-ph]}
  \BibitemShut {NoStop}%
\bibitem [{\citenamefont {Peters}\ \emph {et~al.}(1998)\citenamefont {Peters},
  \citenamefont {Post}, \citenamefont {Lenske}, \citenamefont {Leupold},\ and\
  \citenamefont {Mosel}}]{Peters:1997va}%
  \BibitemOpen
  \bibfield  {author} {\bibinfo {author} {\bibfnamefont {W.}~\bibnamefont
  {Peters}}, \bibinfo {author} {\bibfnamefont {M.}~\bibnamefont {Post}},
  \bibinfo {author} {\bibfnamefont {H.}~\bibnamefont {Lenske}}, \bibinfo
  {author} {\bibfnamefont {S.}~\bibnamefont {Leupold}}, \ and\ \bibinfo
  {author} {\bibfnamefont {U.}~\bibnamefont {Mosel}},\ }\href {\doibase
  10.1016/S0375-9474(98)00803-3} {\bibfield  {journal} {\bibinfo  {journal}
  {Nucl. Phys.}\ }\textbf {\bibinfo {volume} {A632}},\ \bibinfo {pages} {109}
  (\bibinfo {year} {1998})},\ \Eprint {http://arxiv.org/abs/nucl-th/9708004}
  {arXiv:nucl-th/9708004 [nucl-th]} \BibitemShut {NoStop}%
\bibitem [{\citenamefont {Post}\ \emph {et~al.}(2001)\citenamefont {Post},
  \citenamefont {Leupold},\ and\ \citenamefont {Mosel}}]{Post:2000qi}%
  \BibitemOpen
  \bibfield  {author} {\bibinfo {author} {\bibfnamefont {M.}~\bibnamefont
  {Post}}, \bibinfo {author} {\bibfnamefont {S.}~\bibnamefont {Leupold}}, \
  and\ \bibinfo {author} {\bibfnamefont {U.}~\bibnamefont {Mosel}},\ }\href
  {\doibase 10.1016/S0375-9474(00)00613-8} {\bibfield  {journal} {\bibinfo
  {journal} {Nucl. Phys.}\ }\textbf {\bibinfo {volume} {A689}},\ \bibinfo
  {pages} {753} (\bibinfo {year} {2001})},\ \Eprint
  {http://arxiv.org/abs/nucl-th/0008027} {arXiv:nucl-th/0008027 [nucl-th]}
  \BibitemShut {NoStop}%
\bibitem [{\citenamefont {Post}\ \emph {et~al.}(2004)\citenamefont {Post},
  \citenamefont {Leupold},\ and\ \citenamefont {Mosel}}]{Post:2003hu}%
  \BibitemOpen
  \bibfield  {author} {\bibinfo {author} {\bibfnamefont {M.}~\bibnamefont
  {Post}}, \bibinfo {author} {\bibfnamefont {S.}~\bibnamefont {Leupold}}, \
  and\ \bibinfo {author} {\bibfnamefont {U.}~\bibnamefont {Mosel}},\ }\href
  {\doibase 10.1016/j.nuclphysa.2004.05.016} {\bibfield  {journal} {\bibinfo
  {journal} {Nucl. Phys. A}\ }\textbf {\bibinfo {volume} {741}},\ \bibinfo
  {pages} {81} (\bibinfo {year} {2004})},\ \Eprint
  {http://arxiv.org/abs/nucl-th/0309085} {arXiv:nucl-th/0309085} \BibitemShut
  {NoStop}%
\bibitem [{\citenamefont {Muehlich}\ \emph {et~al.}(2006)\citenamefont
  {Muehlich}, \citenamefont {Shklyar}, \citenamefont {Leupold}, \citenamefont
  {Mosel},\ and\ \citenamefont {Post}}]{Muehlich:2006nn}%
  \BibitemOpen
  \bibfield  {author} {\bibinfo {author} {\bibfnamefont {P.}~\bibnamefont
  {Muehlich}}, \bibinfo {author} {\bibfnamefont {V.}~\bibnamefont {Shklyar}},
  \bibinfo {author} {\bibfnamefont {S.}~\bibnamefont {Leupold}}, \bibinfo
  {author} {\bibfnamefont {U.}~\bibnamefont {Mosel}}, \ and\ \bibinfo {author}
  {\bibfnamefont {M.}~\bibnamefont {Post}},\ }\href {\doibase
  10.1016/j.nuclphysa.2006.10.037} {\bibfield  {journal} {\bibinfo  {journal}
  {Nucl. Phys. A}\ }\textbf {\bibinfo {volume} {780}},\ \bibinfo {pages} {187}
  (\bibinfo {year} {2006})},\ \Eprint {http://arxiv.org/abs/nucl-th/0607061}
  {arXiv:nucl-th/0607061} \BibitemShut {NoStop}%
\bibitem [{\citenamefont {Jung}\ \emph {et~al.}(2017)\citenamefont {Jung},
  \citenamefont {Rennecke}, \citenamefont {Tripolt}, \citenamefont {von
  Smekal},\ and\ \citenamefont {Wambach}}]{Jung:2016yxl}%
  \BibitemOpen
  \bibfield  {author} {\bibinfo {author} {\bibfnamefont {C.}~\bibnamefont
  {Jung}}, \bibinfo {author} {\bibfnamefont {F.}~\bibnamefont {Rennecke}},
  \bibinfo {author} {\bibfnamefont {R.-A.}\ \bibnamefont {Tripolt}}, \bibinfo
  {author} {\bibfnamefont {L.}~\bibnamefont {von Smekal}}, \ and\ \bibinfo
  {author} {\bibfnamefont {J.}~\bibnamefont {Wambach}},\ }\href {\doibase
  10.1103/PhysRevD.95.036020} {\bibfield  {journal} {\bibinfo  {journal} {Phys.
  Rev. D}\ }\textbf {\bibinfo {volume} {95}},\ \bibinfo {pages} {036020}
  (\bibinfo {year} {2017})},\ \Eprint {http://arxiv.org/abs/1610.08754}
  {arXiv:1610.08754 [hep-ph]} \BibitemShut {NoStop}%
\bibitem [{\citenamefont {Jung}\ and\ \citenamefont {von
  Smekal}(2019)}]{Jung:2019nnr}%
  \BibitemOpen
  \bibfield  {author} {\bibinfo {author} {\bibfnamefont {C.}~\bibnamefont
  {Jung}}\ and\ \bibinfo {author} {\bibfnamefont {L.}~\bibnamefont {von
  Smekal}},\ }\href {\doibase 10.1103/PhysRevD.100.116009} {\bibfield
  {journal} {\bibinfo  {journal} {Phys. Rev. D}\ }\textbf {\bibinfo {volume}
  {100}},\ \bibinfo {pages} {116009} (\bibinfo {year} {2019})},\ \Eprint
  {http://arxiv.org/abs/1909.13712} {arXiv:1909.13712 [hep-ph]} \BibitemShut
  {NoStop}%
\bibitem [{\citenamefont {Rapp}\ and\ \citenamefont
  {Wambach}(2000)}]{Rapp:1999ej}%
  \BibitemOpen
  \bibfield  {author} {\bibinfo {author} {\bibfnamefont {R.}~\bibnamefont
  {Rapp}}\ and\ \bibinfo {author} {\bibfnamefont {J.}~\bibnamefont {Wambach}},\
  }\href {\doibase 10.1007/0-306-47101-9\_1} {\bibfield  {journal} {\bibinfo
  {journal} {Adv. Nucl. Phys.}\ }\textbf {\bibinfo {volume} {25}},\ \bibinfo
  {pages} {1} (\bibinfo {year} {2000})},\ \Eprint
  {http://arxiv.org/abs/hep-ph/9909229} {arXiv:hep-ph/9909229} \BibitemShut
  {NoStop}%
\bibitem [{\citenamefont {Hayano}\ and\ \citenamefont
  {Hatsuda}(2010)}]{Hayano:2008vn}%
  \BibitemOpen
  \bibfield  {author} {\bibinfo {author} {\bibfnamefont {R.~S.}\ \bibnamefont
  {Hayano}}\ and\ \bibinfo {author} {\bibfnamefont {T.}~\bibnamefont
  {Hatsuda}},\ }\href {\doibase 10.1103/RevModPhys.82.2949} {\bibfield
  {journal} {\bibinfo  {journal} {Rev. Mod. Phys.}\ }\textbf {\bibinfo {volume}
  {82}},\ \bibinfo {pages} {2949} (\bibinfo {year} {2010})},\ \Eprint
  {http://arxiv.org/abs/0812.1702} {arXiv:0812.1702 [nucl-ex]} \BibitemShut
  {NoStop}%
\bibitem [{\citenamefont {Leupold}\ \emph {et~al.}(2010)\citenamefont
  {Leupold}, \citenamefont {Metag},\ and\ \citenamefont
  {Mosel}}]{Leupold:2009kz}%
  \BibitemOpen
  \bibfield  {author} {\bibinfo {author} {\bibfnamefont {S.}~\bibnamefont
  {Leupold}}, \bibinfo {author} {\bibfnamefont {V.}~\bibnamefont {Metag}}, \
  and\ \bibinfo {author} {\bibfnamefont {U.}~\bibnamefont {Mosel}},\ }\href
  {\doibase 10.1142/S0218301310014728} {\bibfield  {journal} {\bibinfo
  {journal} {Int. J. Mod. Phys. E}\ }\textbf {\bibinfo {volume} {19}},\
  \bibinfo {pages} {147} (\bibinfo {year} {2010})},\ \Eprint
  {http://arxiv.org/abs/0907.2388} {arXiv:0907.2388 [nucl-th]} \BibitemShut
  {NoStop}%
\bibitem [{\citenamefont {Salabura}\ and\ \citenamefont
  {Stroth}(2020)}]{Salabura:2020tou}%
  \BibitemOpen
  \bibfield  {author} {\bibinfo {author} {\bibfnamefont {P.}~\bibnamefont
  {Salabura}}\ and\ \bibinfo {author} {\bibfnamefont {J.}~\bibnamefont
  {Stroth}},\ }\href@noop {} {\  (\bibinfo {year} {2020})},\ \Eprint
  {http://arxiv.org/abs/2005.14589} {arXiv:2005.14589 [nucl-ex]} \BibitemShut
  {NoStop}%
\bibitem [{\citenamefont {Agakichiev}\ \emph {et~al.}(2007)\citenamefont
  {Agakichiev} \emph {et~al.}}]{Agakichiev:2006tg}%
  \BibitemOpen
  \bibfield  {author} {\bibinfo {author} {\bibfnamefont {G.}~\bibnamefont
  {Agakichiev}} \emph {et~al.} (\bibinfo {collaboration} {HADES}),\ }\href
  {\doibase 10.1103/PhysRevLett.98.052302} {\bibfield  {journal} {\bibinfo
  {journal} {Phys. Rev. Lett.}\ }\textbf {\bibinfo {volume} {98}},\ \bibinfo
  {pages} {052302} (\bibinfo {year} {2007})},\ \Eprint
  {http://arxiv.org/abs/nucl-ex/0608031} {arXiv:nucl-ex/0608031 [nucl-ex]}
  \BibitemShut {NoStop}%
\bibitem [{\citenamefont {Agakishiev}\ \emph {et~al.}(2008)\citenamefont
  {Agakishiev} \emph {et~al.}}]{Agakishiev:2007ts}%
  \BibitemOpen
  \bibfield  {author} {\bibinfo {author} {\bibfnamefont {G.}~\bibnamefont
  {Agakishiev}} \emph {et~al.} (\bibinfo {collaboration} {HADES}),\ }\href
  {\doibase 10.1016/j.physletb.2008.03.062} {\bibfield  {journal} {\bibinfo
  {journal} {Phys. Lett.}\ }\textbf {\bibinfo {volume} {B663}},\ \bibinfo
  {pages} {43} (\bibinfo {year} {2008})},\ \Eprint
  {http://arxiv.org/abs/0711.4281} {arXiv:0711.4281 [nucl-ex]} \BibitemShut
  {NoStop}%
\bibitem [{\citenamefont {Agakishiev}\ \emph {et~al.}(2011)\citenamefont
  {Agakishiev} \emph {et~al.}}]{Agakishiev:2011vf}%
  \BibitemOpen
  \bibfield  {author} {\bibinfo {author} {\bibfnamefont {G.}~\bibnamefont
  {Agakishiev}} \emph {et~al.} (\bibinfo {collaboration} {HADES}),\ }\href
  {\doibase 10.1103/PhysRevC.84.014902} {\bibfield  {journal} {\bibinfo
  {journal} {Phys. Rev.}\ }\textbf {\bibinfo {volume} {C84}},\ \bibinfo {pages}
  {014902} (\bibinfo {year} {2011})},\ \Eprint {http://arxiv.org/abs/1103.0876}
  {arXiv:1103.0876 [nucl-ex]} \BibitemShut {NoStop}%
\bibitem [{\citenamefont {Adamczewski-Musch}\ \emph
  {et~al.}(2019{\natexlab{a}})\citenamefont {Adamczewski-Musch} \emph
  {et~al.}}]{Adamczewski-Musch:2019byl}%
  \BibitemOpen
  \bibfield  {author} {\bibinfo {author} {\bibfnamefont {J.}~\bibnamefont
  {Adamczewski-Musch}} \emph {et~al.} (\bibinfo {collaboration} {HADES}),\
  }\href {\doibase 10.1038/s41567-019-0583-8} {\bibfield  {journal} {\bibinfo
  {journal} {Nature Phys.}\ }\textbf {\bibinfo {volume} {15}},\ \bibinfo
  {pages} {1040} (\bibinfo {year} {2019}{\natexlab{a}})}\BibitemShut {NoStop}%
\bibitem [{\citenamefont {Huovinen}\ \emph {et~al.}(2002)\citenamefont
  {Huovinen}, \citenamefont {Belkacem}, \citenamefont {Ellis},\ and\
  \citenamefont {Kapusta}}]{Huovinen:2002im}%
  \BibitemOpen
  \bibfield  {author} {\bibinfo {author} {\bibfnamefont {P.}~\bibnamefont
  {Huovinen}}, \bibinfo {author} {\bibfnamefont {M.}~\bibnamefont {Belkacem}},
  \bibinfo {author} {\bibfnamefont {P.}~\bibnamefont {Ellis}}, \ and\ \bibinfo
  {author} {\bibfnamefont {J.~I.}\ \bibnamefont {Kapusta}},\ }\href {\doibase
  10.1103/PhysRevC.66.014903} {\bibfield  {journal} {\bibinfo  {journal} {Phys.
  Rev. C}\ }\textbf {\bibinfo {volume} {66}},\ \bibinfo {pages} {014903}
  (\bibinfo {year} {2002})},\ \Eprint {http://arxiv.org/abs/nucl-th/0203023}
  {arXiv:nucl-th/0203023} \BibitemShut {NoStop}%
\bibitem [{\citenamefont {Endres}\ \emph {et~al.}(2015)\citenamefont {Endres},
  \citenamefont {van Hees}, \citenamefont {Weil},\ and\ \citenamefont
  {Bleicher}}]{Endres:2015fna}%
  \BibitemOpen
  \bibfield  {author} {\bibinfo {author} {\bibfnamefont {S.}~\bibnamefont
  {Endres}}, \bibinfo {author} {\bibfnamefont {H.}~\bibnamefont {van Hees}},
  \bibinfo {author} {\bibfnamefont {J.}~\bibnamefont {Weil}}, \ and\ \bibinfo
  {author} {\bibfnamefont {M.}~\bibnamefont {Bleicher}},\ }\href {\doibase
  10.1103/PhysRevC.92.014911} {\bibfield  {journal} {\bibinfo  {journal} {Phys.
  Rev. C}\ }\textbf {\bibinfo {volume} {92}},\ \bibinfo {pages} {014911}
  (\bibinfo {year} {2015})},\ \Eprint {http://arxiv.org/abs/1505.06131}
  {arXiv:1505.06131 [nucl-th]} \BibitemShut {NoStop}%
\bibitem [{\citenamefont {Galatyuk}\ \emph {et~al.}(2016)\citenamefont
  {Galatyuk}, \citenamefont {Hohler}, \citenamefont {Rapp}, \citenamefont
  {Seck},\ and\ \citenamefont {Stroth}}]{Galatyuk:2015pkq}%
  \BibitemOpen
  \bibfield  {author} {\bibinfo {author} {\bibfnamefont {T.}~\bibnamefont
  {Galatyuk}}, \bibinfo {author} {\bibfnamefont {P.~M.}\ \bibnamefont
  {Hohler}}, \bibinfo {author} {\bibfnamefont {R.}~\bibnamefont {Rapp}},
  \bibinfo {author} {\bibfnamefont {F.}~\bibnamefont {Seck}}, \ and\ \bibinfo
  {author} {\bibfnamefont {J.}~\bibnamefont {Stroth}},\ }\href {\doibase
  10.1140/epja/i2016-16131-1} {\bibfield  {journal} {\bibinfo  {journal} {Eur.
  Phys. J. A}\ }\textbf {\bibinfo {volume} {52}},\ \bibinfo {pages} {131}
  (\bibinfo {year} {2016})},\ \Eprint {http://arxiv.org/abs/1512.08688}
  {arXiv:1512.08688 [nucl-th]} \BibitemShut {NoStop}%
\bibitem [{\citenamefont {Staudenmaier}\ \emph {et~al.}(2018)\citenamefont
  {Staudenmaier}, \citenamefont {Weil}, \citenamefont {Steinberg},
  \citenamefont {Endres},\ and\ \citenamefont
  {Petersen}}]{Staudenmaier:2017vtq}%
  \BibitemOpen
  \bibfield  {author} {\bibinfo {author} {\bibfnamefont {J.}~\bibnamefont
  {Staudenmaier}}, \bibinfo {author} {\bibfnamefont {J.}~\bibnamefont {Weil}},
  \bibinfo {author} {\bibfnamefont {V.}~\bibnamefont {Steinberg}}, \bibinfo
  {author} {\bibfnamefont {S.}~\bibnamefont {Endres}}, \ and\ \bibinfo {author}
  {\bibfnamefont {H.}~\bibnamefont {Petersen}},\ }\href {\doibase
  10.1103/PhysRevC.98.054908} {\bibfield  {journal} {\bibinfo  {journal} {Phys.
  Rev. C}\ }\textbf {\bibinfo {volume} {98}},\ \bibinfo {pages} {054908}
  (\bibinfo {year} {2018})},\ \Eprint {http://arxiv.org/abs/1711.10297}
  {arXiv:1711.10297 [nucl-th]} \BibitemShut {NoStop}%
\bibitem [{\citenamefont {Seck}\ \emph {et~al.}(2018)\citenamefont {Seck},
  \citenamefont {Galatyuk}, \citenamefont {Rapp},\ and\ \citenamefont
  {Stroth}}]{Seck:2017zjr}%
  \BibitemOpen
  \bibfield  {author} {\bibinfo {author} {\bibfnamefont {F.}~\bibnamefont
  {Seck}}, \bibinfo {author} {\bibfnamefont {T.}~\bibnamefont {Galatyuk}},
  \bibinfo {author} {\bibfnamefont {R.}~\bibnamefont {Rapp}}, \ and\ \bibinfo
  {author} {\bibfnamefont {J.}~\bibnamefont {Stroth}},\ }\href {\doibase
  10.1088/1742-6596/1024/1/012011} {\bibfield  {journal} {\bibinfo  {journal}
  {J. Phys. Conf. Ser.}\ }\textbf {\bibinfo {volume} {1024}},\ \bibinfo {pages}
  {012011} (\bibinfo {year} {2018})},\ \Eprint
  {http://arxiv.org/abs/1710.06256} {arXiv:1710.06256 [nucl-th]} \BibitemShut
  {NoStop}%
\bibitem [{\citenamefont {Lang}\ \emph {et~al.}(1991)\citenamefont {Lang},
  \citenamefont {Blaettel}, \citenamefont {Cassing}, \citenamefont {Koch},
  \citenamefont {Mosel},\ and\ \citenamefont {Weber}}]{Lang:1991qa}%
  \BibitemOpen
  \bibfield  {author} {\bibinfo {author} {\bibfnamefont {A.}~\bibnamefont
  {Lang}}, \bibinfo {author} {\bibfnamefont {B.}~\bibnamefont {Blaettel}},
  \bibinfo {author} {\bibfnamefont {W.}~\bibnamefont {Cassing}}, \bibinfo
  {author} {\bibfnamefont {V.}~\bibnamefont {Koch}}, \bibinfo {author}
  {\bibfnamefont {U.}~\bibnamefont {Mosel}}, \ and\ \bibinfo {author}
  {\bibfnamefont {K.}~\bibnamefont {Weber}},\ }\href {\doibase
  10.1007/BF01294677} {\bibfield  {journal} {\bibinfo  {journal} {Z. Phys. A}\
  }\textbf {\bibinfo {volume} {340}},\ \bibinfo {pages} {287} (\bibinfo {year}
  {1991})}\BibitemShut {NoStop}%
\bibitem [{\citenamefont {Rapp}(2019)}]{Rapp:2019rr}%
  \BibitemOpen
  \bibfield  {author} {\bibinfo {author} {\bibfnamefont {R.}~\bibnamefont
  {Rapp}},\ }\href {\doibase 10.1038/s41567-019-0614-5} {\bibfield  {journal}
  {\bibinfo  {journal} {Nat. Phys.}\ }\textbf {\bibinfo {volume} {15}},\
  \bibinfo {pages} {990} (\bibinfo {year} {2019})}\BibitemShut {NoStop}%
\bibitem [{\citenamefont {Buss}\ \emph {et~al.}(2012)\citenamefont {Buss},
  \citenamefont {Gaitanos}, \citenamefont {Gallmeister}, \citenamefont {van
  Hees}, \citenamefont {Kaskulov}, \citenamefont {Lalakulich}, \citenamefont
  {Larionov}, \citenamefont {Leitner}, \citenamefont {Weil},\ and\
  \citenamefont {Mosel}}]{Buss:2011mx}%
  \BibitemOpen
  \bibfield  {author} {\bibinfo {author} {\bibfnamefont {O.}~\bibnamefont
  {Buss}}, \bibinfo {author} {\bibfnamefont {T.}~\bibnamefont {Gaitanos}},
  \bibinfo {author} {\bibfnamefont {K.}~\bibnamefont {Gallmeister}}, \bibinfo
  {author} {\bibfnamefont {H.}~\bibnamefont {van Hees}}, \bibinfo {author}
  {\bibfnamefont {M.}~\bibnamefont {Kaskulov}}, \bibinfo {author}
  {\bibfnamefont {O.}~\bibnamefont {Lalakulich}}, \bibinfo {author}
  {\bibfnamefont {A.~B.}\ \bibnamefont {Larionov}}, \bibinfo {author}
  {\bibfnamefont {T.}~\bibnamefont {Leitner}}, \bibinfo {author} {\bibfnamefont
  {J.}~\bibnamefont {Weil}}, \ and\ \bibinfo {author} {\bibfnamefont
  {U.}~\bibnamefont {Mosel}},\ }\href {\doibase 10.1016/j.physrep.2011.12.001}
  {\bibfield  {journal} {\bibinfo  {journal} {Phys. Rept.}\ }\textbf {\bibinfo
  {volume} {512}},\ \bibinfo {pages} {1} (\bibinfo {year} {2012})},\ \Eprint
  {http://arxiv.org/abs/1106.1344} {arXiv:1106.1344 [hep-ph]} \BibitemShut
  {NoStop}%
\bibitem [{\citenamefont {Blaettel}\ \emph {et~al.}(1993)\citenamefont
  {Blaettel}, \citenamefont {Koch},\ and\ \citenamefont
  {Mosel}}]{Blaettel:1993uz}%
  \BibitemOpen
  \bibfield  {author} {\bibinfo {author} {\bibfnamefont {B.}~\bibnamefont
  {Blaettel}}, \bibinfo {author} {\bibfnamefont {V.}~\bibnamefont {Koch}}, \
  and\ \bibinfo {author} {\bibfnamefont {U.}~\bibnamefont {Mosel}},\ }\href
  {\doibase 10.1088/0034-4885/56/1/001} {\bibfield  {journal} {\bibinfo
  {journal} {Rept. Prog. Phys.}\ }\textbf {\bibinfo {volume} {56}},\ \bibinfo
  {pages} {1} (\bibinfo {year} {1993})}\BibitemShut {NoStop}%
\bibitem [{\citenamefont {Ehehalt}\ \emph {et~al.}(1993)\citenamefont
  {Ehehalt}, \citenamefont {Cassing}, \citenamefont {Engel}, \citenamefont
  {Mosel},\ and\ \citenamefont {Wolf}}]{Ehehalt:1993px}%
  \BibitemOpen
  \bibfield  {author} {\bibinfo {author} {\bibfnamefont {W.}~\bibnamefont
  {Ehehalt}}, \bibinfo {author} {\bibfnamefont {W.}~\bibnamefont {Cassing}},
  \bibinfo {author} {\bibfnamefont {A.}~\bibnamefont {Engel}}, \bibinfo
  {author} {\bibfnamefont {U.}~\bibnamefont {Mosel}}, \ and\ \bibinfo {author}
  {\bibfnamefont {G.}~\bibnamefont {Wolf}},\ }\href {\doibase
  10.1016/0370-2693(93)91701-N} {\bibfield  {journal} {\bibinfo  {journal}
  {Phys. Lett. B}\ }\textbf {\bibinfo {volume} {298}},\ \bibinfo {pages} {31}
  (\bibinfo {year} {1993})}\BibitemShut {NoStop}%
\bibitem [{\citenamefont {Lang}\ \emph {et~al.}(1992)\citenamefont {Lang},
  \citenamefont {Cassing}, \citenamefont {Mosel},\ and\ \citenamefont
  {Weber}}]{Lang:1992jz}%
  \BibitemOpen
  \bibfield  {author} {\bibinfo {author} {\bibfnamefont {A.}~\bibnamefont
  {Lang}}, \bibinfo {author} {\bibfnamefont {W.}~\bibnamefont {Cassing}},
  \bibinfo {author} {\bibfnamefont {U.}~\bibnamefont {Mosel}}, \ and\ \bibinfo
  {author} {\bibfnamefont {K.}~\bibnamefont {Weber}},\ }\href {\doibase
  10.1016/0375-9474(92)90189-Q} {\bibfield  {journal} {\bibinfo  {journal}
  {Nucl. Phys. A}\ }\textbf {\bibinfo {volume} {541}},\ \bibinfo {pages} {507}
  (\bibinfo {year} {1992})}\BibitemShut {NoStop}%
\bibitem [{\citenamefont {Effenberger}\ and\ \citenamefont
  {Mosel}(1999)}]{Effenberger:1999uv}%
  \BibitemOpen
  \bibfield  {author} {\bibinfo {author} {\bibfnamefont {M.}~\bibnamefont
  {Effenberger}}\ and\ \bibinfo {author} {\bibfnamefont {U.}~\bibnamefont
  {Mosel}},\ }\href {\doibase 10.1103/PhysRevC.60.051901} {\bibfield  {journal}
  {\bibinfo  {journal} {Phys. Rev. C}\ }\textbf {\bibinfo {volume} {60}},\
  \bibinfo {pages} {051901} (\bibinfo {year} {1999})},\ \Eprint
  {http://arxiv.org/abs/nucl-th/9906085} {arXiv:nucl-th/9906085} \BibitemShut
  {NoStop}%
\bibitem [{\citenamefont {Cassing}\ and\ \citenamefont
  {Juchem}(2000{\natexlab{a}})}]{Cassing:1999wx}%
  \BibitemOpen
  \bibfield  {author} {\bibinfo {author} {\bibfnamefont {W.}~\bibnamefont
  {Cassing}}\ and\ \bibinfo {author} {\bibfnamefont {S.}~\bibnamefont
  {Juchem}},\ }\href {\doibase 10.1016/S0375-9474(99)00393-0} {\bibfield
  {journal} {\bibinfo  {journal} {Nucl. Phys. A}\ }\textbf {\bibinfo {volume}
  {665}},\ \bibinfo {pages} {377} (\bibinfo {year} {2000}{\natexlab{a}})},\
  \Eprint {http://arxiv.org/abs/nucl-th/9903070} {arXiv:nucl-th/9903070}
  \BibitemShut {NoStop}%
\bibitem [{\citenamefont {Cassing}\ and\ \citenamefont
  {Juchem}(2000{\natexlab{b}})}]{Cassing:1999mh}%
  \BibitemOpen
  \bibfield  {author} {\bibinfo {author} {\bibfnamefont {W.}~\bibnamefont
  {Cassing}}\ and\ \bibinfo {author} {\bibfnamefont {S.}~\bibnamefont
  {Juchem}},\ }\href {\doibase 10.1016/S0375-9474(00)00050-6} {\bibfield
  {journal} {\bibinfo  {journal} {Nucl. Phys. A}\ }\textbf {\bibinfo {volume}
  {672}},\ \bibinfo {pages} {417} (\bibinfo {year} {2000}{\natexlab{b}})},\
  \Eprint {http://arxiv.org/abs/nucl-th/9910052} {arXiv:nucl-th/9910052}
  \BibitemShut {NoStop}%
\bibitem [{\citenamefont {Leupold}(2000)}]{Leupold:1999ga}%
  \BibitemOpen
  \bibfield  {author} {\bibinfo {author} {\bibfnamefont {S.}~\bibnamefont
  {Leupold}},\ }\href {\doibase 10.1016/S0375-9474(00)00057-9} {\bibfield
  {journal} {\bibinfo  {journal} {Nucl. Phys. A}\ }\textbf {\bibinfo {volume}
  {672}},\ \bibinfo {pages} {475} (\bibinfo {year} {2000})},\ \Eprint
  {http://arxiv.org/abs/nucl-th/9909080} {arXiv:nucl-th/9909080} \BibitemShut
  {NoStop}%
\bibitem [{\citenamefont {Effenberger}\ \emph {et~al.}(1999)\citenamefont
  {Effenberger}, \citenamefont {Bratkovskaya},\ and\ \citenamefont
  {Mosel}}]{Effenberger:1999ay}%
  \BibitemOpen
  \bibfield  {author} {\bibinfo {author} {\bibfnamefont {M.}~\bibnamefont
  {Effenberger}}, \bibinfo {author} {\bibfnamefont {E.}~\bibnamefont
  {Bratkovskaya}}, \ and\ \bibinfo {author} {\bibfnamefont {U.}~\bibnamefont
  {Mosel}},\ }\href {\doibase 10.1103/PhysRevC.60.044614} {\bibfield  {journal}
  {\bibinfo  {journal} {Phys. Rev. C}\ }\textbf {\bibinfo {volume} {60}},\
  \bibinfo {pages} {044614} (\bibinfo {year} {1999})},\ \Eprint
  {http://arxiv.org/abs/nucl-th/9903026} {arXiv:nucl-th/9903026} \BibitemShut
  {NoStop}%
\bibitem [{\citenamefont {Kodama}\ \emph {et~al.}(1984)\citenamefont {Kodama},
  \citenamefont {Duarte}, \citenamefont {Chung}, \citenamefont {Donangelo},\
  and\ \citenamefont {Nazareth}}]{Kodama:1983yk}%
  \BibitemOpen
  \bibfield  {author} {\bibinfo {author} {\bibfnamefont {T.}~\bibnamefont
  {Kodama}}, \bibinfo {author} {\bibfnamefont {S.}~\bibnamefont {Duarte}},
  \bibinfo {author} {\bibfnamefont {K.}~\bibnamefont {Chung}}, \bibinfo
  {author} {\bibfnamefont {R.}~\bibnamefont {Donangelo}}, \ and\ \bibinfo
  {author} {\bibfnamefont {R.}~\bibnamefont {Nazareth}},\ }\href {\doibase
  10.1103/PhysRevC.29.2146} {\bibfield  {journal} {\bibinfo  {journal} {Phys.
  Rev. C}\ }\textbf {\bibinfo {volume} {29}},\ \bibinfo {pages} {2146}
  (\bibinfo {year} {1984})}\BibitemShut {NoStop}%
\bibitem [{\citenamefont {Wolf}(1993)}]{Wolf:1993cj}%
  \BibitemOpen
  \bibfield  {author} {\bibinfo {author} {\bibfnamefont {G.}~\bibnamefont
  {Wolf}},\ }in\ \href@noop {} {\emph {\bibinfo {booktitle} {{2nd TAPS
  Workshop}}}}\ (\bibinfo {year} {1993})\ pp.\ \bibinfo {pages}
  {418--467}\BibitemShut {NoStop}%
\bibitem [{\citenamefont {Effenberger}(1999)}]{EffenbergerPhD}%
  \BibitemOpen
  \bibfield  {author} {\bibinfo {author} {\bibfnamefont {M.}~\bibnamefont
  {Effenberger}},\ }\emph {\bibinfo {title} {{Eigenschaften von Hadronen in
  Kernmaterie in einem vereinheitlichten Transportmodell}}},\ \href@noop {}
  {Ph.D. thesis},\ \bibinfo  {school} {Giessen U.} (\bibinfo {year}
  {1999})\BibitemShut {NoStop}%
\bibitem [{\citenamefont {Sjostrand}\ \emph {et~al.}(2006)\citenamefont
  {Sjostrand}, \citenamefont {Mrenna},\ and\ \citenamefont
  {Skands}}]{Sjostrand:2006za}%
  \BibitemOpen
  \bibfield  {author} {\bibinfo {author} {\bibfnamefont {T.}~\bibnamefont
  {Sjostrand}}, \bibinfo {author} {\bibfnamefont {S.}~\bibnamefont {Mrenna}}, \
  and\ \bibinfo {author} {\bibfnamefont {P.~Z.}\ \bibnamefont {Skands}},\
  }\href {\doibase 10.1088/1126-6708/2006/05/026} {\bibfield  {journal}
  {\bibinfo  {journal} {JHEP}\ }\textbf {\bibinfo {volume} {05}},\ \bibinfo
  {pages} {026} (\bibinfo {year} {2006})},\ \Eprint
  {http://arxiv.org/abs/hep-ph/0603175} {arXiv:hep-ph/0603175} \BibitemShut
  {NoStop}%
\bibitem [{\citenamefont {Nambu}\ and\ \citenamefont
  {Sakurai}(1962)}]{Nambu:1962zz}%
  \BibitemOpen
  \bibfield  {author} {\bibinfo {author} {\bibfnamefont {Y.}~\bibnamefont
  {Nambu}}\ and\ \bibinfo {author} {\bibfnamefont {J.~J.}\ \bibnamefont
  {Sakurai}},\ }\href {\doibase 10.1103/PhysRevLett.8.79} {\bibfield  {journal}
  {\bibinfo  {journal} {Phys. Rev. Lett.}\ }\textbf {\bibinfo {volume} {8}},\
  \bibinfo {pages} {79} (\bibinfo {year} {1962})}\BibitemShut {NoStop}%
\bibitem [{\citenamefont {Dumbrajs}\ \emph {et~al.}(1983)\citenamefont
  {Dumbrajs}, \citenamefont {Koch}, \citenamefont {Pilkuhn}, \citenamefont
  {Oades}, \citenamefont {Behrens}, \citenamefont {De~Swart},\ and\
  \citenamefont {Kroll}}]{Dumbrajs:1983jd}%
  \BibitemOpen
  \bibfield  {author} {\bibinfo {author} {\bibfnamefont {O.}~\bibnamefont
  {Dumbrajs}}, \bibinfo {author} {\bibfnamefont {R.}~\bibnamefont {Koch}},
  \bibinfo {author} {\bibfnamefont {H.}~\bibnamefont {Pilkuhn}}, \bibinfo
  {author} {\bibfnamefont {G.~c.}\ \bibnamefont {Oades}}, \bibinfo {author}
  {\bibfnamefont {H.}~\bibnamefont {Behrens}}, \bibinfo {author} {\bibfnamefont
  {J.~j.}\ \bibnamefont {De~Swart}}, \ and\ \bibinfo {author} {\bibfnamefont
  {P.}~\bibnamefont {Kroll}},\ }\href {\doibase 10.1016/0550-3213(83)90288-2}
  {\bibfield  {journal} {\bibinfo  {journal} {Nucl. Phys.}\ }\textbf {\bibinfo
  {volume} {B216}},\ \bibinfo {pages} {277} (\bibinfo {year}
  {1983})}\BibitemShut {NoStop}%
\bibitem [{\citenamefont {Weil}\ \emph {et~al.}(2012)\citenamefont {Weil},
  \citenamefont {van Hees},\ and\ \citenamefont {Mosel}}]{Weil:2012ji}%
  \BibitemOpen
  \bibfield  {author} {\bibinfo {author} {\bibfnamefont {J.}~\bibnamefont
  {Weil}}, \bibinfo {author} {\bibfnamefont {H.}~\bibnamefont {van Hees}}, \
  and\ \bibinfo {author} {\bibfnamefont {U.}~\bibnamefont {Mosel}},\ }\href
  {\doibase 10.1140/epja/i2012-12111-9, 10.1140/epja/i2012-12150-2} {\bibfield
  {journal} {\bibinfo  {journal} {Eur. Phys. J.}\ }\textbf {\bibinfo {volume}
  {A48}},\ \bibinfo {pages} {111} (\bibinfo {year} {2012})},\ \bibinfo {note}
  {[Erratum: Eur. Phys. J.A48,150(2012)]},\ \Eprint
  {http://arxiv.org/abs/1203.3557} {arXiv:1203.3557 [nucl-th]} \BibitemShut
  {NoStop}%
\bibitem [{\citenamefont {Tanabashi}\ \emph {et~al.}(2018)\citenamefont
  {Tanabashi} \emph {et~al.}}]{Tanabashi:2018oca}%
  \BibitemOpen
  \bibfield  {author} {\bibinfo {author} {\bibfnamefont {M.}~\bibnamefont
  {Tanabashi}} \emph {et~al.} (\bibinfo {collaboration} {Particle Data
  Group}),\ }\href {\doibase 10.1103/PhysRevD.98.030001} {\bibfield  {journal}
  {\bibinfo  {journal} {Phys. Rev. D}\ }\textbf {\bibinfo {volume} {98}},\
  \bibinfo {pages} {030001} (\bibinfo {year} {2018})}\BibitemShut {NoStop}%
\bibitem [{\citenamefont {Krivoruchenko}\ and\ \citenamefont
  {Faessler}(2002)}]{Krivoruchenko:2001hs}%
  \BibitemOpen
  \bibfield  {author} {\bibinfo {author} {\bibfnamefont {M.}~\bibnamefont
  {Krivoruchenko}}\ and\ \bibinfo {author} {\bibfnamefont {A.}~\bibnamefont
  {Faessler}},\ }\href {\doibase 10.1103/PhysRevD.65.017502} {\bibfield
  {journal} {\bibinfo  {journal} {Phys. Rev. D}\ }\textbf {\bibinfo {volume}
  {65}},\ \bibinfo {pages} {017502} (\bibinfo {year} {2002})},\ \Eprint
  {http://arxiv.org/abs/nucl-th/0104045} {arXiv:nucl-th/0104045} \BibitemShut
  {NoStop}%
\bibitem [{\citenamefont {Landsberg}(1985)}]{Landsberg:1986fd}%
  \BibitemOpen
  \bibfield  {author} {\bibinfo {author} {\bibfnamefont {L.}~\bibnamefont
  {Landsberg}},\ }\href {\doibase 10.1016/0370-1573(85)90129-2} {\bibfield
  {journal} {\bibinfo  {journal} {Phys. Rept.}\ }\textbf {\bibinfo {volume}
  {128}},\ \bibinfo {pages} {301} (\bibinfo {year} {1985})}\BibitemShut
  {NoStop}%
\bibitem [{\citenamefont {Dalitz}(1951)}]{Dalitz:1951aj}%
  \BibitemOpen
  \bibfield  {author} {\bibinfo {author} {\bibfnamefont {R.}~\bibnamefont
  {Dalitz}},\ }\href {\doibase 10.1088/0370-1298/64/7/115} {\bibfield
  {journal} {\bibinfo  {journal} {Proc. Phys. Soc. A}\ }\textbf {\bibinfo
  {volume} {64}},\ \bibinfo {pages} {667} (\bibinfo {year} {1951})}\BibitemShut
  {NoStop}%
\bibitem [{\citenamefont {Faessler}\ \emph {et~al.}(2000)\citenamefont
  {Faessler}, \citenamefont {Fuchs},\ and\ \citenamefont
  {Krivoruchenko}}]{Faessler:1999de}%
  \BibitemOpen
  \bibfield  {author} {\bibinfo {author} {\bibfnamefont {A.}~\bibnamefont
  {Faessler}}, \bibinfo {author} {\bibfnamefont {C.}~\bibnamefont {Fuchs}}, \
  and\ \bibinfo {author} {\bibfnamefont {M.}~\bibnamefont {Krivoruchenko}},\
  }\href {\doibase 10.1103/PhysRevC.61.035206} {\bibfield  {journal} {\bibinfo
  {journal} {Phys. Rev. C}\ }\textbf {\bibinfo {volume} {61}},\ \bibinfo
  {pages} {035206} (\bibinfo {year} {2000})},\ \Eprint
  {http://arxiv.org/abs/nucl-th/9904024} {arXiv:nucl-th/9904024} \BibitemShut
  {NoStop}%
\bibitem [{\citenamefont {Arnaldi}\ \emph
  {et~al.}(2009{\natexlab{b}})\citenamefont {Arnaldi} \emph
  {et~al.}}]{Arnaldi:2009aa}%
  \BibitemOpen
  \bibfield  {author} {\bibinfo {author} {\bibfnamefont {R.}~\bibnamefont
  {Arnaldi}} \emph {et~al.} (\bibinfo {collaboration} {NA60}),\ }\href
  {\doibase 10.1016/j.physletb.2009.05.029} {\bibfield  {journal} {\bibinfo
  {journal} {Phys. Lett. B}\ }\textbf {\bibinfo {volume} {677}},\ \bibinfo
  {pages} {260} (\bibinfo {year} {2009}{\natexlab{b}})},\ \Eprint
  {http://arxiv.org/abs/0902.2547} {arXiv:0902.2547 [hep-ph]} \BibitemShut
  {NoStop}%
\bibitem [{\citenamefont {Bratkovskaya}\ and\ \citenamefont
  {Cassing}(1997)}]{Bratkovskaya:1996qe}%
  \BibitemOpen
  \bibfield  {author} {\bibinfo {author} {\bibfnamefont {E.}~\bibnamefont
  {Bratkovskaya}}\ and\ \bibinfo {author} {\bibfnamefont {W.}~\bibnamefont
  {Cassing}},\ }\href {\doibase 10.1016/S0375-9474(97)00140-1} {\bibfield
  {journal} {\bibinfo  {journal} {Nucl. Phys. A}\ }\textbf {\bibinfo {volume}
  {619}},\ \bibinfo {pages} {413} (\bibinfo {year} {1997})},\ \Eprint
  {http://arxiv.org/abs/nucl-th/9611042} {arXiv:nucl-th/9611042} \BibitemShut
  {NoStop}%
\bibitem [{\citenamefont {Ramalho}\ \emph {et~al.}(2016)\citenamefont
  {Ramalho}, \citenamefont {Pena}, \citenamefont {Weil}, \citenamefont {van
  Hees},\ and\ \citenamefont {Mosel}}]{Ramalho:2015qna}%
  \BibitemOpen
  \bibfield  {author} {\bibinfo {author} {\bibfnamefont {G.}~\bibnamefont
  {Ramalho}}, \bibinfo {author} {\bibfnamefont {M.~T.}\ \bibnamefont {Pena}},
  \bibinfo {author} {\bibfnamefont {J.}~\bibnamefont {Weil}}, \bibinfo {author}
  {\bibfnamefont {H.}~\bibnamefont {van Hees}}, \ and\ \bibinfo {author}
  {\bibfnamefont {U.}~\bibnamefont {Mosel}},\ }\href {\doibase
  10.1103/PhysRevD.93.033004} {\bibfield  {journal} {\bibinfo  {journal} {Phys.
  Rev. D}\ }\textbf {\bibinfo {volume} {93}},\ \bibinfo {pages} {033004}
  (\bibinfo {year} {2016})},\ \Eprint {http://arxiv.org/abs/1512.03764}
  {arXiv:1512.03764 [hep-ph]} \BibitemShut {NoStop}%
\bibitem [{\citenamefont {Shyam}\ and\ \citenamefont
  {Mosel}(2010)}]{Shyam:2010vr}%
  \BibitemOpen
  \bibfield  {author} {\bibinfo {author} {\bibfnamefont {R.}~\bibnamefont
  {Shyam}}\ and\ \bibinfo {author} {\bibfnamefont {U.}~\bibnamefont {Mosel}},\
  }\href {\doibase 10.1103/PhysRevC.82.062201} {\bibfield  {journal} {\bibinfo
  {journal} {Phys. Rev. C}\ }\textbf {\bibinfo {volume} {82}},\ \bibinfo
  {pages} {062201} (\bibinfo {year} {2010})},\ \Eprint
  {http://arxiv.org/abs/1006.3873} {arXiv:1006.3873 [hep-ph]} \BibitemShut
  {NoStop}%
\bibitem [{\citenamefont {Gale}\ and\ \citenamefont
  {Kapusta}(1987)}]{Gale:1987ki}%
  \BibitemOpen
  \bibfield  {author} {\bibinfo {author} {\bibfnamefont {C.}~\bibnamefont
  {Gale}}\ and\ \bibinfo {author} {\bibfnamefont {J.~I.}\ \bibnamefont
  {Kapusta}},\ }\href {\doibase 10.1103/PhysRevC.35.2107} {\bibfield  {journal}
  {\bibinfo  {journal} {Phys. Rev. C}\ }\textbf {\bibinfo {volume} {35}},\
  \bibinfo {pages} {2107} (\bibinfo {year} {1987})}\BibitemShut {NoStop}%
\bibitem [{\citenamefont {Wolf}\ \emph {et~al.}(1990)\citenamefont {Wolf},
  \citenamefont {Batko}, \citenamefont {Cassing}, \citenamefont {Mosel},
  \citenamefont {Niita},\ and\ \citenamefont {Schaefer}}]{Wolf:1990ur}%
  \BibitemOpen
  \bibfield  {author} {\bibinfo {author} {\bibfnamefont {G.}~\bibnamefont
  {Wolf}}, \bibinfo {author} {\bibfnamefont {G.}~\bibnamefont {Batko}},
  \bibinfo {author} {\bibfnamefont {W.}~\bibnamefont {Cassing}}, \bibinfo
  {author} {\bibfnamefont {U.}~\bibnamefont {Mosel}}, \bibinfo {author}
  {\bibfnamefont {K.}~\bibnamefont {Niita}}, \ and\ \bibinfo {author}
  {\bibfnamefont {M.}~\bibnamefont {Schaefer}},\ }\href {\doibase
  10.1016/0375-9474(90)90222-8} {\bibfield  {journal} {\bibinfo  {journal}
  {Nucl. Phys. A}\ }\textbf {\bibinfo {volume} {517}},\ \bibinfo {pages} {615}
  (\bibinfo {year} {1990})}\BibitemShut {NoStop}%
\bibitem [{\citenamefont {Weil}(2013)}]{Weil:2013mya}%
  \BibitemOpen
  \bibfield  {author} {\bibinfo {author} {\bibfnamefont {J.}~\bibnamefont
  {Weil}},\ }\emph {\bibinfo {title} {{Vector Mesons in Medium in a Transport
  Approach}}},\ \href@noop {} {Ph.D. thesis},\ \bibinfo  {school} {Giessen U.}
  (\bibinfo {year} {2013}),\ \bibinfo {note}
  {\url{https://www.uni-giessen.de/fbz/fb07/fachgebiete/physik/institute/theorie/inst/theses/dissertation}}\BibitemShut
  {NoStop}%
\bibitem [{\citenamefont {Bratkovskaya}\ \emph {et~al.}(1995)\citenamefont
  {Bratkovskaya}, \citenamefont {Teryaev},\ and\ \citenamefont
  {Toneev}}]{Bratkovskaya:1995kh}%
  \BibitemOpen
  \bibfield  {author} {\bibinfo {author} {\bibfnamefont {E.}~\bibnamefont
  {Bratkovskaya}}, \bibinfo {author} {\bibfnamefont {O.}~\bibnamefont
  {Teryaev}}, \ and\ \bibinfo {author} {\bibfnamefont {V.}~\bibnamefont
  {Toneev}},\ }\href {\doibase 10.1016/0370-2693(95)00164-G} {\bibfield
  {journal} {\bibinfo  {journal} {Phys. Lett. B}\ }\textbf {\bibinfo {volume}
  {348}},\ \bibinfo {pages} {283} (\bibinfo {year} {1995})}\BibitemShut
  {NoStop}%
\bibitem [{\citenamefont {Manley}\ and\ \citenamefont
  {Saleski}(1992)}]{Manley:1992yb}%
  \BibitemOpen
  \bibfield  {author} {\bibinfo {author} {\bibfnamefont {D.~M.}\ \bibnamefont
  {Manley}}\ and\ \bibinfo {author} {\bibfnamefont {E.~M.}\ \bibnamefont
  {Saleski}},\ }\href {\doibase 10.1103/PhysRevD.45.4002} {\bibfield  {journal}
  {\bibinfo  {journal} {Phys. Rev.}\ }\textbf {\bibinfo {volume} {D45}},\
  \bibinfo {pages} {4002} (\bibinfo {year} {1992})}\BibitemShut {NoStop}%
\bibitem [{\citenamefont {Larionov}\ \emph {et~al.}(2012)\citenamefont
  {Larionov}, \citenamefont {Gaitanos},\ and\ \citenamefont
  {Mosel}}]{Larionov:2011fs}%
  \BibitemOpen
  \bibfield  {author} {\bibinfo {author} {\bibfnamefont {A.~B.}\ \bibnamefont
  {Larionov}}, \bibinfo {author} {\bibfnamefont {T.}~\bibnamefont {Gaitanos}},
  \ and\ \bibinfo {author} {\bibfnamefont {U.}~\bibnamefont {Mosel}},\ }\href
  {\doibase 10.1103/PhysRevC.85.024614} {\bibfield  {journal} {\bibinfo
  {journal} {Phys. Rev.}\ }\textbf {\bibinfo {volume} {C85}},\ \bibinfo {pages}
  {024614} (\bibinfo {year} {2012})},\ \Eprint {http://arxiv.org/abs/1107.2326}
  {arXiv:1107.2326 [nucl-th]} \BibitemShut {NoStop}%
\bibitem [{\citenamefont {Manley}\ \emph {et~al.}(1984)\citenamefont {Manley},
  \citenamefont {Arndt}, \citenamefont {Goradia},\ and\ \citenamefont
  {Teplitz}}]{Manley:1984jz}%
  \BibitemOpen
  \bibfield  {author} {\bibinfo {author} {\bibfnamefont {D.}~\bibnamefont
  {Manley}}, \bibinfo {author} {\bibfnamefont {R.~A.}\ \bibnamefont {Arndt}},
  \bibinfo {author} {\bibfnamefont {Y.}~\bibnamefont {Goradia}}, \ and\
  \bibinfo {author} {\bibfnamefont {V.~L.}\ \bibnamefont {Teplitz}},\ }\href
  {\doibase 10.1103/PhysRevD.30.904} {\bibfield  {journal} {\bibinfo  {journal}
  {Phys. Rev. D}\ }\textbf {\bibinfo {volume} {30}},\ \bibinfo {pages} {904}
  (\bibinfo {year} {1984})}\BibitemShut {NoStop}%
\bibitem [{\citenamefont {Donnachie}\ and\ \citenamefont
  {Landshoff}(1992)}]{Donnachie:1992ny}%
  \BibitemOpen
  \bibfield  {author} {\bibinfo {author} {\bibfnamefont {A.}~\bibnamefont
  {Donnachie}}\ and\ \bibinfo {author} {\bibfnamefont {P.}~\bibnamefont
  {Landshoff}},\ }\href {\doibase 10.1016/0370-2693(92)90832-O} {\bibfield
  {journal} {\bibinfo  {journal} {Phys. Lett. B}\ }\textbf {\bibinfo {volume}
  {296}},\ \bibinfo {pages} {227} (\bibinfo {year} {1992})},\ \Eprint
  {http://arxiv.org/abs/hep-ph/9209205} {arXiv:hep-ph/9209205} \BibitemShut
  {NoStop}%
\bibitem [{\citenamefont {Sjostrand}\ \emph {et~al.}(2001)\citenamefont
  {Sjostrand}, \citenamefont {Lonnblad},\ and\ \citenamefont
  {Mrenna}}]{Sjostrand:2001yu}%
  \BibitemOpen
  \bibfield  {author} {\bibinfo {author} {\bibfnamefont {T.}~\bibnamefont
  {Sjostrand}}, \bibinfo {author} {\bibfnamefont {L.}~\bibnamefont {Lonnblad}},
  \ and\ \bibinfo {author} {\bibfnamefont {S.}~\bibnamefont {Mrenna}},\
  }\href@noop {} {\  (\bibinfo {year} {2001})},\ \Eprint
  {http://arxiv.org/abs/hep-ph/0108264} {arXiv:hep-ph/0108264} \BibitemShut
  {NoStop}%
\bibitem [{\citenamefont {Falter}\ \emph {et~al.}(2004)\citenamefont {Falter},
  \citenamefont {Cassing}, \citenamefont {Gallmeister},\ and\ \citenamefont
  {Mosel}}]{Falter:2004uc}%
  \BibitemOpen
  \bibfield  {author} {\bibinfo {author} {\bibfnamefont {T.}~\bibnamefont
  {Falter}}, \bibinfo {author} {\bibfnamefont {W.}~\bibnamefont {Cassing}},
  \bibinfo {author} {\bibfnamefont {K.}~\bibnamefont {Gallmeister}}, \ and\
  \bibinfo {author} {\bibfnamefont {U.}~\bibnamefont {Mosel}},\ }\href
  {\doibase 10.1103/PhysRevC.70.054609} {\bibfield  {journal} {\bibinfo
  {journal} {Phys. Rev. C}\ }\textbf {\bibinfo {volume} {70}},\ \bibinfo
  {pages} {054609} (\bibinfo {year} {2004})},\ \Eprint
  {http://arxiv.org/abs/nucl-th/0406023} {arXiv:nucl-th/0406023} \BibitemShut
  {NoStop}%
\bibitem [{\citenamefont {Hunt}\ and\ \citenamefont
  {Manley}(2019)}]{Hunt:2018wqz}%
  \BibitemOpen
  \bibfield  {author} {\bibinfo {author} {\bibfnamefont {B.}~\bibnamefont
  {Hunt}}\ and\ \bibinfo {author} {\bibfnamefont {D.}~\bibnamefont {Manley}},\
  }\href {\doibase 10.1103/PhysRevC.99.055205} {\bibfield  {journal} {\bibinfo
  {journal} {Phys. Rev. C}\ }\textbf {\bibinfo {volume} {99}},\ \bibinfo
  {pages} {055205} (\bibinfo {year} {2019})},\ \Eprint
  {http://arxiv.org/abs/1810.13086} {arXiv:1810.13086 [nucl-ex]} \BibitemShut
  {NoStop}%
\bibitem [{\citenamefont {Bratkovskaya}\ and\ \citenamefont
  {Cassing}(2008)}]{Bratkovskaya:2007jk}%
  \BibitemOpen
  \bibfield  {author} {\bibinfo {author} {\bibfnamefont {E.}~\bibnamefont
  {Bratkovskaya}}\ and\ \bibinfo {author} {\bibfnamefont {W.}~\bibnamefont
  {Cassing}},\ }\href {\doibase 10.1016/j.nuclphysa.2008.04.004} {\bibfield
  {journal} {\bibinfo  {journal} {Nucl. Phys. A}\ }\textbf {\bibinfo {volume}
  {807}},\ \bibinfo {pages} {214} (\bibinfo {year} {2008})},\ \Eprint
  {http://arxiv.org/abs/0712.0635} {arXiv:0712.0635 [nucl-th]} \BibitemShut
  {NoStop}%
\bibitem [{\citenamefont {Agakishiev}\ \emph {et~al.}(2010)\citenamefont
  {Agakishiev} \emph {et~al.}}]{Agakishiev:2009yf}%
  \BibitemOpen
  \bibfield  {author} {\bibinfo {author} {\bibfnamefont {G.}~\bibnamefont
  {Agakishiev}} \emph {et~al.} (\bibinfo {collaboration} {HADES}),\ }\href
  {\doibase 10.1016/j.physletb.2010.05.010} {\bibfield  {journal} {\bibinfo
  {journal} {Phys. Lett.}\ }\textbf {\bibinfo {volume} {B690}},\ \bibinfo
  {pages} {118} (\bibinfo {year} {2010})},\ \Eprint
  {http://arxiv.org/abs/0910.5875} {arXiv:0910.5875 [nucl-ex]} \BibitemShut
  {NoStop}%
\bibitem [{\citenamefont {Agakishiev}\ \emph
  {et~al.}(2012{\natexlab{a}})\citenamefont {Agakishiev} \emph
  {et~al.}}]{Agakishiev:2012tc}%
  \BibitemOpen
  \bibfield  {author} {\bibinfo {author} {\bibfnamefont {G.}~\bibnamefont
  {Agakishiev}} \emph {et~al.} (\bibinfo {collaboration} {HADES}),\ }\href
  {\doibase 10.1103/PhysRevC.85.054005} {\bibfield  {journal} {\bibinfo
  {journal} {Phys. Rev. C}\ }\textbf {\bibinfo {volume} {85}},\ \bibinfo
  {pages} {054005} (\bibinfo {year} {2012}{\natexlab{a}})},\ \Eprint
  {http://arxiv.org/abs/1203.2549} {arXiv:1203.2549 [nucl-ex]} \BibitemShut
  {NoStop}%
\bibitem [{\citenamefont {Agakishiev}\ \emph
  {et~al.}(2012{\natexlab{b}})\citenamefont {Agakishiev} \emph
  {et~al.}}]{HADES:2011ab}%
  \BibitemOpen
  \bibfield  {author} {\bibinfo {author} {\bibfnamefont {G.}~\bibnamefont
  {Agakishiev}} \emph {et~al.} (\bibinfo {collaboration} {HADES}),\ }\href
  {\doibase 10.1140/epja/i2012-12064-y} {\bibfield  {journal} {\bibinfo
  {journal} {Eur. Phys. J.}\ }\textbf {\bibinfo {volume} {A48}},\ \bibinfo
  {pages} {64} (\bibinfo {year} {2012}{\natexlab{b}})},\ \Eprint
  {http://arxiv.org/abs/1112.3607} {arXiv:1112.3607 [nucl-ex]} \BibitemShut
  {NoStop}%
\bibitem [{\citenamefont {Galatyuk}()}]{Galatyuk_priv}%
  \BibitemOpen
  \bibfield  {author} {\bibinfo {author} {\bibfnamefont {T.}~\bibnamefont
  {Galatyuk}},\ }\href@noop {} {}\bibinfo {note} {{private
  communication}}\BibitemShut {NoStop}%
\bibitem [{\citenamefont {Machleidt}\ \emph {et~al.}(1987)\citenamefont
  {Machleidt}, \citenamefont {Holinde},\ and\ \citenamefont
  {Elster}}]{Machleidt:1987hj}%
  \BibitemOpen
  \bibfield  {author} {\bibinfo {author} {\bibfnamefont {R.}~\bibnamefont
  {Machleidt}}, \bibinfo {author} {\bibfnamefont {K.}~\bibnamefont {Holinde}},
  \ and\ \bibinfo {author} {\bibfnamefont {C.}~\bibnamefont {Elster}},\ }\href
  {\doibase 10.1016/S0370-1573(87)80002-9} {\bibfield  {journal} {\bibinfo
  {journal} {Phys. Rept.}\ }\textbf {\bibinfo {volume} {149}},\ \bibinfo
  {pages} {1} (\bibinfo {year} {1987})}\BibitemShut {NoStop}%
\bibitem [{\citenamefont {Frankfurt}\ and\ \citenamefont
  {Strikman}(1979)}]{Frankfurt:1977vc}%
  \BibitemOpen
  \bibfield  {author} {\bibinfo {author} {\bibfnamefont {L.}~\bibnamefont
  {Frankfurt}}\ and\ \bibinfo {author} {\bibfnamefont {M.}~\bibnamefont
  {Strikman}},\ }\href {\doibase 10.1016/0550-3213(79)90018-X} {\bibfield
  {journal} {\bibinfo  {journal} {Nucl. Phys. B}\ }\textbf {\bibinfo {volume}
  {148}},\ \bibinfo {pages} {107} (\bibinfo {year} {1979})}\BibitemShut
  {NoStop}%
\bibitem [{\citenamefont {Frankfurt}\ and\ \citenamefont
  {Strikman}(1981)}]{Frankfurt:1981mk}%
  \BibitemOpen
  \bibfield  {author} {\bibinfo {author} {\bibfnamefont {L.}~\bibnamefont
  {Frankfurt}}\ and\ \bibinfo {author} {\bibfnamefont {M.}~\bibnamefont
  {Strikman}},\ }\href {\doibase 10.1016/0370-1573(81)90129-0} {\bibfield
  {journal} {\bibinfo  {journal} {Phys. Rept.}\ }\textbf {\bibinfo {volume}
  {76}},\ \bibinfo {pages} {215} (\bibinfo {year} {1981})}\BibitemShut
  {NoStop}%
\bibitem [{\citenamefont {Pachmayer}(2008)}]{Pachmayer:2008}%
  \BibitemOpen
  \bibfield  {author} {\bibinfo {author} {\bibfnamefont {Y.}~\bibnamefont
  {Pachmayer}},\ }\emph {\bibinfo {title} {{Dielektronenproduktion in $^{12}$C
  + $^{12}$C Kollisionen bei 1 GeV pro Nukleon}}},\ \href@noop {} {Ph.D.
  thesis},\ \bibinfo  {school} {Frankfurt U.} (\bibinfo {year} {2008}),\
  \bibinfo {note}
  {\url{http://publikationen.ub.uni-frankfurt.de/frontdoor/index/index/docId/5895}}\BibitemShut
  {NoStop}%
\bibitem [{\citenamefont {Sudol}(2007)}]{Sudol:2007}%
  \BibitemOpen
  \bibfield  {author} {\bibinfo {author} {\bibfnamefont {M.}~\bibnamefont
  {Sudol}},\ }\emph {\bibinfo {title} {{Measurement of low-mass $e^+ e^-$ pair
  production in 2 A GeV C-C collisions with HADES}}},\ \href@noop {} {Ph.D.
  thesis},\ \bibinfo  {school} {Frankfurt U.} (\bibinfo {year} {2007}),\
  \bibinfo {note}
  {\url{http://publikationen.ub.uni-frankfurt.de/frontdoor/index/index/docId/5894}}\BibitemShut
  {NoStop}%
\bibitem [{\citenamefont {Bratkovskaya}\ \emph {et~al.}(2013)\citenamefont
  {Bratkovskaya}, \citenamefont {Aichelin}, \citenamefont {Thomere},
  \citenamefont {Vogel},\ and\ \citenamefont {Bleicher}}]{Bratkovskaya:2013vx}%
  \BibitemOpen
  \bibfield  {author} {\bibinfo {author} {\bibfnamefont {E.}~\bibnamefont
  {Bratkovskaya}}, \bibinfo {author} {\bibfnamefont {J.}~\bibnamefont
  {Aichelin}}, \bibinfo {author} {\bibfnamefont {M.}~\bibnamefont {Thomere}},
  \bibinfo {author} {\bibfnamefont {S.}~\bibnamefont {Vogel}}, \ and\ \bibinfo
  {author} {\bibfnamefont {M.}~\bibnamefont {Bleicher}},\ }\href {\doibase
  10.1103/PhysRevC.87.064907} {\bibfield  {journal} {\bibinfo  {journal} {Phys.
  Rev. C}\ }\textbf {\bibinfo {volume} {87}},\ \bibinfo {pages} {064907}
  (\bibinfo {year} {2013})},\ \Eprint {http://arxiv.org/abs/1301.0786}
  {arXiv:1301.0786 [nucl-th]} \BibitemShut {NoStop}%
\bibitem [{\citenamefont {Endres}\ \emph {et~al.}(2016)\citenamefont {Endres},
  \citenamefont {van Hees},\ and\ \citenamefont {Bleicher}}]{Endres:2015egk}%
  \BibitemOpen
  \bibfield  {author} {\bibinfo {author} {\bibfnamefont {S.}~\bibnamefont
  {Endres}}, \bibinfo {author} {\bibfnamefont {H.}~\bibnamefont {van Hees}}, \
  and\ \bibinfo {author} {\bibfnamefont {M.}~\bibnamefont {Bleicher}},\ }\href
  {\doibase 10.1103/PhysRevC.93.054901} {\bibfield  {journal} {\bibinfo
  {journal} {Phys. Rev. C}\ }\textbf {\bibinfo {volume} {93}},\ \bibinfo
  {pages} {054901} (\bibinfo {year} {2016})},\ \Eprint
  {http://arxiv.org/abs/1512.06549} {arXiv:1512.06549 [nucl-th]} \BibitemShut
  {NoStop}%
\bibitem [{\citenamefont {Adamczewski-Musch}\ \emph {et~al.}(2020)\citenamefont
  {Adamczewski-Musch} \emph {et~al.}}]{Adamczewski-Musch:2020vrg}%
  \BibitemOpen
  \bibfield  {author} {\bibinfo {author} {\bibfnamefont {J.}~\bibnamefont
  {Adamczewski-Musch}} \emph {et~al.} (\bibinfo {collaboration} {HADES}),\
  }\href {\doibase 10.1140/epja/s10050-020-00237-2} {\bibfield  {journal}
  {\bibinfo  {journal} {Eur. Phys. J. A}\ }\textbf {\bibinfo {volume} {56}},\
  \bibinfo {pages} {259} (\bibinfo {year} {2020})},\ \Eprint
  {http://arxiv.org/abs/2005.08774} {arXiv:2005.08774 [nucl-ex]} \BibitemShut
  {NoStop}%
\bibitem [{\citenamefont {Staudenmaier}\ \emph {et~al.}(2020)\citenamefont
  {Staudenmaier}, \citenamefont {K\"ubler},\ and\ \citenamefont
  {Elfner}}]{Staudenmaier:2020xqr}%
  \BibitemOpen
  \bibfield  {author} {\bibinfo {author} {\bibfnamefont {J.}~\bibnamefont
  {Staudenmaier}}, \bibinfo {author} {\bibfnamefont {N.}~\bibnamefont
  {K\"ubler}}, \ and\ \bibinfo {author} {\bibfnamefont {H.}~\bibnamefont
  {Elfner}},\ }\href@noop {} {\  (\bibinfo {year} {2020})},\ \Eprint
  {http://arxiv.org/abs/2008.05813} {arXiv:2008.05813 [hep-ph]} \BibitemShut
  {NoStop}%
\bibitem [{\citenamefont {Adamczewski-Musch}\ \emph {et~al.}(2017)\citenamefont
  {Adamczewski-Musch} \emph {et~al.}}]{Adamczewski-Musch:2017oij}%
  \BibitemOpen
  \bibfield  {author} {\bibinfo {author} {\bibfnamefont {J.}~\bibnamefont
  {Adamczewski-Musch}} \emph {et~al.} (\bibinfo {collaboration} {HADES}),\
  }\href {\doibase 10.1140/epja/i2017-12341-3} {\bibfield  {journal} {\bibinfo
  {journal} {Eur. Phys. J. A}\ }\textbf {\bibinfo {volume} {53}},\ \bibinfo
  {pages} {149} (\bibinfo {year} {2017})},\ \Eprint
  {http://arxiv.org/abs/1703.08575} {arXiv:1703.08575 [nucl-ex]} \BibitemShut
  {NoStop}%
\bibitem [{\citenamefont {Agakishiev}\ \emph {et~al.}(2009)\citenamefont
  {Agakishiev} \emph {et~al.}}]{Agakishiev:2009zv}%
  \BibitemOpen
  \bibfield  {author} {\bibinfo {author} {\bibfnamefont {G.}~\bibnamefont
  {Agakishiev}} \emph {et~al.} (\bibinfo {collaboration} {HADES}),\ }\href
  {\doibase 10.1140/epja/i2008-10746-7} {\bibfield  {journal} {\bibinfo
  {journal} {Eur. Phys. J. A}\ }\textbf {\bibinfo {volume} {40}},\ \bibinfo
  {pages} {45} (\bibinfo {year} {2009})},\ \Eprint
  {http://arxiv.org/abs/0902.4377} {arXiv:0902.4377 [nucl-ex]} \BibitemShut
  {NoStop}%
\bibitem [{\citenamefont {Song}\ and\ \citenamefont {Ko}(2015)}]{Song:2015hua}%
  \BibitemOpen
  \bibfield  {author} {\bibinfo {author} {\bibfnamefont {T.}~\bibnamefont
  {Song}}\ and\ \bibinfo {author} {\bibfnamefont {C.~M.}\ \bibnamefont {Ko}},\
  }\href {\doibase 10.1103/PhysRevC.91.014901} {\bibfield  {journal} {\bibinfo
  {journal} {Phys. Rev. C}\ }\textbf {\bibinfo {volume} {91}},\ \bibinfo
  {pages} {014901} (\bibinfo {year} {2015})}\BibitemShut {NoStop}%
\bibitem [{\citenamefont {Adamczewski-Musch}\ \emph
  {et~al.}(2018{\natexlab{a}})\citenamefont {Adamczewski-Musch} \emph
  {et~al.}}]{Adamczewski-Musch:2017sdk}%
  \BibitemOpen
  \bibfield  {author} {\bibinfo {author} {\bibfnamefont {J.}~\bibnamefont
  {Adamczewski-Musch}} \emph {et~al.} (\bibinfo {collaboration} {HADES}),\
  }\href {\doibase 10.1140/epja/i2018-12513-7} {\bibfield  {journal} {\bibinfo
  {journal} {Eur. Phys. J. A}\ }\textbf {\bibinfo {volume} {54}},\ \bibinfo
  {pages} {85} (\bibinfo {year} {2018}{\natexlab{a}})},\ \Eprint
  {http://arxiv.org/abs/1712.07993} {arXiv:1712.07993 [nucl-ex]} \BibitemShut
  {NoStop}%
\bibitem [{\citenamefont {Harabasz}\ \emph {et~al.}(2020)\citenamefont
  {Harabasz}, \citenamefont {Florkowski}, \citenamefont {Galatyuk},
  \citenamefont {Gumberidze}, \citenamefont {Ryblewski}, \citenamefont
  {Salabura},\ and\ \citenamefont {Stroth}}]{Harabasz:2020sei}%
  \BibitemOpen
  \bibfield  {author} {\bibinfo {author} {\bibfnamefont {S.}~\bibnamefont
  {Harabasz}}, \bibinfo {author} {\bibfnamefont {W.}~\bibnamefont
  {Florkowski}}, \bibinfo {author} {\bibfnamefont {T.}~\bibnamefont
  {Galatyuk}}, \bibinfo {author} {\bibfnamefont {M.}~\bibnamefont
  {Gumberidze}}, \bibinfo {author} {\bibfnamefont {R.}~\bibnamefont
  {Ryblewski}}, \bibinfo {author} {\bibfnamefont {P.}~\bibnamefont {Salabura}},
  \ and\ \bibinfo {author} {\bibfnamefont {J.}~\bibnamefont {Stroth}},\
  }\href@noop {} {\  (\bibinfo {year} {2020})},\ \Eprint
  {http://arxiv.org/abs/2003.12992} {arXiv:2003.12992 [nucl-th]} \BibitemShut
  {NoStop}%
\bibitem [{\citenamefont {Neubert}\ and\ \citenamefont
  {Botvina}(2000)}]{Neubert:1999sv}%
  \BibitemOpen
  \bibfield  {author} {\bibinfo {author} {\bibfnamefont {W.}~\bibnamefont
  {Neubert}}\ and\ \bibinfo {author} {\bibfnamefont {A.}~\bibnamefont
  {Botvina}},\ }\href {\doibase 10.1007/s100500050016} {\bibfield  {journal}
  {\bibinfo  {journal} {Eur. Phys. J. A}\ }\textbf {\bibinfo {volume} {7}},\
  \bibinfo {pages} {101} (\bibinfo {year} {2000})},\ \Eprint
  {http://arxiv.org/abs/nucl-th/9912019} {arXiv:nucl-th/9912019} \BibitemShut
  {NoStop}%
\bibitem [{\citenamefont {Behnke}(2016)}]{Behnke:2016}%
  \BibitemOpen
  \bibfield  {author} {\bibinfo {author} {\bibfnamefont {C.}~\bibnamefont
  {Behnke}},\ }\emph {\bibinfo {title} {{Reconstruction of $\pi^0$ and $\eta$
  Mesons via Conversion in $^{197}$Au + $^{197}$Au at 1.23 GeV/u with the HADES
  Spectrometer}}},\ \href@noop {} {Ph.D. thesis},\ \bibinfo  {school}
  {Frankfurt U.} (\bibinfo {year} {2016}),\ \bibinfo {note}
  {\url{http://publikationen.ub.uni-frankfurt.de/frontdoor/index/index/docId/44136}}\BibitemShut
  {NoStop}%
\bibitem [{\citenamefont {Adamczewski-Musch}\ \emph
  {et~al.}(2019{\natexlab{b}})\citenamefont {Adamczewski-Musch} \emph
  {et~al.}}]{Adamczewski-Musch:2018xwg}%
  \BibitemOpen
  \bibfield  {author} {\bibinfo {author} {\bibfnamefont {J.}~\bibnamefont
  {Adamczewski-Musch}} \emph {et~al.} (\bibinfo {collaboration} {HADES}),\
  }\href {\doibase 10.1016/j.physletb.2019.03.065} {\bibfield  {journal}
  {\bibinfo  {journal} {Phys. Lett. B}\ }\textbf {\bibinfo {volume} {793}},\
  \bibinfo {pages} {457} (\bibinfo {year} {2019}{\natexlab{b}})},\ \Eprint
  {http://arxiv.org/abs/1812.07304} {arXiv:1812.07304 [nucl-ex]} \BibitemShut
  {NoStop}%
\bibitem [{\citenamefont {Adamczewski-Musch}\ \emph
  {et~al.}(2018{\natexlab{b}})\citenamefont {Adamczewski-Musch} \emph
  {et~al.}}]{Adamczewski-Musch:2017rtf}%
  \BibitemOpen
  \bibfield  {author} {\bibinfo {author} {\bibfnamefont {J.}~\bibnamefont
  {Adamczewski-Musch}} \emph {et~al.} (\bibinfo {collaboration} {HADES}),\
  }\href {\doibase 10.1016/j.physletb.2018.01.048} {\bibfield  {journal}
  {\bibinfo  {journal} {Phys. Lett. B}\ }\textbf {\bibinfo {volume} {778}},\
  \bibinfo {pages} {403} (\bibinfo {year} {2018}{\natexlab{b}})},\ \Eprint
  {http://arxiv.org/abs/1703.08418} {arXiv:1703.08418 [nucl-ex]} \BibitemShut
  {NoStop}%
\bibitem [{\citenamefont {Welke}\ \emph {et~al.}(1988)\citenamefont {Welke},
  \citenamefont {Prakash}, \citenamefont {Kuo}, \citenamefont {Das~Gupta},\
  and\ \citenamefont {Gale}}]{Welke:1988zz}%
  \BibitemOpen
  \bibfield  {author} {\bibinfo {author} {\bibfnamefont {G.}~\bibnamefont
  {Welke}}, \bibinfo {author} {\bibfnamefont {M.}~\bibnamefont {Prakash}},
  \bibinfo {author} {\bibfnamefont {T.}~\bibnamefont {Kuo}}, \bibinfo {author}
  {\bibfnamefont {S.}~\bibnamefont {Das~Gupta}}, \ and\ \bibinfo {author}
  {\bibfnamefont {C.}~\bibnamefont {Gale}},\ }\href {\doibase
  10.1103/PhysRevC.38.2101} {\bibfield  {journal} {\bibinfo  {journal} {Phys.
  Rev. C}\ }\textbf {\bibinfo {volume} {38}},\ \bibinfo {pages} {2101}
  (\bibinfo {year} {1988})}\BibitemShut {NoStop}%
\bibitem [{\citenamefont {Ericson}\ and\ \citenamefont
  {Weise}(1988)}]{Ericson:1988gk}%
  \BibitemOpen
  \bibfield  {author} {\bibinfo {author} {\bibfnamefont {T.~E.~O.}\
  \bibnamefont {Ericson}}\ and\ \bibinfo {author} {\bibfnamefont
  {W.}~\bibnamefont {Weise}},\ }\href@noop {} {\emph {\bibinfo {title} {{Pions
  and Nuclei}}}},\ Vol.~\bibinfo {volume} {74}\ (\bibinfo  {publisher}
  {Clarendon Press},\ \bibinfo {address} {Oxford, UK},\ \bibinfo {year}
  {1988})\BibitemShut {NoStop}%
\bibitem [{\citenamefont {Oset}\ and\ \citenamefont
  {Salcedo}(1987)}]{Oset:1987re}%
  \BibitemOpen
  \bibfield  {author} {\bibinfo {author} {\bibfnamefont {E.}~\bibnamefont
  {Oset}}\ and\ \bibinfo {author} {\bibfnamefont {L.}~\bibnamefont {Salcedo}},\
  }\href {\doibase 10.1016/0375-9474(87)90185-0} {\bibfield  {journal}
  {\bibinfo  {journal} {Nucl. Phys. A}\ }\textbf {\bibinfo {volume} {468}},\
  \bibinfo {pages} {631} (\bibinfo {year} {1987})}\BibitemShut {NoStop}%
\end{thebibliography}%

\end{document}